\documentclass[preprint,twocolumn] {emulateapj}
\usepackage{graphicx,amssymb,natbib}
\usepackage{times,datetime}

\shorttitle{$\alpha.40$: the Galaxy Population}
\shortauthors{S. Huang et al.}

\begin{document}

\title{The Arecibo Legacy Fast ALFA Survey: The Galaxy Population Detected by ALFALFA}
\author{Shan Huang, Martha P. Haynes, Riccardo Giovanelli}
\affil{Center for Radiophysics and Space Research, Space Sciences Building, Cornell University, Ithaca, NY 14853.}
\email{shan@astro.cornell.edu, haynes@astro.cornell.edu, riccardo@astro.cornell.edu}
\author{Jarle Brinchmann}\affil{Sterrewacht Leiden, Leiden University, NL-2300 RA Leiden, 
       The Netherlands} \email{jarle@strw.leidenuniv.nl}


\begin{abstract} 
Making use of HI 21 cm line measurements from the ALFALFA survey ($\alpha.40$)
and photometry from the Sloan Digital Sky Survey (SDSS) 
and GALEX, we investigate the global scaling relations and 
fundamental planes linking stars and gas for a sample of 9417 common galaxies: 
the $\alpha.40$-SDSS-GALEX sample. In addition to their 
HI properties derived from the ALFALFA dataset, stellar masses ($M_*$) 
and star formation rates (SFRs) are derived from fitting the 
UV-optical spectral energy distributions. 96\% of the $\alpha.40$-SDSS-GALEX 
galaxies belong to the blue cloud, with the average 
gas fraction $f_{HI}\equiv M_{HI}/M_* \sim 1.5$. A transition in SF properties
is found whereby below $M_* \sim 10^{9.5}M_\odot$,
the slope of the star forming sequence changes, the dispersion in the specific 
star formation rate (SSFR) distribution increases and the star formation 
efficiency (SFE) mildly increases with $M_*$. 
The evolutionary track in the SSFR--$M_*$ diagram, as well as that in the color 
magnitude diagram are linked to the HI content; below this transition
mass, the star formation is regulated strongly by the HI. Comparison of 
HI- and optically-selected samples over the same restricted volume
shows that the HI-selected population is less evolved 
and has overall higher SFR and SSFR at a given stellar mass, 
but lower SFE and extinction, suggesting
either that a bottleneck exists in the HI to H$_2$ conversion, or that the 
process of SF in the very HI-dominated galaxies obeys an unusual, low
efficiency star formation law. A trend is found that, for a given stellar mass,
high gas fraction galaxies reside preferentially in dark matter halos 
with high spin parameters. Because it represents a full census of HI-bearing
galaxies at $z\sim0$, the scaling relations and fundamental planes derived
for the ALFALFA population can be used to assess the HI detection rate by future
blind HI surveys and intensity mapping experiments at higher redshift.
\end{abstract}


\section{Introduction}

In the last decade, the galaxy catalogs contributed by legacy programs like
the Sloan Digital Sky Survey (SDSS) and the Galaxy Evolution Explorer (GALEX) satellite 
extragalactic surveys
have enabled us to quantify properties associated with the stellar populations 
of galaxies in the local universe. Through their statistically-based 
insight into the stellar component and the interrelationship of the physical 
parameters, these surveys have provided quantitative clues of importance to our understanding of 
the formation and evolution of galaxies \citep[e.g.][]{Brinchmann2004, Salim2007}. 
The bimodal distribution evident in the 
color-magnitude diagram \citep{Baldry2004, Schiminovich2007} suggests
a likely evolution scenario whereby galaxies in the blue cloud form stars vigorously
and grow through mergers and later, after depleting their gas reservoirs,
then migrate to the red sequence. This picture is also supported by tracers of
the star formation history (SFH): e.g., the specific star formation rate (SSFR), 
the star formation rate (SFR) per unit stellar mass, is seen to vary with the total stellar mass 
\citep{Brinchmann2004}. The star-forming sequence (high SSFR) is associated with 
actively star forming blue cloud galaxies. Indeed, the 
stellar mass appears to be the crucial quantity governing the star formation (SF) along 
this sequence. In the absence of mergers or other events that trigger a starburst, blue 
galaxies on the sequence evolve towards higher stellar mass and lower SSFR, and 
eventually become red and dead. 

In statistical terms, all surveys are biased by the properties that define them. For
example, optical surveys are biased in terms of optical flux and possibly surface brightness.
The SDSS legacy galaxy redshift sample has an apparent $r$-band Petrosian magnitude limit of 17.77, 
as well as a surface brightness limit of 23.0 mag arcsec$^{-2}$ at the half light radius in $r$
\citep{Strauss2002}. In contrast, blind HI surveys are unbiased by optical characteristics 
but have their own limitations in terms of HI line emission sensitivity, usually as a function 
of HI line width \citep{Martin2010, Haynes2011}. Because
$M_{HI}/L_{opt}$ increases with decreasing $L_{opt}$, HI-selected samples are more inclusive 
of star-forming galaxies than optical samples of similar depth. Indeed, since almost all 
star-forming galaxies contain neutral gas, an HI-selected sample can approach a full census 
of star-forming galaxies. 
For example, \citet{West2010} demonstrated the HI-selection identifies galaxies with lower 
surface brightness, smaller absolute magnitudes, bluer colors and smaller stellar masses 
than those used in typical SDSS studies. 
However, the limitation with an HI-selected sample is that it will miss the early type galaxies, 
which contain very little neutral gas \citep{Garcia-Appadoo2009}. Furthermore, analysis of
the spatial correlation function $\xi(r)$ shows that the HI-selected galaxies represent
the least clustered population on small scales \citep{Martin2012}, a fact important for
interpreting the results of future HI intensity mapping experiments.

Exploiting the mapping capability of the Arecibo L-band feed array (ALFA) and the sensitivity
of the Arecibo 305~m antenna,
the Arecibo Legacy Fast ALFA (ALFALFA) survey \citep{Giovanelli2005a, Giovanelli2005b} is an ongoing blind HI survey, 
aimed at mapping $\sim$7000 square degrees of high galactic latitude 
sky between $0^\circ$ and $+36^\circ$ in declination. When complete, the survey will detect 
more than 30,000 galaxies out to redshift of 0.06 with a median recessional velocity, $cz$, 
of $\sim 8200~{\rm km~s^{-1}}$ \citep{Haynes2011}.
Compared to the HI Parkes All Sky Survey \citep[HIPASS:][]{Barnes2001, Meyer2004, Wong2006},
ALFALFA is 8 times more sensitive, its spectral coverage extends over 1.6 times the
bandwidth of HIPASS, and the angular resolution of ALFA is 4 times better than that
of the Parkes multifeed receiver. The combination of sensitivity and spectral bandwidth enables
ALFALFA to detect thousands of massive HI disks with $M_{HI} > 10^{10}~{\rm M_\odot}$.
In addition, the Arecibo spectral backend yields a finer velocity resolution than characterized
HIPASS, making it possible to detect sources with HI line widths as narrow 
as $\sim 15~{\rm km~s^{-1}}$. As discussed in \citet{Haynes2011} and earlier papers in the
ALFALFA series, the median centroiding accuracy of the HI sources is $\sim 20^{\prime\prime}$
allowing the identification of most probable optical counterparts in 97\% of cases. 
ALFALFA provides the first full census of HI-bearing objects over a cosmologically significant 
volume of the local universe \citep{Martin2010} so that its population includes the very rare 
objects missing from the smaller volumes sampled by previous blind HI surveys.
It enables the study of the characteristics of HI-selected galaxies 
in comparison with the galaxy populations included in the SDSS and GALEX surveys. 

The 2011 ALFALFA catalog, `$\alpha.40$', covers $\sim 40 \%$ of the final targeted sky area
\citep{Haynes2011}, giving HI masses, systemic velocities and HI line widths for $\sim$16000 
high quality detections (\S\ref{HI}). In addition to the HI
line parameters, the $\alpha.40$ catalog also includes an assignment of the most probable optical 
counterpart (OC) to each HI line detection. \citet{Haynes2011} discuss the process of assigning 
OCs to the HI sources, and where the footprints of the surveys
overlap, the HI sources are cross-referenced to the SDSS Data Release 7 \citep[DR7:][]{Abazajian2009},
permitting the derivation of properties associated with the stellar components of the
HI-bearing galaxies detected by ALFALFA. As shown by \citet{Haynes2011}, the $\alpha.40$ population
is highly biased against red-sequence objects. 

In this paper, we investigate further the nature of the
stellar counterparts of the $\alpha.40$ HI line detections, adding to the optical SDSS data
photometric measures from the GALEX catalog (\S\ref{xmatch}). 
By combining the measurements of the gaseous and stellar components, we characterize the 
$\alpha.40$ galaxy population and study in \S\ref{SED} global trends within it. 
SED-fitting to the seven bands from the UV to optical is applied to derive the principal
stellar properties, including the stellar masses (\S\ref{MHIMs}) and SF (\S\ref{SFS}). 
To understand better the characteristics of the gas-bearing galaxies and the 
potential bias associated with HI-selection, we define a volume-limited
sub-sample extracted from the $\alpha.40$ catalog with a similar one extracted from the SDSS-DR7 
(\S\ref{sample}) and compare the two in \S\ref{comp}
in terms of survey depth (\S\ref{basic}), extinction (\S\ref{ext}), color (\S\ref{CMD}), 
SF behavior (\S\ref{SF}), etc. 
The empirical distribution of the halo spin parameter is derived in \S\ref{lambda}, which 
suggests that the HI-selected galaxies favor high spin parameter halos. 
Our conclusions are summarized in \S\ref{con}. 

Throughout this paper, we adopt a reduced Hubble constant $h = H_0/(100~{\rm km~s^{-1}~Mpc^{-1}}) = 0.7$. 
A \citet{Chabrier2003} IMF is adopted. 

\section{Sample and Data}

\subsection{ALFALFA parent sample}
\label{HI}

The HI parent sample used here is drawn from the $\alpha.40$ catalog presented
in \citet{Haynes2011}. The $\alpha.40$ catalog covers
two regions in the Spring sky (i.e., the Virgo direction, $\rm 7^h 30^m < RA < 16^h30^m$, 
$\rm 4^\circ < Dec < 16^\circ$ and $\rm 24^\circ < Dec < 32^\circ$) and two in the
the Fall sky (i.e., the anti-Virgo direction, $\rm 22^h < RA < 3^h$, $\rm 14^\circ < Dec < 16^\circ$ 
and $\rm 24^\circ < Dec < 28^\circ$). As discussed in \citet{Haynes2011}, ALFALFA HI line 
detections are categorized
according to their signal-to-noise ratio (S/N) and corresponding reliability. Code 1
sources have S/N $\gtrsim$ 6.5 and are highly reliable. Another set of entries, designated
as ``code 2'' sources, or ``priors'', have lower S/N but coincide with a likely OC 
at the same redshift. Most of these sources are likely to be real \citep{Haynes2011}.
The catalog includes 15041 extragalactic HI sources,
11941 with code 1 and 3100 with code 2. 
In this paper, we consider both the code 1 and 2 $\alpha.40$ detections and adopt
the HI measures, distances and HI masses presented in the $\alpha.40$
presented by \citet{Haynes2011}. 
It is important to note that $\sim$ 70\% of the ALFALFA sources are new HI detections
proving that previous targeted surveys based on optical selection 
(magnitude, size and morphology), most notably the extensive collection
contained in the Cornell digital HI archive \citep{Springob2005}, missed large segments
of the local gas-bearing population. 

In particular, one of the most surprising results of ALFALFA to date is the richness of 
the high HI mass galaxy population \citep{Martin2010}. 
With its combination of sensitivity and depth, ALFALFA reveals that there still exists at $z\sim0$ 
a population of massive galaxies which retain massive HI ($M_{HI} >10^{10}~{\rm M_\odot}$) disks. 
Some, in fact, contain a dominant fraction of their baryons in HI gas. As the most HI massive local
galaxies, they are the $z\sim0$ analogs of the massive disks detected at $z\sim0.2$ \citep{Catinella2008} 
and those that will dominate the deep surveys being planned for even higher $z$ 
with the EVLA, APERTIF, ASKAP and MeerKAT and eventually the Square Kilometre Array (SKA). 
By design, ALFALFA provides a census
of HI bearing galaxies within a cosmologically significant volume over a wider
dynamic range of HI masses than previous studies. It thus serves as the reference
$z\sim0$ HI-selected population.

\subsection{Optical and UV counterparts of ALFALFA HI sources}
\label{xmatch}

In addition to the HI measurements, \citet{Haynes2011} attempt to identity the
most probable optical counterpart (OC) of each HI line source in the $\alpha.40$
catalog. Ancillary information such as redshift coincidence, angular size, color 
and morphology is used in making the OC assignment.
As discussed by \citet{Haynes2011}, the process is not perfect, particularly
for low signal-to-noise sources for which the centroiding accuracy can exceed
30$^{\prime\prime}$ and in regions of source confusion. Nonetheless, the
vast majority of OC assignments are probably valid and thus permit the comparison of the
stellar and gaseous components of the ALFALFA population. 

\subsubsection{SDSS photometry}
Towards this aim, \citet{Haynes2011} also provide a cross reference of the assigned OCs 
with the SDSS-DR7 \citep{Abazajian2009} where the two surveys have overlapping
sky footprints. The northern Fall region is not covered by the SDSS 
legacy imaging survey DR7. Of the
15041 HI source included in the $\alpha.40$
catalog, 
201 have no OC assigned by 
\citet{Haynes2011}, 
2310 lie outside the DR7 footprint, and 60 appear to be in the region of the SDSS imaging survey
but cannot be associated with an object in the SDSS photometric database.
In most of the latter cases, the OC is evident in the SDSS images but is projected
close to a bright foreground star or contaminated by its glare. 

As discussed by numerous authors e.g., \citet{West2010}, gas rich nearby galaxies 
are often blue and patchy with the result that their overall optical emission is shredded
among several photometric objects by the SDSS pipeline measurements.
The ALFALFA-SDSS cross-reference given in Table 3 of \citet{Haynes2011} 
includes a photometric code that identifies
objects with suspicious SDSS photometry, and we exclude such objects (293/12470) 
from the SED-fitting.  For the others, we require that model magnitudes are available
in all five SDSS bands ({\it u, g, r, i, z}) and retrieve the SDSS photometric parameters from the
standard SDSS-DR7 database. 

The current $\alpha.40$ catalog used here relies on the optical identifications and photometry
derived from the SDSS-DR7 database as cross-referenced in Table 3 of \citet{Haynes2011}. 
\citet{Blanton2011} have presented a new method for background
subtracting the SDSS imaging which they apply to the SDSS-III DR8 images. As ALFALFA progresses,
we will migrate to use of the improved SDSS pipeline. Especially for purposes of comparison with 
results obtained by other authors who have likewise relied on DR7, we retain the use of DR7
in the present work.

\subsubsection{GALEX photometry}

Using a similar approach, we have conducted a separate cross-match of the ALFALFA
OCs to the GALEX UV photometric catalog.
The imaging mode of the GALEX instrument surveys the sky simultaneously in two broad
bands, one in the FUV (effective wavelength of 1516$\rm \AA$) and a second in the NUV 
(effective wavelength of 2267$\rm \AA$). The GALEX field of view is $\sim1.2^\circ$ in 
diameter \citep{Morrissey2007}, although image quality deteriorates in the outer annulus
beyond a radius of $0.55^\circ$. We use the GALEX GR6 data release, with 
its improvements to flat fielding, adjustment to the photometric zero-point, etc. 
Given the poorer image resolution ($\sim 4.5^{\prime\prime}$ FWHM), compared to that
of the SDSS, as well as the lower UV source density, visual inspection 
shows that the GALEX DR6 pipeline measurements corresponding to the ALFALFA galaxies
suffer less from shredding issues. We have also validated that the GALEX pipeline photometry
is in close agreement with our own photometric reprocessing
of the GALEX images for a sample of dwarf galaxies detected by ALFALFA \citep{Huang2012}. 

The GALEX mission includes several survey modes that differ in their exposure time per tile: 
The All-sky Imaging Survey(AIS) is the shallowest ($\sim$100 sec), while the
Medium Imaging Survey (MIS, $\sim$1500 sec) is designed to maximize the coverage 
of the sky that is included in the SDSS. The latter also includes the Nearby Galaxy
Survey (NGS), a selected set of targeted fields of similar depth as the standard MIS. A third 
Deep Imaging Survey (DIS) is much deeper but covers only a small solid angle. In adopting
the GALEX counterparts, we give preference to sources extracted from the MIS, NGS or DIS, but
make use of the AIS for those objects not included in the deeper surveys.

To cross-match the ALFALFA OCs to the GALEX catalog, we first search the position of the
ALFALFA OCs for all GALEX neighbors within 36$^{\prime\prime}$. Objects close
to the GALEX field edge, i.e. with a distance from the field center $> 0.55^\circ$, 
are dropped to avoid duplications with objects in overlapping tiles and known
GALEX imaging artifact effects. To permit homogeneous SED-fitting to all 7 bands
({\it FUV, NUV, ugriz}), we require that UV images of comparable depth must exist 
for both of the UV bands; this criterion results in the adoption of AIS measurements 
where the MIS catalog is incomplete, e.g. because of the failure of the FUV detector. 
In other cases, matches are missed because the FUV and 
NUV sources may fail to coincide because of, e.g. astrometry error or shredding in one 
or both bands. Among the 14840 extragalactic ALFALFA sources with OCs
1828 (12.3\%) have no GALEX counterpart returned within 36$^{\prime\prime}$, either
because they lie outside of the GALEX footprint or have no detected UV emission in both bands.
516 (3.5\%) are excluded because they lie too close to a GALEX field edge and
1317 (8.9\%) are not matched because all neighbors are detected only in one band but not the other.
The remaining 11179 OCs are matched to the nearest neighbor in the GALEX catalog with
7752 (52.2\%) matched to UV sources found in the AIS and 
3427 (23.1\%) to ones in the MIS. The median separation between the coordinates of the OC
and the cross-matched GALEX object is only 1.6$^{\prime\prime}$. 

Although the addition of the GALEX UV photometry sets better constraints on the SF 
properties and dust extinction \citep{Salim2005}, the requirement that UV sources must be detected 
in both bands introduces an additional bias against non star-forming galaxies.
SED-fitting to the SDSS bands only \citep{Huang2012} demonstrates that 
13.5\% of the 
12156 galaxies in the $\alpha.40$-SDSS DR7 cross-match have 
$\log SSFR < 10^{-11}~{\rm yr^{-1}}$, and therefore belong to the quiescent population. 
This fraction drops to  
3.9\% of the 9417 $\alpha.40$ galaxies that have counterparts 
in both the SDSS and GALEX photometric catalogs. As we discuss in \S\ref{HIO}, the 
HI-selection produces the stronger bias against the red sequence.

\subsubsection{UV-to-optical colors of ALFALFA galaxies}
\label{NUV_r}

\begin{figure*}
\center{
\includegraphics[scale=1]{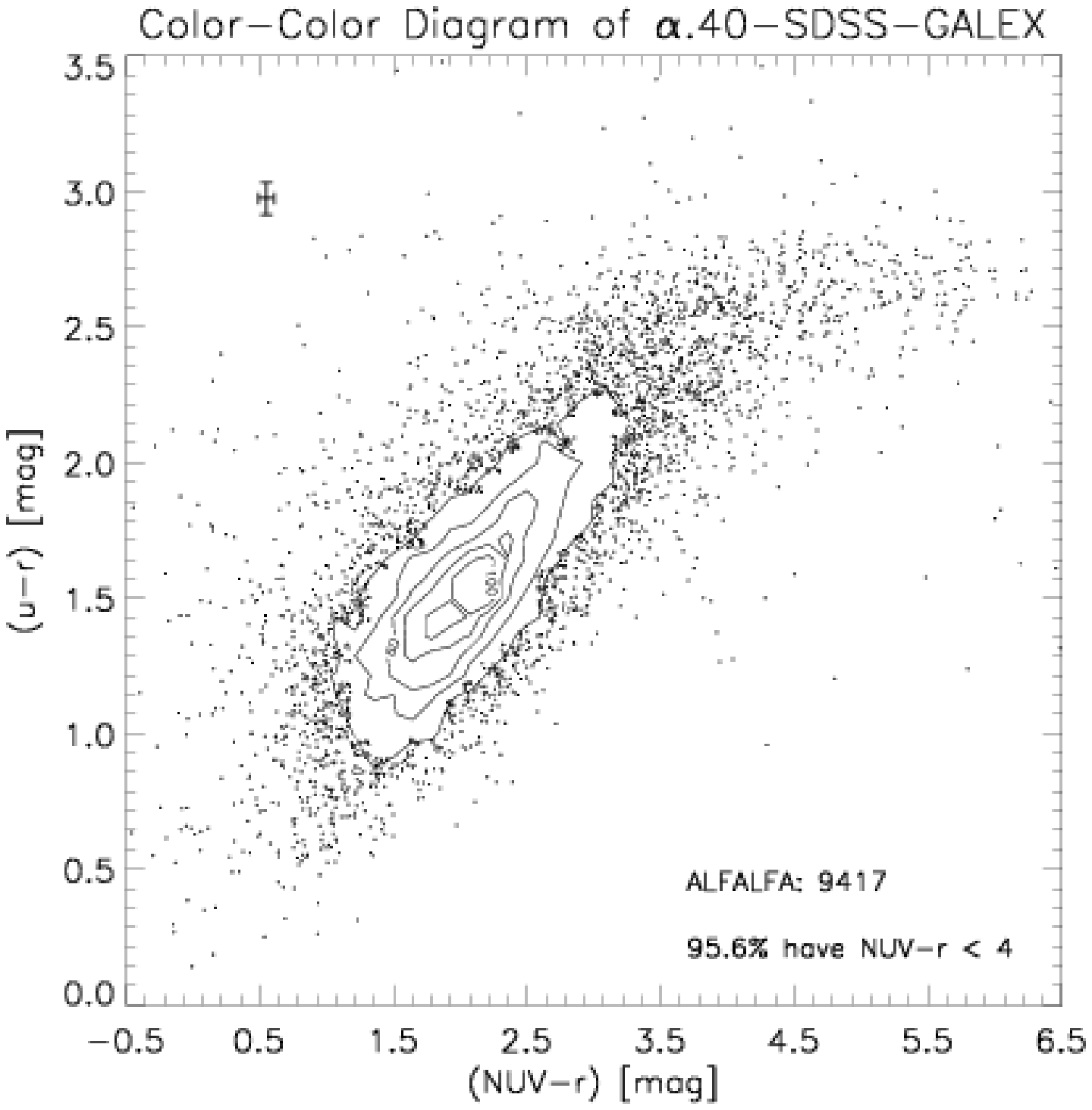}
}
\caption[]{UV-optical color-color diagram. 
Contours and points represent the $\alpha.40$-SDSS-GALEX common sample. 
The number density of galaxies in each gird cell is labeled on selected contours, 
e.g., the lowest contour level beyond which scatter points are plotted is 20 galaxies per grid cell. 
The grid size is shown by the interval of minor ticks on both axes. 
Among these 9417 galaxies, 96\% have $(NUV-r)<4$ and thus lie in the blue cloud by the criteria of 
\citet{Salim2007}. The ($NUV-r$) and ($u-r$) colors are well correlated among the blue galaxies, 
with a slope of 
$\delta(u-r)/\delta(NUV-r) \sim 0.6$, but the distribution flattens among the reddest population.
The ($NUV-r$) color serves as a stronger diagnostic of SFH than colors 
derived only from the optical bands. 
}
\label{fig:CCD}
\end{figure*}

Figure \ref{fig:CCD} presents the optical-UV color--color diagram derived for the 
9417 galaxies which have complete entries in all three of the
ALFALFA-$\alpha.40$, the SDSS-DR7 and GALEX-DR6 catalogs and for which the 7-band SED-fitting
produces a valid result (see \S\ref{SED}), denoted as the $\alpha.40$-SDSS-GALEX sample hereafter. 
Contours and points depict the distribution for the $\alpha.40$-SDSS-GALEX sample in
high and low number density regions respectively, 
with typical error bars shown in the upper left corner (pipeline magnitude errors). 
Colors hereafter are all corrected for Galactic extinction 
but not internal extinction (see \S\ref{ext}).
In the optical, the SDSS pipeline extinction-corrected values are used, while at
UV wavelengths, we adopt the $E(B-V)$ values based on the maps of \citet{Schlegel1998}, 
the \citet{Cardelli1989} extinction law with $R_V=A_V/E(B-V)=3.1$, 
and $A(\lambda)/E(B-V)=8.24$ for the FUV and 8.2 for the NUV, following \citet{Wyder2007}. 
Because of the bandwidth limit of ALFALFA, only low redshift galaxies with 
small K-correction are included, 
and we ignore this term in computing colors, i.e., no K-correction is applied. 

Because it contrasts the recent SF, as indicated by the UV light, with
the total past SF, as indicated by the optical light, the UV-to-optical color 
is a stronger diagnostic of SFH than colors derived from 
the optical bands only. This result is also evident in Figure \ref{fig:CCD}. 
Based on the distribution of their GALEX-SDSS matched catalog, \citet{Salim2007} 
define their blue cloud galaxies as those with $(NUV-r)<4$. Using the same cutoff, 
96\% of the $\alpha.40$-SDSS-GALEX galaxies lie on the blue side, suggesting that HI-selection
induces a strong preference for blue star-forming galaxies, or conversely, 
a strong bias against red sequence galaxies.
Bluewards of this division, the two colors are well correlated with a slope of 
$\delta(u-r)/\delta(NUV-r) \sim 0.6$, which is comparable to what was found by 
\citet{Wyder2007},  $\delta(u-r)/\delta(NUV-r) \sim 0.5$. 
Although only a small population (411) of galaxies appear redward of $(NUV-r)=4$, 
the ($u-r$) colors of the red objects increase less quickly with ($NUV-r$) and the
distribution of the reddest tail is nearly flat. As discussed in \S\ref{ext} 
the degeneracy of ($u-r$) among the red populations
is even more pronounced in an optically-selected sample with more red galaxies.
Therefore, the SED-fitting to the SDSS bands only is sufficient to constrain the SF for 
the HI-selected blue galaxy population in general \citep[e.g.][]{Huang2012} 
but gives systematic overestimates of SFRs for optically-selected red galaxies such as those 
in the Virgo cluster known to have quenched SF \citep[e.g.][]{Hallenbeck2012}, 
i.e., it is crucial to include the UV bands in the SED-fitting to infer the SF of the red population. 
As a result, we adopt the UV-optical color ($NUV-r$) rather than the optical-only 
($u-r$) in the analysis of SF and gas properties below. 

\section{Global Properties of the ALFALFA Galaxy Population}
\label{SED}

To derive the global properties of the stellar components of the ALFALFA galaxies,
we adopt the methodology of \citet{Salim2007}. 
In particular, stellar masses and SFRs are derived from SED-fitting the seven GALEX/SDSS bands. 
Further details of the method and fitting
quality as applied to the $\alpha.40$ sample are found in \citet{Huang2012} which focuses
on the lowest HI mass dwarf population.
In addition, the Gaussian prior distribution of the effective optical depth in $V$ band, $\tau_V$, 
is applied, with the mean predicted by \citet{Giovanelli1997} and a standard deviation of 0.55~dex. 
Such an improvement reduces the overestimate of internal extinction and SFR 
with decreasing stellar mass, as identified by \citet{Salim2007} (see \S\ref{ext} for more details), 
but still accounts for the effect of dust in disk systems. 
In this section, we discuss the results for the full $\alpha.40$-SDSS-GALEX overlap sample
(9417 galaxies).

\subsection{Gas and stars}
\label{MHIMs}

Current understanding interprets the standard SDSS color magnitude diagram (CMD) 
in terms of an evolutionary
scenario under which galaxies migrate from the blue cloud to the red sequence as they 
assemble their mass. This picture is further reinforced by the presence of the 
star-forming sequence in the SSFR vs. stellar mass diagram; more massive
galaxies show lower SSFRs. Consistent with this picture, one would expect
galaxies to grow increasingly gas poor and thus having lower HI fractions
$f_{HI}$ (defined throughout this work as $f_{HI} \equiv M_{HI}/M_*$)  
as they assemble their mass. 
Therefore, blue galaxies with high gas fractions indicate disks 
Which are stable against collapse, making their SF much less efficient \citep{West2009}. 

\subsubsection{HI versus stellar mass}

In the last decades, many studies have investigated how the HI content varies with 
stellar properties in galaxies, such as morphology, luminosity, size and SF activity 
\citep{Gavazzi1996, Boselli2001, Kannappan2004, Disney2008, Garcia-Appadoo2009, West2009, West2010, Toribio2011}. 
Despite the complex interplay of dynamics, SF, chemical enrichment and feedback etc., 
the stellar and HI components, as well as the dark matter halo, exhibit correlations 
with each other. 
However, many of these studies have relied on relatively small and/or inhomogeneous samples 
limited to the very nearby universe. 
Although the main scaling relations were known, constraints 
on the accuracy of these relationships, as well as the quantification of their scatter 
are still not well determined.  
Based on an H$\alpha$ narrow-band imaging survey of $\sim$400 
galaxies selected from ALFALFA, 
\citet{Gavazzi2012a, Gavazzi2012b} also investigate the relationships between 
HI and newly-formed stars, emphasizing the study of environment effects. 
Here we focus on the nature of the population detected by the ALFALFA survey. 

\begin{figure*}
\center{
\includegraphics[scale=0.8]{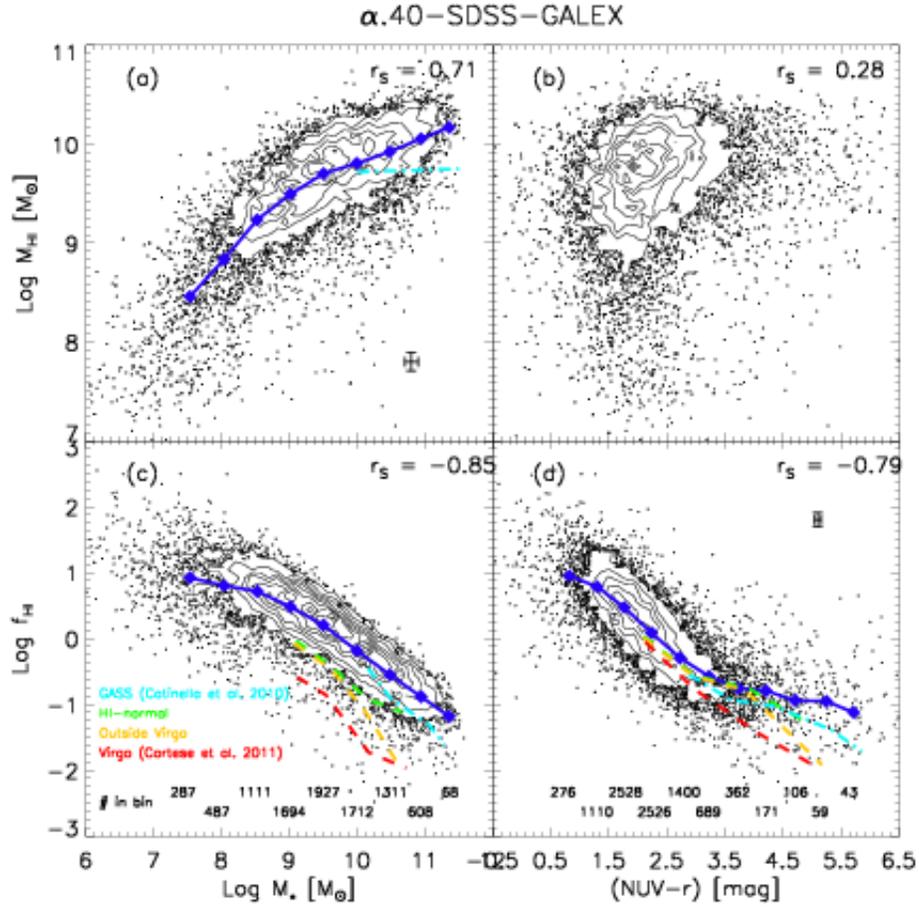}
}
\caption[]{Scaling relations between HI mass, stellar mass and color. 
Contours and points follow the definitions in Figure \ref{fig:CCD}. 
Blue diamonds and solid lines indicate the $\langle \log y \rangle$ values in each $\log x$ bin. 
The number of galaxies in each $\log x$ bin is listed at the bottom of panels (c, d). 
Cyan dash-dotted lines in panels (c, d) denote the average values of the GASS galaxies 
with $M_* > 10^{10}~{\rm M_\odot}$ \citep{Catinella2010}. 
Dashed lines in the same panels are from \citet{Cortese2011} derived for galaxies belong to different 
environments: `HI-normal' (green on top), `outside Virgo' (yellow in middle) 
and `inside Virgo' (red at bottom). 
Typical error bars of individual galaxies are given in the corner of panels (a, d). 
Spearman's rank correlation coefficients, $r_S$, are listed in all panels. 
}
\label{fig:gas}
\end{figure*}

Figure \ref{fig:gas} illustrates the relationships of the HI mass and $f_{HI}$
(the vertical axes) with the
stellar mass and color (horizontal axes). The contours and points outline
the distributions of the galaxies in the $\alpha.40$-SDSS-GALEX sample; 
the blue diamonds and solid lines trace the average values $\langle \log y \rangle$ and $\langle \log x \rangle$ 
in bins of $\log x$, with a bin size of 0.5 dex in panels (a), (c) and (d).
The number of galaxies in each stellar mass or ($NUV-r$) color bin is given at the bottom of 
panels (c, d). 
Typical error bars of individual galaxies are shown in the corners of panels (a) and (d). 
The Spearman's rank correlation coefficients of the relation, $r_S$, are shown in the upper right corner of all panels. 
Compared to similar studies that have previously probed the global 
scaling relations involving $M_{HI}$ \citep[e.g.][]{Bothwell2009, Catinella2010}, the $\alpha.40$ 
sample offers a more complete statistical sampling of the full range of HI and stellar masses.
As discussed in \citet{Haynes2011}, ALFALFA's combination of sensitivity, sky coverage and bandwidth
yields a sample that probes a wide dynamical range in HI mass (7-11 dex with a mean
of 9.56 dex), from the most massive giant spirals 
with $M_{HI} > 10^{10}~{\rm M_\odot}$ to the lowest HI mass dwarfs with log $M_{HI} < 10^{7.5}~{\rm M_\odot}$. 
In fact, the stellar mass range that is probed is slightly wider: 6--11.5 dex, with a mean of 
9.43 dex. As an HI-selected sample, $\alpha.40$-SDSS-GALEX demonstrates the ability 
to recover galaxies with small $M_*$. 

Figure \ref{fig:gas}(a) shows the distribution of 
$M_{HI}$ with $M_*$. The cyan dash-dotted line traces the linear fit to the 
GASS sample of high stellar mass galaxies 
\citep[$M_{*} > 10^{10}~{\rm M_\odot}$;][]{Catinella2010}: 
\begin{eqnarray*}
	\log \langle M_{HI}(M_*) \rangle = 0.02 \log M_* + 9.52. 
\end{eqnarray*}
Note that those authors chose to calculate $\log \langle M_{HI} \rangle$ rather than $\langle \log M_{HI} \rangle$
because the $\langle \log M_{HI} \rangle$ value is depressed by the contribution of gas-poor galaxies
in their $M_*$-selected sample; a similar effect results in their adoption of
$\log \langle f_{HI} \rangle$ rather than $\langle \log f_{HI} \rangle$. In contrast, an HI-selected sample such
as ours does not sample the low HI fraction massive objects so that, as a function of 
$\langle \log M_* \rangle$, $\langle \log M_{HI} \rangle$ and $\langle \log f_{HI} \rangle$ adequately trace the main distribution. 
Moreover, as pointed out in \citet{Cortese2011}, the distribution of $f_{HI}$ is closer to 
log-normal than Gaussian, and thus they also prefer $\langle \log f_{HI} \rangle$ to $\log \langle f_{HI} \rangle$. 

We confirm the previous findings that $M_{HI}$ increases with $M_*$. 
However, the correlation does not appear to be a simple linear one, 
i.e. $\delta M_{HI}/\delta M_*$ is smaller at the high mass end. 
The linear fit to the blue diamonds in Figure \ref{fig:gas}(a) is 
\begin{eqnarray}
\label{eqa:MHIMs}
\langle \log M_{HI} \rangle =
\left\{
\begin{array}{rl}
	0.712 \langle \log M_* \rangle + 3.117,~\log M_* \leq 9; \\
			0.276 \langle \log M_* \rangle + 7.042,~\log M_* > 9.
\end{array}
\right.
\end{eqnarray}
This trend is consistent with the idea that 
once AGNs are turned on in massive galaxies, gas is lost due to AGN feedback. 
The fact that $f_{HI}$ is lower in massive SF/AGN composites than in purely SF galaxies 
of the same mass may be the cause of a similar break in slope of the star-forming sequence (see \S\ref{SFS}, 
at a slightly higher transition mass, $\log M_{*} \sim 9.5$). 
Furthermore, compared to the high stellar mass GASS galaxies, the 
ALFALFA population is overall more gas-rich for the same stellar mass (log $M_* > 10$) and 
traces a steeper slope in the $M_{HI}$ vs. $M_*$ scaling relation, i.e. there is a systematically 
larger discrepancy in the typical HI content of the ALFALFA and GASS populations in the largest 
$M_*$ bins. 
Besides the change in slope, there appears to be an increased scatter in the $M_{HI}$ distribution 
below log $M_{*} \sim 9$, a regime only poorly sampled by other studies. In fact, \citet{Huang2012}
point out that at the lowest HI masses, ALFALFA detects a population of dwarf galaxies with
low $f_{HI}$ for their $M_{*}$; some of these objects are dwarf ellipticals/spheroidals (dE/dSph) galaxies in the
Virgo cluster and may have accreted their current gas supply only recently \citep{Hallenbeck2012}. 
The HI gas can be easily removed in low mass systems due to their shallow potential wells, 
so that the galaxy migrates onto the red sequence as its SF quenches. 

Figure \ref{fig:gas}(c) shows how the HI fraction $f_{HI}$ depends
on $M_*$.  
The cyan dash-dotted line again traces the GASS result for the high stellar masses \citep{Catinella2010},
while the green (upper), red (lower) and yellow (middle) dashed lines 
trace the separate samples of HI-normal galaxies, ones in Virgo and
outside-Virgo respectively from \citet{Cortese2011} who looked for trends among galaxies
in different environments. Again, the known trend that higher $M_{*}$ galaxies have lower $f_{HI}$
is clearly evident, with a correlation coefficient $r_S=-0.85$.
$M_*$ depends more strongly on $f_{HI}$ than on $M_{HI}$ ($r_S=0.71$) 
partly because the same measure of the $M_*$ enters also in the computation of $f_{HI}$. 
Compared to other findings, the ALFALFA population uniformly includes galaxies which are 
more gas rich for a given $M_*$. 
Their extraordinarily high $f_{HI}$ indicates little integrated past SF, while their blue 
colors may be attributed to a SFH that steadily rises to the present day 
or a truly young stellar component \citep{Garcia-Appadoo2009}. 

Both the GASS and \citet{Cortese2011} samples include
galaxies that have lower $f_{HI}$ and lie below the ALFALFA HI detection threshold.
For example, the Virgo cluster is well known to contain a significant population
of HI deficient galaxies \citep{Davies1973, Giovanelli1985, Solanes2002} whose
HI line flux densities are too low for them to be detected by the short ALFALFA observations;
their detections were made using longer duration, target Arecibo observations.
The offset of the ALFALFA
population from the other samples is therefore as expected. However, it is interesting to
note that the scaling of $f_{HI}$ with log $M_{*}$ derived here and by
\citet{Cortese2011} for the HI-normal galaxies, while they do not coincide
in amplitude, do show comparable slopes at intermediate masses,
and perhaps the same is true for all samples at log $M_{*} < 9.7$. 
A ``fast'', shallow survey like ALFALFA derives the same trend as one which relies largely on
much deeper, pointed observations. 
The flattening off of $f_{HI}$ at log $M_{*} \lesssim 9$ is traced only by the ALFALFA dwarfs \citep{Huang2012}. 

It is important to note that the nearby, low mass galaxies are the ones most susceptible
to shredding by the SDSS pipeline so that, statistically, their stellar masses are more
likely to be underestimated, resulting in an extreme tail of galaxies with unrealistically 
high $f_{HI}$. By exclusion of objects with suspect SDSS photometry as noted by \citet{Haynes2011}, 
the most egregious cases have been excluded from this analysis. 
Similar problems with the use of the SDSS pipeline measurements 
have much less effect on the main distribution. At the same time, 
source confusion within the ALFA beam (FWHM $\sim 3^\prime.5$) is more likely
among more distant systems so that the $M_{HI}$ (and $f_{HI}$) of some high $M_{HI}$ sources 
may be overestimated. 
However, other than cases of major mergers, the highest $M_{HI}$ galaxies are always significantly
more massive than their small companions, so that the change in $f_{HI}$, if the 
contribution from companions is removed, would only be small. Overall, the
trend of falling $f_{HI}$ with increasing $M_*$ seen in the ALFALFA 
galaxies is well defined. For the ALFALFA population overall,
the median 
$f_{HI}\sim 1.5$. HI-selected galaxies
are uniformly gas rich for their stellar mass following a scaling relation over
the range of {\it stellar mass}  $8.0 < \log M_{*} < 11.0$. 

In Figure \ref{fig:gas}(c), the number density of points drops 
sharply on the upper edge of the main distribution: there is a real cutoff in the
galaxy population with even higher $f_{HI}$ than ALFALFA detects. 
The increased dispersion in the contours on the lower edge of the $f_{HI}$
distribution with substantial numbers of outliers with lower $f_{HI}$ than the 
main population confirms that, because of its HI-selection, ALFALFA misses
much of the gas-poor galaxy population. Longer integration times would
obviously detect galaxies of lower $M_{HI}$ and thus lower $f_{HI}$ at
constant $M_*$. The GASS observing strategy \citep{Catinella2010}
is specifically designed to probe to constant $f_{HI}$
by conducting significantly longer but targeted HI observations. 
The GASS program thus characterizes the overall population of galaxies selected
by stellar mass at the high mass end. The ALFALFA survey, on the other
hand, samples well the full dynamical range of the HI-rich (for their stellar
mass) population

Figures \ref{fig:gas}(b, d) explore the variation of $M_{HI}$ and $f_{HI}$
with ($NUV-r$). Definitions of diamonds and lines in panel (d) are the same as in panel (c).
As noted in Figure \ref{fig:CCD}, nearly all ALFALFA galaxies are blue, and the population is
highly biased against the red sequence. While there is a wide spread in $M_{HI}$, there is little
trend of $M_{HI}$ with color ($r_S=0.28$).  In fact, there
are 128 $\alpha.40$-SDSS-GALEX red galaxies ($NUV-r > 4$) with $\log M_{HI} > 10$,  
including the early type galaxies with quenched SF but unusually high HI masses (e.g. AGC~260442), 
the edge on galaxies with significant internal extinction (e.g. UGC~6312 has a dust lane evident in the SDSS image), 
and even the ``red spirals'' found in the Galaxy Zoo \citep{Masters2010}, e.g. UGC~9624 and UGC~9283. 
There are 116 red galaxies ($NUV-r > 4$) with $\log M_{HI} < 9.5$; 
most are early type ``dead'' galaxies. \citet{Cortese2011} found that the blue cloud 
galaxies have the same $f_{HI}$ 
regardless of environment, whereas for the red galaxies, Virgo members are significantly 
gas poorer than HI-normal systems (see dashed lines in Figures \ref{fig:gas}d). 
In particular, their fit for HI-normal galaxies (green) agrees well with the main 
trend for the $\alpha.40$ galaxies. 

In contrast,  $f_{HI}$ is a strong function of color among the ALFALFA population ($r_S=-0.79$ in panel d), 
at least among the blue cloud galaxies, i.e. the bluer galaxies tend to have higher HI fractions. 
However, this trend gradually flattens for the very red galaxies ($NUV-r \gtrsim 3.5$), 
i.e. the very red galaxies among the ALFALFA population have higher $f_{HI}$ than would
be predicted by extrapolation of the trend traced by the blue galaxies. 
Compared to the $f_{HI}$ vs. ($NUV-r$) trends derived by \citet{Catinella2010} or \citet{Cortese2011}, 
traced by the dash-dotted and dashed curves respectively in panel (d), the offset of blue diamonds 
on the blue side is small, but the deviation becomes systematically larger in the redder bins. 
Such a change in behavior can be partly explained by the fact that ALFALFA detects only a very small
subset of these red galaxies. The presence of HI in
this small population of otherwise ``red and dead'' galaxies is most easily explained if
their HI gas has been acquired only recently, as has been invoked previously to explain the
HI in ellipticals \citep[e.g.,][]{Wardle1986, Morganti2006}, 
and the annular HI distributions seen in many S0s \citep[e.g.,][]{vanDriel1988, Donovan2009}. 
Deep HI synthesis imaging of the SAURON and ATLAS$^{3D}$ samples of early-type galaxies shows that HI is commonly
detected in galaxies which do not reside in cluster cores \citep[e.g.,][]{Oosterloo2010, Serra2012}.
A significant fraction of non-cluster early-types contain some cool HI gas, with the large spread in HI content
likely due to differences in their accretion histories.

\subsubsection{Predictors of HI gas fraction}

\begin{figure*}
\center{
\includegraphics[scale=0.8]{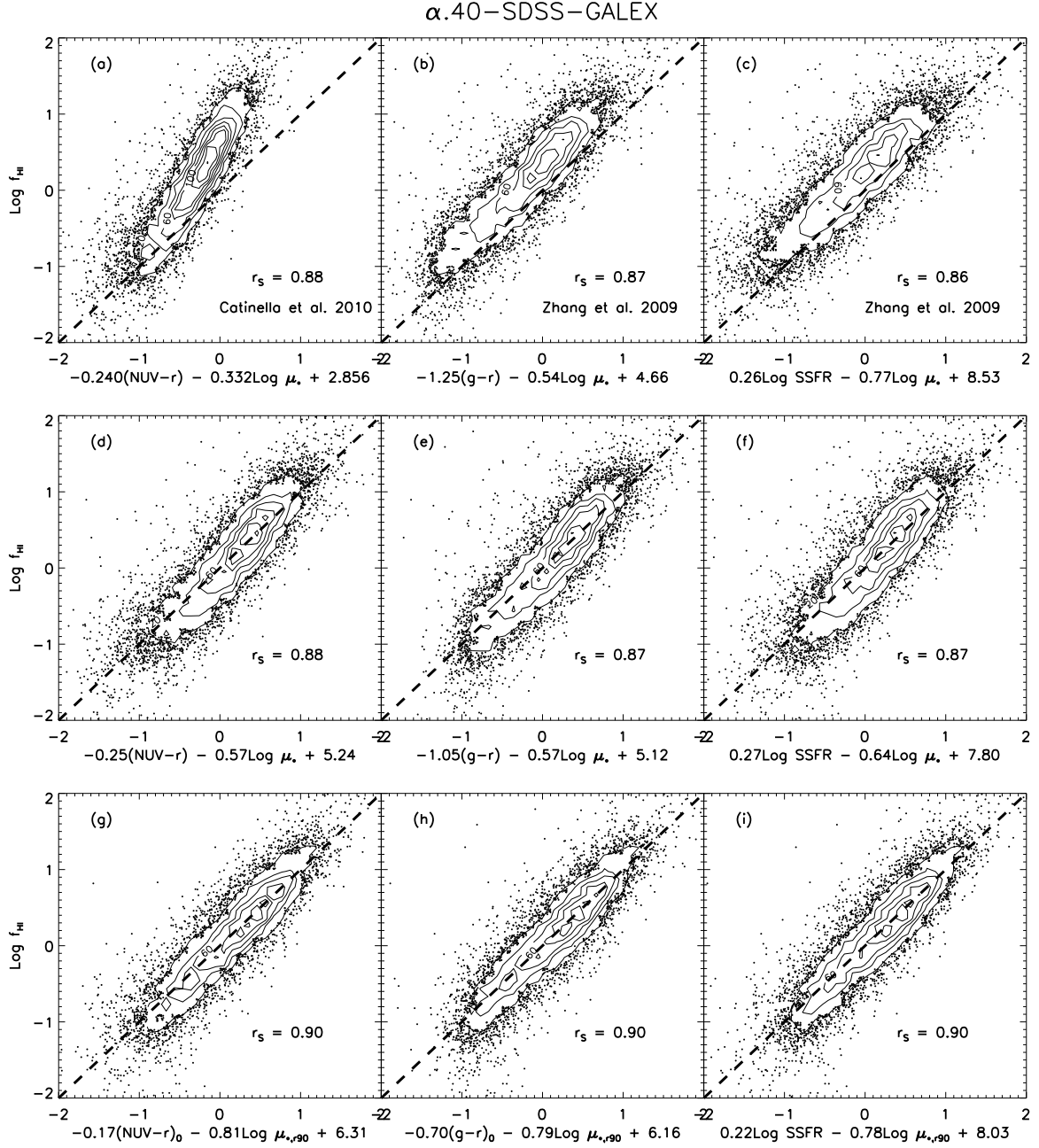}
}
\caption[]{HI fraction estimators. Dashed lines are the one-to-one lines. 
Contours and points follow the definitions in Figure \ref{fig:CCD}. 
Spearman's rank correlation coefficients are shown in the lower right corners of all panels.
{\it Upper panels -} The ALFALFA-observed $f_{HI}$ versus the predicted values based on the fundamental planes 
	calibrated from the GASS sample \citep{Catinella2010} in panel (a), as well as from the SDSS-selected sample 
	\citep{Zhang2009} in panels (b) and (c). 
{\it Middle panels -} Observed $f_{HI}$ versus the values predicted by the best fit to similar planes in this work, 
	given the $\alpha.40$-SDSS-GALEX sample. The systematic offsets are removed and the correlations are tighter. 
{\it Bottom panels -} Compared to the predictors in the middle row, 
	the colors have been corrected for internal extinction and the stellar mass surface density is based on $r_{90, r}$, 
	so that the scatter is reduced. 
}
\label{fig:fgas}
\end{figure*}

The tight correlation between $f_{HI}$ and ($NUV-r$) can be used as a predictor of $M_{HI}$ 
given measures of color, i.e., the `photometric gas fraction' technique \citep{Kannappan2004}. 
Furthermore, a `fundamental plane' of $f_{HI}$--$(NUV-r)$--$\mu_*$ 
has been identified by the GASS survey \citep{Catinella2010}, where the stellar mass surface 
density is defined as 
	$\mu_*~[{\rm M_\odot~kpc^{-2}}] = 0.5 M_*~[{\rm M_\odot]}/\pi(r_{50, z}~[{\rm kpc}] )^2$ 
and $r_{50, z}$ is the radius containing 50\% of the Petrosian flux in the $z$-band. 
Their best fit `plane' is 
\begin{eqnarray*}
	\log f_{HI} = -0.240(NUV-r)-0.332\log \mu_*+2.856, 
\end{eqnarray*}
and the scatter of such a $f_{HI}$ predictor is reduced relative to the $f_{HI}$--$(NUV-r)$ correlation 
with the additional parameter $\mu_*$. 
Because colors essentially trace the SSFRs (see also \S\ref{SFS}), similar predictors are 
calibrated by \citet{Zhang2009} for an optically-selected sample as: 
\begin{eqnarray*}
	\log f_{HI} = 
	\left\{
	\begin{array}{lr}
		-1.25(g-r)-0.54\log \mu_*+4.66; \\
		0.26\log SSFR-0.77\log \mu_*+8.53.
	\end{array}
	\right.
\end{eqnarray*}
It should be noted that no correction for internal extinction is applied in
those analyses, although it is well
known that the inner disks of spirals are optically thick \citep[e.g.,][]{Giovanelli1995}.  
We discuss the need for an internal extinction correction below in \S\ref{ext}.
The $f_{HI}$ predicted by these formulae  
are plotted on the horizontal axes in Figures \ref{fig:fgas}(a-c), respectively. 
Compared to the ALFALFA measurements of the $f_{HI}$ (vertical axes), they all predict 
systematically smaller $f_{HI}$. 
This reaffirms that the HI-selected sample is biased towards the gas rich population. 
The deviation from the one-to-one dashed line increases with $f_{HI}$ 
in the case of the GASS calibration (Figure \ref{fig:fgas}a). 
The \citet{Zhang2009} estimators (Figure \ref{fig:fgas}b and c) systematically underpredict $f_{HI}$ 
of the $\alpha.40$-GALEX-SDSS galaxies by $\sim$0.3 dex. 

Exactly which scaling relation to use to predict the properties of a population
depends of course on what the scientific objective is.
For example, the scaling relations for an optically-selected sample may be valid for a
stellar mass selected population. However, the relations derived for ALFALFA give better
predictions for the HI detection rate for future SKA surveys which will likewise be
HI-selected. Previous simulations of SKA detection rate are mostly based on the HIPASS 
results locally 
\citep[e.g., ][]{Abdalla2005, Obreschkow2009}. But HIPASS suffered from limitations in
its volume sensitivity, and in fact, ALFALFA detects more HI sources at the high $M_{HI}$
end \citep{Martin2010}. 
Based on the $\alpha.40$-GALEX-SDSS galaxies, linear regression gives: 
\begin{eqnarray}
	\label{eqa:fgas1}
	\log f_{HI} = -0.25(NUV-r)-0.57\log \mu_*+5.24; \\
		\label{eqa:fgas2}
		=-1.05(g-r)-0.57\log \mu_*+5.12; \\ 
		\label{eqa:fgas3}
		=0.27\log SSFR-0.64\log \mu_*+7.80.
\end{eqnarray}
Note that for comparison with other authors, the colors used here are corrected for 
Galactic extinction but not for internal extinction.
In Figures \ref{fig:fgas}(d-f), the ALFALFA $f_{HI}$ measures are plotted against the values predicted by equations 
(\ref{eqa:fgas1})-(\ref{eqa:fgas3}), with the correlation coefficients $r_S=$ 0.88, 0.87 and 0.87 respectively. 
The systematic offset is removed according to our best fit and the correlations are as tight as the 
\citet{Catinella2010} ($r_S=$ 0.88) and the two different \citet{Zhang2009} ($r_S=$ 0.87 and 0.86) results. 
Among the three planes, the $f_{HI}$--$(NUV-r)$--$\mu_*$ correlation has the least scatter and it is 
also tighter than 
the $f_{HI}$--$(NUV-r)$ correlation ($r_S=-0.79$). 

Note that the fundamental plane found here is noticeably different from the GASS relationship 
\citep{Catinella2010}. The main trend in Figure \ref{fig:fgas}(d) reveals a break in slope at 
$\log f_{HI} \sim -0.5$, whereas the GASS relation is confined to only below this critical $f_{HI}$. 
Although the GASS sample is complete in the massive $M_*$ domain, 
it does not probe the lower stellar mass, gas rich systems. Blindly applying the 
fundamental plane defined by the massive gas-poor galaxies through extrapolation into the 
gas-rich regime results in serious underprediction of $f_{HI}$. 
Because of the change in slope, the deviation from \citep{Catinella2010} is systematically 
larger in the high $f_{HI}$ galaxies. 

Because the HI population is so overwhelmingly dominated by blue cloud dwellers
and since HI is presumably a constituent of a galaxy's disk, not its bulge or halo, 
it seems appropriate to explore scaling relations which are tied more heavily to the 
galaxy's disk. Hence, we define a disk stellar mass surface density
$\mu_{*,r90}~[{\rm M_\odot~kpc^{-2}}] \equiv 0.9 M_*~[{\rm M_\odot]}/\pi(r_{90, r}~[{\rm kpc}] )^2$, 
where $r_{90, r}$ is the radius containing 90\% of the Petrosian flux in the $r$-band. 
Compared to $\mu_*$, $\mu_{*,r90}$ is based on the $r$-band flux with higher S/N and
less bulge contribution. 
In addition, adopting colors corrected for internal extinction (see \S\ref{ext}), 
we derive improved predictors as follows: 
\begin{eqnarray}
	\label{eqa:fgas4}
	\log f_{HI} = -0.17(NUV-r)_0-0.81\log \mu_{*,r90}+6.31; \\
		\label{eqa:fgas5}
		=-0.70(g-r)_0-0.79\log \mu_{*,r90}+6.16; \\ 
		\label{eqa:fgas6}
		=0.22\log SSFR-0.78\log \mu_{*,r90}+8.03.
\end{eqnarray}
The values given by equation (\ref{eqa:fgas4})-(\ref{eqa:fgas6}) are plotted on the horizontal axes 
in Figures \ref{fig:fgas}(g-i). 
They show less scatter from the one-to-one dashed line with better $r_S=$ 0.90. 
We suggest that scaling laws which incorporate properties which reflect the 
disk nature of the HI distribution, and specifically the above relations, provide 
the most appropriate approach to predicting the characteristics of HI-selected populations.

\subsubsection{Assessing the molecular gas H$_2$ contribution}

ALFALFA is an extragalactic HI line survey and, as such, probes only the neutral ISM.
Yet, the process of SF in most galaxies is more directly coupled to the molecular gas,
and the question of which gas component -- HI, H$_2$ or total gas -- correlates best with SF 
is still debated. To account for the full gas content, we
thus need to assess the expected contribution of molecular gas to the total gas mass in
the ALFALFA population galaxies. 

An outgrowth of the GASS survey of high stellar mass galaxies, COLD GASS is a legacy 
survey which has measured the CO(1-0) line of $\sim$350 randomly selected GASS sample
galaxies (0.025 $< z <$ 0.05) with the IRAM 30m telescope. COLD GASS
has uncovered the existence of sharp thresholds in galaxy structural parameters such as $\mu_*$, 
concentration index and ($NUV-r$) color, above which the detection rate of the CO line drops suddenly. 
These thresholds correspond approximately to the transition between the blue cloud and 
red sequence \citep{Saintonge2011b}. Even though \citet{Catinella2010} found some red sequence 
galaxies  with a surprisingly large HI component, none of the 68 galaxies in the first installment
of COLD GASS with $(NUV-r)>5$ are securely detected in CO. At the same time,
only 1.4\% of the $\alpha.40$-GALEX-SDSS galaxies have $(NUV-r)>5$, so that
the HI-selected galaxies should have a high detection rate in CO. 

Under the assumption that molecular gas forms out of lower density 
clouds of atomic gas, one might naively expect a tight correlation between $M_{HI}$ and $M_{H_2}$. 
However, within the subsample of galaxies detected both in HI and CO by COLD GASS, the fraction 
$(M_{H_2}/M_{HI})$ varies greatly, from 0.037 up to 4.09; the two quantities are only weakly 
correlated \citep{Saintonge2011}. 
The relative proportions of molecular and dense atomic gas in giant molecular clouds depend on the 
cloud column density and metallicity \citep{Krumholz2008}, and the clouds 
could even be primarily atomic if the metallicity is sufficiently low \citep{Ostriker2010}. 

Of all the parameters that \citet{Saintonge2011b} investigated, 
the mean molecular gas fraction ($f_{H_2}\equiv M_{H_2}/M_*$) among the COLD GASS galaxies
correlates most strongly with their ($NUV-r$) color, with 
\begin{eqnarray*}
	\log f_{H_2} = -0.219(NUV-r)-0.596, 
\end{eqnarray*}
although it is weaker than the $f_{HI}$--$(NUV-r)$ correlation, probably because 
H$_2$ resides in the inner region 
where extinction is higher, whereas HI dominates in the outer disks \citep{Saintonge2011}.   
At the same time, they find that $f_{H_2}$ is only a weak decreasing function of $M_*$. 
As a result, although ALFALFA probes a lower stellar mass range than COLD GASS does,
the $f_{H_2}$--$(NUV-r)$ correlation above can still be roughly applied to 
the $\alpha.40$-GALEX-SDSS galaxies. Specifically,
since the $\alpha.40$-GALEX-SDSS galaxies have a mean ($NUV-r$) of 2.24, the results of COLD GASS
predict a mean $f_{H_2}$ of 0.082 for the HI-selected population, higher than the 0.066 
of the COLD GASS detections. 

We note that although $f_{H_2}$ is only a weak decreasing function of $M_*$, $f_{HI}$ clearly 
decreases with increasing $M_*$, i.e., the $M_{H_2}/M_{HI}$ fraction appears to decline 
in less massive galaxies \citep[see also][]{Blanton2009}. 
For luminous galaxies, a substantial fraction of the gas is sometimes in molecular form, but 
the detection of CO in low mass galaxies has been shown to be very difficult \citep[e.g.,][]{Leroy2009}. 
Therefore, we conclude that $f_{H_2}\lesssim0.1$ for the ALFALFA population and thus ignore its contribution 
to the total gas fraction, focusing instead on the well-determined atomic gas fraction $f_{HI}$.

\subsection{Star formation properties}
\label{SFS}

In addition to the stellar mass,
SED-fitting also yields an estimate of the current SFR averaged over 
the last 100~Myr.  \citet{Salim2005} have shown the importance of including the GALEX UV
bands, especially the FUV, to reduce the uncertainties in SFRs derived from SED-fitting.
In addition to the SFR itself, several other quantities of physical interest also
become available. 
For example, the SSFR, defined as $SFR/M_*$, compares the current SFR with that in the past 
(as measured by $M_*$), and thus is well correlated with the birthrate-, or $b-$ parameter, 
defined as $SFR/\langle SFR \rangle$. 
Both the SSFR and the $b-$parameter describe the SFH. 
At the same time, normalization of the SFR by $M_{HI}$ instead of $M_*$
yields the star formation efficiency (SFE), defined as $SFR/M_{HI}$. The SFE
compares the current SFR with its potential 
in the future, the latter regulated by $M_{HI}$, the available fuel. The reciprocal of 
the SFE is the Roberts time, $t_{R}=M_{HI}/SFR$ \citep{Roberts1963, Sandage1986}, 
the timescale for depletion of the
HI gas reservoir, assuming a constant SFR at the current level. 

Figure \ref{fig:SF} shows a montage illustrating how the SF related properties,
$SFR$, $SSFR$, $SFE$ (y axes) vary with $M_*$, $M_{HI}$ and the ($NUV-r$) color (x axes).
As before, contours and points trace the $\alpha.40$-SDSS-GALEX population. 
Spearman's rank correlation coefficients are shown in the lower left corners of all panels. 
Typical error bars of individual galaxy estimates are plotted in the lower right corners of panels (a, e, i). 
In the bottom
row, tracing the SFE, the cyan dashed line shows the average value obtained by \citet{Schiminovich2010}
for the high stellar mass GASS sample, while the green dash-dotted line marks the value corresponding
to the Hubble timescale. 

\subsubsection{The SFR and SSFR in HI-selected Galaxies}

\begin{figure*}
\center{
\includegraphics[scale=0.8]{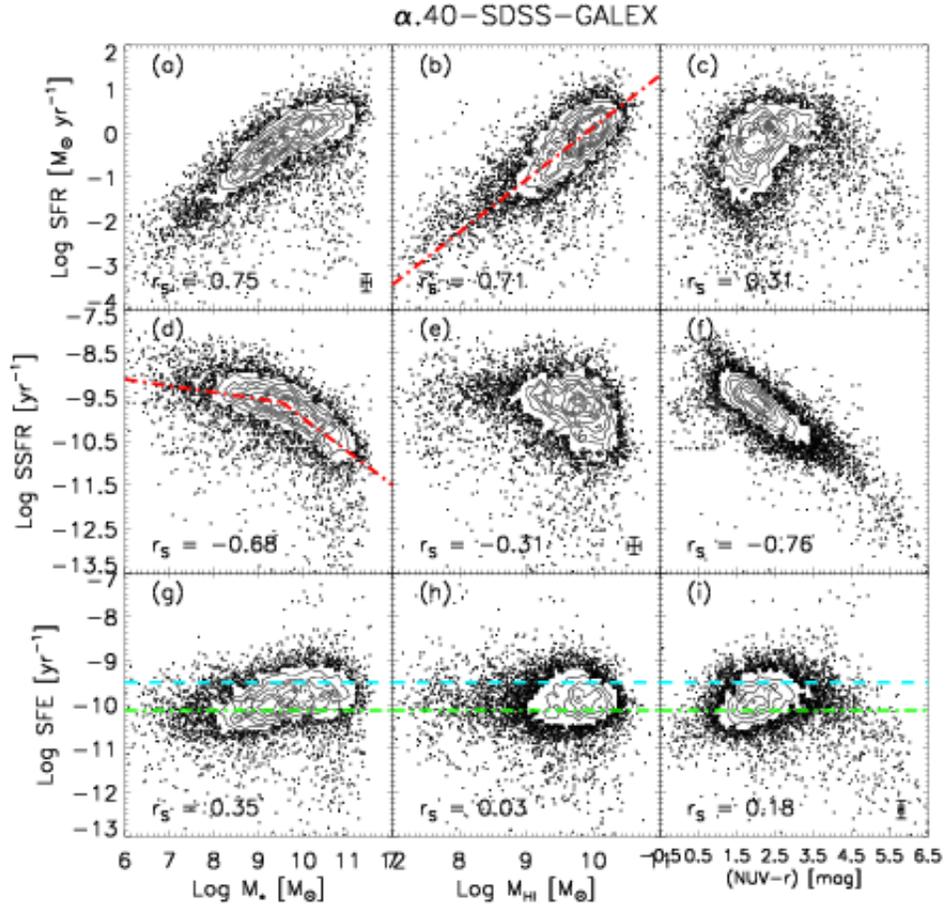}
}
\caption[]{SF properties of the $\alpha.40$-SDSS-GALEX population. 
Contours and points follow the definitions in Figure \ref{fig:CCD}. 
Spearman's rank correlation coefficients are shown in the lower left corners of all panels. 
Colors here are not corrected for internal extinction.
Typical error bars of individual galaxy estimates are plotted in the lower right corners of panels (a, e, i). 
The red dash-dotted line in panel (b), with a slope of 1.19, represents a global, atomic, volumetric 
SFL defined by the $\alpha.40$-SDSS-GALEX galaxies. 
The red dash-dotted line in panel (d) represents the linear fit to the star-forming sequence 
given in equation (\ref{eqa:SFS}). 
In the bottom row, tracing the SFE, the cyan dashed line shows the average value obtained by 
\citet{Schiminovich2010} for the GASS sample, while the green dash-dotted line 
corresponds to the Hubble timescale. 
}
\label{fig:SF}
\end{figure*}

Among the ALFALFA population and in agreement with previous studies, e.g. \citet{Salim2007},
SFRs generally increase but SSFRs decrease with increasing stellar mass 
(Figure \ref{fig:SF}a, d). Similar trends are also evident 
with $M_{HI}$ (Figure \ref{fig:SF}b, e), albeit with larger scatter, especially in the $SSFR$ vs. $M_{HI}$ distribution ($r_S=-0.31$).
The trend of decreasing SSFR with increasing stellar mass suggests 
the ``downsizing'' scenario of structure formation \citep{Cowie1996}, 
such that the high $M_*$ galaxies form most of their stars in the 
first $\sim3$~Gyr after their formation \citep{Bell2003} and today exhibit
relatively suppressed SF. In contrast, the low mass systems in such a picture
remain active in SF throughout their histories. 
Under a hierarchical dark matter halo assembly scenario in which the low mass structures form
first and then merge to form massive galaxies, the ``downsizing'' concept suggests a late epoch of 
gas replenishment and regrowth in low mass systems. 

Although an HI-selected sample like ALFALFA is biased against massive galaxies with
low SFRs and low SSFRs (see also \S\ref{SF}), there is a hint in Figures \ref{fig:SF}(a, d) that
the number density of such galaxies increases in the
$M_* \gtrsim 10^{10}~{\rm M_\odot}$ regime in comparison to the intermediate mass 
range ($10^{8}~{\rm M_\odot} \lesssim M_* \lesssim 10^{10}~{\rm M_\odot}$). The
presence of some points in the lower right corner of the $SFR$ vs. $M_*$ plot suggests 
that at least some objects with large stellar masses and detectable HI but very low 
SFRs are included in the ALFALFA population. More importantly however, 
there is not a comparably rich population of massive HI disks with low SFRs, i.e., 
the number density of galaxies in the lower right corner of Figure \ref{fig:SF}b is lower than 
that in the lower right corner of \ref{fig:SF}a. Where there is a lot of HI, there is always some SF.

As evident in Figure \ref{fig:SF}(c), 
the expected correlation between $SFR$ and ($NUV-r$), that 
galaxies bluewards of $(NUV-r) \simeq 4$  have higher SFRs than ones 
redwards of that value \citep[e.g.][]{Salim2005}, is not so well defined by the ALFALFA population 
($r_S=0.31$), mainly because ALFALFA detects only a few very red galaxies. In particular,
we lack sufficient dynamic range in ($NUV-r$) to 
probe the trend along the red sequence seen in optically-selected samples that galaxies bluer in ($NUV-r$) 
have higher SFRs, especially if colors after extinction correction are plotted. 

On the other hand, the $SSFR$ is a strong function of the ($NUV-r$) color (Figure \ref{fig:SF}f), 
with the Spearman's rank correlation coefficient 
$r_S=-0.76$. 
This is also the tightest among all the correlations shown in Figure \ref{fig:SF}. 
It is natural to expect that ($NUV-r$) is closely tied to the $SSFR$. 
Since the $NUV$ luminosity largely reflects the SF and the
$r$-band luminosity the stellar mass, the ($NUV-r$) 
color, as the ratio of the two, serves as a proxy for the $SSFR$. 
Given the fact that $NUV$ better characterizes the SFR than the $u$-band, one may also expect 
the $SSFR$ to correlate better with ($NUV-r$) color than with ($u-r$). 
However, because $NUV$ suffers more from internal extinction and the associated corrections
can be highly uncertain, extra scatter is introduced when the $NUV$ is used with
accounting for the impact of extinction. In fact, as demonstrated in
(\S\ref{ext}) and Figure \ref{fig:SFS6b}(b), a shift towards blue colors and
an even tighter correlation between $SSFR$ and ($NUV-r$) 
become apparent when extinction-corrected colors are used.

\subsubsection{The star formation law in HI-selected galaxies}
\label{SFL}

The underlying question linking gas to stars in galaxies, the ``star formation law'' (SFL),
is what limits SF: the formation of molecular gas out of HI or the 
efficiency at which the available molecular gas is converted into stars \citep{Schruba2011}. 
Various forms of the SFL are studied, perhaps most common among them the empirical law
describing how the SF surface density ($\Sigma_{SFR}$) is regulated by the gaseous 
surface density \citep[e.g., $\Sigma_{HI+H_2}$ in][]{Kennicutt1998}. 

It should be noted that
since most galaxies are unresolved by the ALFA 3.5\arcmin ~beam, ALFALFA
measures only the global HI content. Our estimate of the SFL will thus be globally averaged.
Numerous recent studies focusing on 
more detailed observations of smaller yet representative samples have demonstrated
the regulation of SF by molecular gas.  For example,
the HI Nearby Galaxy Survey \citep[THINGS;][]{Walter2008} and the HERA CO Line 
Extragalactic Survey \citep[HERACLES;][]{Leroy2009} provide measurements of 
the surface densities of total gas, atomic and molecular gas, and SFR in 
$\sim$kpc-sized regions within a number of nearby galaxies. 
Measurements of the azimuthally averaged gas and SFR profiles show that the 
SFR correlates better with the molecular hydrogen component than with the total gas density 
within the optical disk 
\citep[e.g.][]{Bigiel2008}, suggesting that the SFR is controlled by the amount of gas in 
gravitationally bound clouds and that H$_2$ is directly important for cooling. 
\citet{Krumholz2011} collated observations of the relationship between gas and 
SFR from resolved observations of Milky Way molecular clouds, from kpc-scale observations 
of Local Group galaxies, and from unresolved observations of both disk and starburst galaxies 
in the local universe and at high redshift. Those authors showed that the data are 
consistent with a simple, local, volumetric SFL and that the SFR is simply $\sim$1\% of the molecular 
gas mass per local free-fall time. 
Furthermore, \citet{Schruba2011} found a tight and roughly linear relationship between IR 
(inferring $\Sigma_{SFR}$) and CO (inferred $\Sigma_{H_2}$) intensity, 
with $\Sigma_{H_2}/\Sigma_{SFR} \sim 1.8~{\rm Gyr}$. 
This relation does not show any notable break between regions that are dominated by 
molecular gas and those dominated by atomic gas, 
although there are galaxy-to-galaxy variations in the sense that less massive galaxies exhibit 
larger ratios of SFR-to-CO, an effect
which may due to depressed CO relative to H$_2$ in low metallicity galaxies. 
Similarly, \citet{Bigiel2011} demonstrated a roughly constant H$_2$ consumption time 
($\Sigma_{H_2}/\Sigma_{SFR} \sim 2.35~{\rm Gyr}$). 

However, other works show that the relationship between SF and gas varies systematically 
depending on the local environment. 
\citet{Bigiel2010, Bigiel2010b}, found an evident correlation between SF and HI in the outer disks 
of spirals and in dwarf galaxies where HI is likely to dominate the ISM. 
Given the poor correlation between HI and SFR found inside star-forming disks, 
this finding strongly hints that different physics governs the formation of star-forming clouds, 
and that the HI column is perhaps the key 
environmental factor in setting the SFR \citep{Bigiel2010}. 
Furthermore, the SFL is likely to have a distinct form in the atomic-gas-dominated regime 
\citep[e.g. $\Sigma_{SFR} \propto \Sigma_{HI+H_2} \sqrt{\rho_{sd}}$, theoretically by]
[where $\rho_{sd}$ is the midplane density of stars plus dark matter]{Ostriker2010}. 
Therefore, we may expect a steeper dependence of $\Sigma_{SFR}$ on $\Sigma_{HI+H_2}$
if there is a dropoff in the stellar and dark matter density with radius. 
There is no single universal slope predicted for the SFL in the diffuse-gas-dominated regime. 
In low gas surface density or low metallicity regions where gas is significantly atomic, 
thermal and chemical processes become dominant in determining where stars can form, 
and the gravitational potential of the stars and dark matter may have a significant effect. 
Similarly, the model developed by \citet{Krumholz2009b} suggests $\Sigma_{SFR}$ 
becomes a steep function of $\Sigma_{HI+H_2}$ when complexes of gas become primarily atomic, 
for low ISM surface density. 
Observations also confirm steeper slopes for the low density outer HI-dominated regions of spiral galaxies, 
as well as dwarf galaxies, compared to the inner molecular-dominated regions of spirals \citep{Bigiel2010}. 

The increasing SFR with HI mass evident in Figure \ref{fig:SF}(b) suggests the 
regulation of SF by the HI gas, with a correlation coefficient of 
$r_S=0.71$. 
The red dash-dotted line in Figure \ref{fig:SF}(b) shows the linear fit to the 
$\alpha.40$-SDSS-GALEX galaxies, with a slope of
1.19, suggesting a global, atomic, volumetric SFL. 
The close to unit slope indicates a SFE close to constant as a function of $M_{HI}$. 
The correlation between SFR and $M_{HI}$ appears to be in conflict with the earlier finding that 
most galaxies show little or no correlation between $\Sigma_{SFR}$ and $\Sigma_{HI}$ \citep{Bigiel2008}. 
The high $f_{HI}$ galaxies represented by the ALFALFA population appear to obey an unusual SFL 
which may not only depend on the H$_2$, but also on the HI, stellar and dark matter properties. 
Additional observational and theoretical work is needed to evaluate how the SF
efficiency of bound clouds depends on the relative amounts of cold HI versus molecular gas. 

\subsubsection{The SFE in HI-selected galaxies}

Figures \ref{fig:SF}(g-i) 
illustrate the distribution of the $SFE$ with $M_*$, $M_{HI}$ and the ($NUV-r$) color.
The timescale for atomic gas depletion for the majority of the ALFALFA galaxies is shorter than 
the Hubble time $t_{H}$, 
and comparable to it for many of those with low stellar masses, $M_* < 10^{9}~{\rm M_\odot}$.  

For the high stellar mass GASS population,
\citet{Schiminovich2010} found that, unlike the SSFR which decreases with 
increasing stellar mass, the SFE remains relatively constant with a value close 
to $SFE=10^{-9.5}$~yr $^{-1}$, or equivalently $t_{R} \sim 3$~Gyr. 
This value is longer than the molecular gas depletion timescale (see \S\ref{SFL}). 
Furthermore, those authors 
also found little variation in the SFE with stellar mass surface density $\mu_*$, 
the ($NUV-r$) color or the concentration index, a result 
which they interpreted as an indication that external processes or feedback mechanisms which control 
the gas supply are important for regulating SF in massive galaxies. 
Considering that $\log SSFR = \log SFE + \log f_{HI}$, an interesting implication of the weak 
correlation between the SFE and the stellar mass is that the fit to 
the $\log SSFR$ versus $\log M_*$ distribution 
would have a similar slope to that of the $\log f_{HI}$ vs. $\log M_*$, 
specifically 
$-0.288$ for $\log M_* \leq 9$ and 
$-0.724$ for $\log M_* > 9$ 
(see equations \ref{eqa:MHIMs} in \S\ref{MHIMs}). 
However, the red dash-dotted line in Figure \ref{fig:SF}(d) 
shows the linear fit to the `star-forming sequence' defined by the $\alpha.40$-SDSS-GALEX 
galaxies: 
\begin{eqnarray}
	\label{eqa:SFS}
	\log SSFR =
	\left\{
	\begin{array}{lr}
		-0.149\log M_* - 8.207,~\log M_* \leq 9.5 \\
		-0.759\log M_*-2.402,~\log M_* > 9.5. 
	\end{array}
		\right.
\end{eqnarray}
The differences in the slopes suggest that $\langle SFE \rangle$ is a weak increasing function of $\langle M_* \rangle$ 
in the low $M_*$ range but remains relatively constant above $\log M_* \sim 9.5$;
this trend is also evident in the bottom row of panels($r_S=0.35$). 
The mild trend of increasing SFE with stellar mass seen in Figure \ref{fig:SF}(g) 
was not evident in the 
GASS study \citep{Schiminovich2010} because the GASS sample includes only galaxies with  
$M_*>10^{10}~{\rm M_\odot}$.

Rather than a simple continuous scaling relation, the change of slope given in equation 
(\ref{eqa:SFS}) and evident in Figure \ref{fig:SF}(d), suggests that a transition mass exists at
$M_*\sim10^{9.5}~{\rm M_\odot}$ in the way in which star formation scales with total $M_*$.
A similar transition mass at $M_*\sim10^{9.4}~{\rm M_\odot}$ in SSFR is adopted by \citet{Salim2007} 
for their blue galaxies with $(NUV-r)<4$. 
Those authors suggested that the lower SSFR is a consequence, at the high $M_*$ end, of a
population of systems which are both star-forming and have AGN, thereby yielding lower 
SSFRs than pure SF galaxies of the same mass. 
Similarly, \citet{Kannappan2009} identified a ``threshold'' stellar mass of several times 
$10^{9}~{\rm M_\odot}$, 
below which the number of blue sequence E/S0 galaxies sharply rises. 
Those authors matched the threshold to the mass scale below which the mean 
HI content of low-$z$ galaxies increases substantially 
both on the red sequence and within the blue cloud. 
Abrupt shifts in the SFE and gas richness near the ``threshold mass'' have been linked to the 
interplay of gas infall, supernova-driven winds, and changes in mass surface density. 
However, it is important to note that such a threshold falls below the ``transition'' mass
characteristic of the ''green valley'', identified in the, e.g., $f_{HI}$ versus $M_*$ and 
$\mu_*$ relations, at a stellar mass $M_* \sim (2-3) \times 10^{10}~{\rm M_\odot}$ proposed
in many other works \citep[e.g.][]{Kauffmann2003, Baldry2004, Bothwell2009, Catinella2010}
and suggested to be
indicative of the SF quenching in massive galaxies as they migrate from the blue cloud
to the red sequence.

Perhaps surprisingly, the ALFALFA galaxies have on average {\it lower} SFE, or 
equivalently, longer $t_R$, compared 
to the optically-selected population, with a
mean of $SFE = 10^{-9.95}$~yr $^{-1}$, or equivalently $t_R=8.9$~Gyr, 
compared to the $t_{R} \sim 3$~Gyr derived for the GASS galaxies. 
We note that the average $t_{R}$ value was volume corrected in \citet{Schiminovich2010}, 
but not in Figure \ref{fig:SF}(g-i). However, we confirm that the volume correction 
(see \S\ref{sample}) results in only subtle changes in the mean $t_{R}$ as a function of $M_*$ for the 
$\alpha.40$-SDSS-GALEX galaxies, and it is still longer than $t_{R} \sim 3$~Gyr. 
As we demonstrate in \S\ref{SF}, 
the HI-selected galaxies have, on average, higher SFRs at a fixed stellar mass, so that the lower SFEs 
must result from their higher HI masses rather than from less active states of
SF. This result reaffirms the general conclusion that 
HI-selected samples are strongly biased towards the most gas-rich galaxies. 
In agreement with the low SFEs characteristic of the ALFALFA population, 
\citet{Bigiel2008} have seen a decrease in SFE in the HI-dominated THINGS galaxies.
Furthermore, \citet{Bigiel2010} found that the SFE decreases with galactocentric radius 
among the THINGS sample across the outer disks beyond the optical radius, 
where HI dominates the ISM, with $t_R$ well above Hubble time. 
In the THINGS dwarf galaxies, the contribution of H$_2$ to the total gas budget 
is generally small even in the inner disks, also corresponding to a low SFE \citep{Bigiel2010}. 
All these results are consistent with the conclusion that
SFEs are low, on average, in HI-rich systems. 

However, we note that the low HI SFE may not be in conflict with the usual H$_2$ SFL. 
The HI-selected high $f_{HI}$ galaxies may still follow the normal behavior of how stars 
form from H$_2$, but rather that a bottleneck exists in the process by which star-forming 
molecular clouds assemble. 
The conversion of HI to H$_2$ depends on environment inside a galaxy and the 
relative abundance of HI and H$_2$ is key to setting the SFR \citep{Bigiel2010b}. 
Although the low HI SFE suggests the inefficiency of HI-to-H$_2$ conversion, 
the HI-to-H$_2$ ratio cannot be arbitrarily high. 
\citet{Ostriker2010} assumed an equilibrium state, in which cooling balances heating 
and pressure balances gravity. This balance can be obtained by a suitable division 
of the gas mass into 
star-forming (gravitational bound) and diffuse components such that their ratio is 
proportional to the vertical gravitational field. If too large a fraction of the total surface 
density is in diffuse gas, the pressure will be too high while the SFR will be too low. 
In this situation, the cooling would exceed heating, and mass would drop out of the 
diffuse gas component to produce additional star-forming gas. 

Close examination of the $SFE$ vs. $M_*$ diagram in Figure \ref{fig:SF}(g) also reveals
a considerable number of outliers with relatively high SFE at the 
high $M_*$ end, falling well above the main distribution.
In general, high SFEs have been measured in starburst galaxies and
interacting systems which are consuming their gas reservoirs on very short timescales. 
However, a close inspection of 13 $\alpha.40$-SDSS-GALEX 
galaxies with $\log SFE > -8$ shows that 9 of them are members of the Virgo Cluster. 
Ram pressure stripping results in strong HI deficiency and very short $t_R$ 
in these extreme outliers. 
In contrast, the outliers below the main distribution can either be 
red, massive, low SSFR galaxies against which ALFALFA is strongly biased, 
or abnormally gas rich (for their stellar mass) but quiescent ones. The latter 
include candidates of recent re-accretion or systems in which the HI gas 
is somehow inhibited from forming stars. 
We return to this point in \S\ref{lambda}.

Figures \ref{fig:SF}(h, i) illustrate that the
$SFE$ barely changes with either $M_{HI}$ or color 
($r_S=0.03$ and $0.18$ respectively) and the scatter in both correlations is large. 

\subsubsection{Linking the gas fraction $f_{HI}$ to SF}

As first discussed by \citet{Roberts1963}, it is not surprising that
SF appears to be regulated by gas content. 
We have already argued that the $\log SSFR$ vs. $\log M_*$ diagram (Figure \ref{fig:SF}d) is similar 
to the $\log f_{HI}$ vs. $\log M_*$ one (Figure \ref{fig:gas}c), both showing similar slopes along the main 
trend; the distributions of $\log SSFR$ and $\log f_{HI}$  in given stellar mass bins 
also become broader at both the high- and low- mass ends. 
Hence, the star-forming sequence in the former diagram can also be understood 
as a sequence of gas-depletion in the latter one. 

\begin{figure*}
\center{
\includegraphics[scale=0.8]{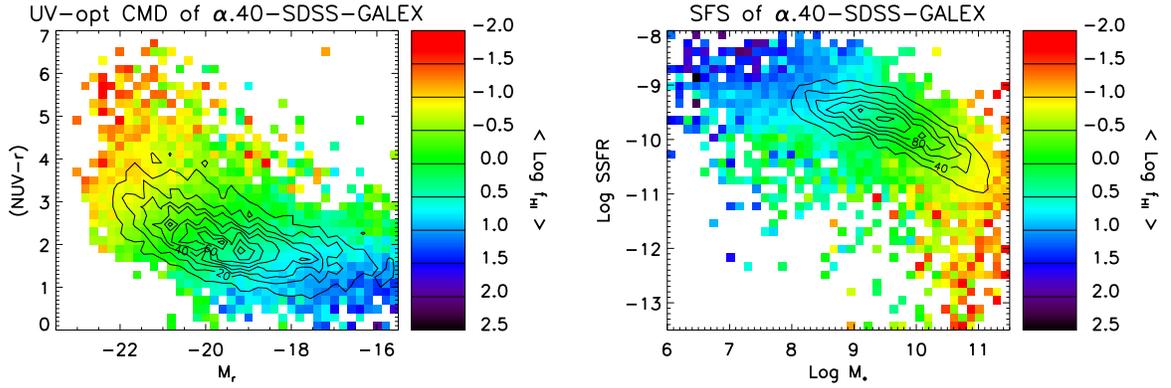}
}
\caption[]{Diagrams of the $\alpha.40$-SDSS-GALEX sample with shade scale showing 
the averaged HI fraction in each grid, while number density is indicated by the contours. 
{\it Left panel -} UV-to-optical CMD. 
The blue cloud galaxies dominate in number and are associated with higher $f_{HI}$ 
on average. At given $M_r$, a redder ($NUV-r$) color on average indicates lower $f_{HI}$; 
this trend is more evident at the faint end. 
{\it Right panel -} The star forming sequence as traced by the contours. 
As galaxies assemble $M_*$ and evolve along the sequence, 
their HI fractions follow a decreasing trend. 
At a fixed $M_*$, galaxies with lower SSFRs on average have lower $f_{HI}$, 
which is also more evident among low mass galaxies with $M_*\lesssim 10^{9.5}~{\rm M_\odot}$ 
The broadenings of the distributions of color, SSFR and $f_{HI}$ at the low mass end are 
correlated to each other. 

(A color version of this figure is available in the online journal.) 
}
\label{fig:ccfgas}
\end{figure*}

Previously, \citet{Kannappan2004} also linked the $f_{HI}$ to bimodalities in galaxy properties. 
She suggested that the bimodality in SFHs may be intimately related to changes in $f_{HI}$, 
and the transition in SF modes at $M_* \sim (2-3) \times 10^{10}~{\rm M_\odot}$,
found by those authors, is not a cause 
but an effect of changing $f_{HI}$, as predicted in cold-mode accretion scenarios. 
Figure \ref{fig:ccfgas} illustrates these relationships by 
showing the averaged $\log f_{HI}$ of galaxies which lie in different loci 
in the CMD (left panel) and the $\log SSFR$ vs. $\log M_*$ diagram (right panel). 
Contours indicate the number density within the $\alpha.40$-SDSS-GALEX sample in each grid
point in the map, while the shade scale traces the mean HI fraction, $\langle \log f_{HI} \rangle$. 
As mentioned in \S\ref{NUV_r}, 
96\% of ALFALFA galaxies lie on the blue side of the optical-NUV CMD, whereas 
a smaller fraction 
(84\%) lie on the blue side of an optical-only CMD (see \S\ref{CMD} below). 
Some of this difference can be attributed to the greater impact of shredding
on the SDSS pipeline magnitudes relative to that of the GALEX photometry. 
Because of the color gradient of galaxies (outer disks are bluer), 
the shredded central redder object is identified as the OC. 
As a result, the adopted photometric object may be
redder in ($u-r$) than the galaxy as a whole actually is.
Moreover, the $u$-band is not as sensitive and thus yields photometry with 
large uncertainties for some of the galaxies. 
Additionally, the ($NUV-r$) color is a stronger diagnostic of the SFH.
For similar reasons, 
the $\langle \log f_{HI} \rangle$ of grid points in regions of low number density should 
be interpreted with caution. However the general trends (1) that the red-sequence
is associated with low $f_{HI}$ and (2) that blue cloud galaxies 
are gas rich are clearly evident.
At a given $M_r$, redder ($NUV-r$) colors, on average, indicate lower $f_{HI}$. 
Furthermore, such a variation of $f_{HI}$ along the ($NUV-r$) axis is more evident 
in the fainter $M_r$ range: $\delta \langle \log f_{HI} \rangle / \delta \langle NUV-r \rangle \simeq 0.75$ at $M_r\simeq-16$, 
whereas $\delta \langle \log f_{HI} \rangle / \delta \langle NUV-r \rangle \simeq 0.25$ at $M_r\simeq-22$. 
Therefore, the correlation of ($NUV-r$) and $f_{HI}$ at a given $M_r$ is hard to see 
in a sample with only massive galaxies, e.g. GASS \citep{Wang2011}. 

Similarly, the right panel of Figure \ref{fig:ccfgas} illustrates
how the HI fraction varies in the $M_*$-SSFR plane.
As galaxies assemble their stellar mass and evolve along the star forming sequence, 
represented by the contours of high number density, their HI fractions decrease. In addition, 
for fixed $M_*$, galaxies with lower SSFRs have, on average, lower $f_{HI}$. Again this 
is more clearly evident at the low stellar mass end, and is consistent with the broadening 
in both the $f_{HI}$ and $SSFR$ distributions at low $M_*$ \citep{Huang2012}. 
In contrast, the variation of $\langle \log f_{HI} \rangle$ along the $\log SSFR$ axis at a given $M_*$ 
is less evident for galaxies with $M_*\gtrsim 10^{9.5}~{\rm M_\odot}$. 
These two trends suggest that, for low $M_*$ systems, high $f_{HI}$ galaxies are more likely to 
be starburst galaxies 
(defined as high SSFR galaxies), whereas galaxies  in the high $M_*$ regime 
selected by high $f_{HI}$ are less likely to be starbursts. 
In summary, Figure \ref{fig:ccfgas} clearly demonstrates that the color, SF and 
gas evolution of galaxies are closely related to one another, as expected. 
Moreover, the regulation of SF by $M_{HI}$ is stronger in the less massive galaxies. 

\section{The Impact of Optical Versus HI-selection}
\label{HIO}

Future surveys of HI in galaxies at intermediate to high redshifts that will
be enabled by the next generation of centimeter-wavelength radio 
telescopes (e.g. the SKA) will aim to
infer the gas evolution from high redshift populations to the local well-studied 
ones. It is important therefore to understand the fundamental properties of local
HI-selected galaxies, as represented by the ALFALFA catalog, and their biases 
relative to the overall galaxy population.
In \S\ref{SED}, we examined the global properties of gas, stars and SF within the ALFALFA 
galaxies themselves. They form an HI-rich, blue and less evolved population with low SFE; 
these characteristics are more pronounced in lower mass systems. 
$M_{HI}$ and $f_{HI}$ are linked to the SF related quantities, demonstrating the 
role that HI plays in the regulation of galaxy evolution along the star forming sequence. 
To understand how the HI-selected population is biased relative to ones selected by
stellar mass or optical flux, 
in this section we construct samples from both the 
$\alpha.40$ and SDSS catalogues and then compare their similarities 
and differences.

\subsection{Construction of control samples}
\label{sample}

In order to ensure the galaxies contained in each of the optically- and HI-selected 
samples are both representative of
their respective population and fair enough to permit comparison with the other,
we construct subsamples of both $\alpha.40$ and the SDSS in the
same sky volume within their common footprint. The volume-limits imposed are
similar to those discussed by \citet{Martin2010}.
Comparable selection criteria are applied 
with a further requirement that acceptable GALEX pipeline photometry must also be available,
so that stellar masses and star formation properties can be derived robustly via
SED-fitting. 

\subsubsection{ALFALFA selected sample, $S_{HI}$}

As discussed by \citet{Martin2010}, radio frequency interference from the 
San Juan airport FAA radar transmitter at 1350 MHz
makes ALFALFA blind to HI signals in a spherical
shell $\sim 10$ Mpc wide centered at a distance of $\sim215$ Mpc.
Therefore, as did those authors,
we exclude 568 galaxies of the sample presented in 
\S\ref{SED} which lie beyond 15000~km~s$^{-1}$ ($D_{sur}$ hereafter).
To maximize the overlap of contiguous sky coverage between the current ALFALFA and 
SDSS DR7, we consider only the two regions in the northern Galactic hemisphere 
($\rm 8^h 00^m < RA < 16^h30^m$, $\rm 4^\circ < Dec < 16^\circ$ and 
$\rm 7^h 40^m < RA < 16^h40^m$, $\rm 24^\circ < Dec < 28^\circ$, 
 see Figure \ref{fig:basic}c). 
Applying these restrictions yields a sample within a sky volume of 
$V_{sur}=1.987\times10^6~{\rm Mpc}^3$, 
a sky area of 1989 deg$^2$ and including 7638 $\alpha.40$-SDSS-GALEX galaxies. 

Next, a weight, $V_{sur}/V_{max}$, is assigned to each galaxy, 
where $V_{max}$ is given by the maximum distance, $D_{max}$, 
at which an HI source can be detected by ALFALFA,
if $D_{max}<D_{sur}$, with $V_{sur}/V_{max}=1$ for the galaxies 
which can be detected all the way outwards to the $D_{sur}$. 
Because the ALFALFA sensitivity depends not only on the 
integrated HI line flux density, $S_{int}$~[Jy~km~s$^{-1}$], but also on 
the HI line profile width, $W_{50}$~[km~s$^{-1}$],  
specifically, the fit to $S_{lim}$, the limiting integrated HI line
flux density that can be detected at S/N 
above 4.5 (code 1 and 2, 25\% complete), as given in \citet{Haynes2011}, is:
\begin{eqnarray*}
{\small
	\log S_{lim} = 
	\left\{
	\begin{array}{lr}
			0.5 \times \log W_{50} - 1.11-0.202, {\rm\ if\ } \log W_{50} < 2.5;\\
			1.0 \times \log W_{50} - 2.36-0.202, {\rm\ if\ } \log W_{50} \geqslant 2.5.
	\end{array}
	\right.
}
\end{eqnarray*}
Then, $D_{max}$ can be calculated given $S_{lim}$ and $M_{HI}$, based on the standard equation 
$M_{HI} = 2.356\times10^5 D_{Mpc}^2 S_{int}$.
In order to characterize the stellar component of the galaxies, we also perform
a cross-match to the SDSS and GALEX databases. The application of such a weight scheme,
or volume correction, is equivalent to resampling the galaxies by their HI properties 
($M_{HI}$ and $W_{50}$) alone, thereby reemphasizing the impact of HI-selection.

To further trim the sample, we drop galaxies whose weight $V_{sur}/V_{max}>60$, 
i.e., we consider only galaxies that could be detected in more than 1.67\% of the survey volume. 
This cutoff corresponds approximately to a lower $M_{HI}$ limit 
of $\sim 10^{8.2}~{\rm M_\odot}$ (see Figure \ref{fig:basic}). 
There is not a hard $M_{HI}$ cutoff because $W_{50}$ also plays a role. 
The galaxies with $V_{sur}/V_{max}>60$ are all relatively nearby 
($D_{max} < 54.7~{\rm km~s^{-1}}$, or assuming Hubble flow, $cz_{max} < 3829~{\rm km~s^{-1}}$, 
$z_{max}<0.0128$), and are low HI mass galaxies less representative of the survey overall. 
Further motivations for applying such a weight cutoff include: (a) for these very local sources, 
distance dependent quantities, e.g. $M_{HI}$, have large uncertainties due to their peculiar velocities; 
(b) such galaxies are also underrepresented in the SDSS redshift sample (see below); 
(c) for resolved, patchy dwarf systems especially, the SDSS pipeline magnitudes can 
    suffer from shredding. The lowest HI mass systems have been considered 
    separately in \citet{Huang2012}. 

The final HI-selected sample referred to as $S_{HI}$ includes  
7157 galaxies. 

\subsubsection{SDSS selected sample, $S_{opt}$}

To construct an optically-selected control sample out of the same sky volume, $V_{sur}$, 
we queried the SDSS DR7 in the same RA and Dec ranges 
for photometric objects which have valid model magnitudes and
were also spectroscopic targets.
We also require them to 
(a) have a spectral classification of ``galaxy''; 
(b) have an SDSS redshift, $z_{SDSS}$, determined with high confidence; 
(c) lie within the same redshift range as $S_{HI}$, $cz_{SDSS} < 15000~{\rm km~s^{-1}}$; 
(d) have Galactic extinction-corrected $r$-band model magnitudes brighter than 17.77. 
24379 galaxies meet these criteria. 
Note the redshifts adopted for this sample use the SDSS measurement, 
whereas that for the $S_{HI}$ comes from the HI line measures. 
Given the $cz_{SDSS}$ and coordinates, distances are estimated 
in the same manner as for the $\alpha.40$ sample using a local flow model for 
$cz < 6000$~km~s$^{-1}$, and Hubble distance otherwise \citep{Haynes2011}. 
Following the same procedure as for $S_{HI}$, we searched for GALEX cross-matches, 
and applied similar SED-fitting to the UV/optical bands. 

To match the weight cutoff of the HI-selected $S_{HI}$ sample, we also calculate weights 
for the SDSS-selected sample but here according to their optical fluxes. 
In this case, $D_{max}$ is the maximum distance at which the object,
given its $r$-band flux, could be included in the SDSS main galaxy redshift sample. 
As for the HI-selected sample, we drop galaxies with weights greater than 60.
Given the magnitude limit of the SDSS redshift survey ($m_r <$ 17.77 mag), such a 
weight cut directly corresponds to an 
$r$-band absolute magnitude limit of $\sim -16$ mag. 
Furthermore, since the mass-to-light ratio varies only mildly with color, 
the luminosity cut approximately translates to a stellar mass lower limit of 
$\sim10^{7.6}~{\rm M_\odot}$. Finally, a small number of galaxies are removed 
because they are included in the $\alpha.40$ catalog but have been previously 
noted by individual inspection to have suspect photometry \citep{Haynes2011}.
The final optically-selected sample referred to as $S_{opt}$ includes 16817 
galaxies, of which 34\% are cross-matched to the $\alpha.40$ catalog (see \S\ref{xmatch}).
The remainder are missed by ALFALFA either because they are (1) gas poor, 
(2) lie at a sufficient distance that 
their HI line flux densities falls below the HI sensitivity limit, or (3) for some other 
reason, e.g. their HI spectrum is contaminated by RFI or was not sampled at all (small
gaps in ALFALFA coverage), or they correspond to one ``child'' of a shredded photometric 
parent object, but another photometric child is favored as the best match to the ALFALFA detection. 

We note that the distributions of weight, $V_{sur}/V_{max}$, for both the samples 
highly peak at 1, and that the weight cut of 60 applied to each dataset is confirmed 
to be high enough to retain the main populations. 
Especially for the $S_{opt}$, 69\% of the galaxies have a unit weight, i.e. can be detected outwards to the edge of $V_{sur}$ as we defined here. The number of galaxies in bins associated 
with a weight above 1 drops more rapidly among the $S_{opt}$ sample, confirming 
that the SDSS is deeper than ALFALFA. 

\subsection{Comparison of control samples}
\label{comp}

\subsubsection{Basic properties}
\label{basic}

\begin{figure*}
\center{
\includegraphics[scale=1]{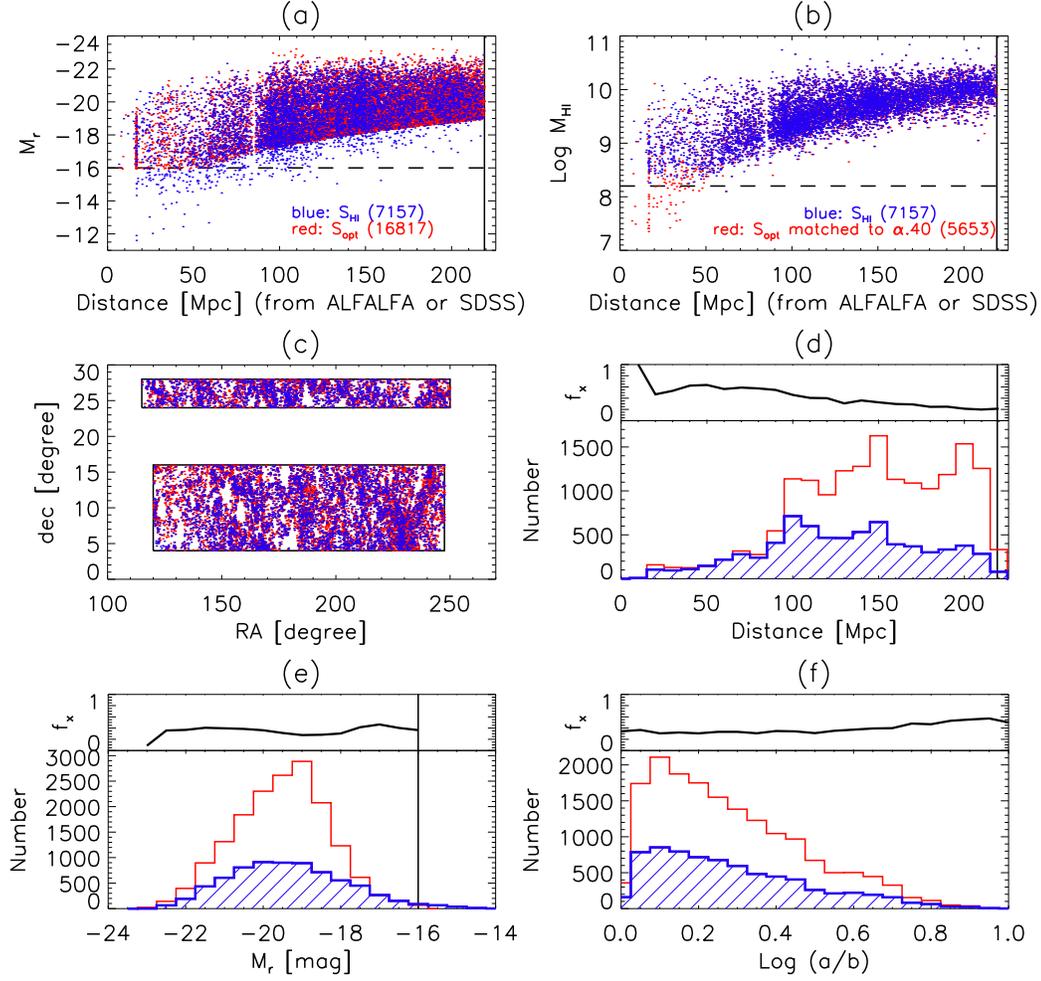}
}
\caption[]{Comparison of the basic properties relevant to the sample selection of the
HI-selected sample
$S_{HI}$ (
7157 galaxies selected from the $\alpha.40$, blue points and blue histograms) and 
the optically-selected one
$S_{opt}$ (16817 galaxies selected from the SDSS, red points and red histograms). 
The panels above the histograms show the fraction of the $S_{opt}$ galaxies 
that are cross-matched to $\alpha.40$ in each bin, similar to an HI detection rate of 
the $S_{opt}$ galaxies by ALFALFA. 
Both samples are extracted from the same sky area (panel c), lie within 
$cz = 15000~{\rm km~s^{-1}}$ (panel d) and are cross-matched to the GALEX catalog. 
Spaenhauer diagrams are in panels (a) and (b), i.e. $r$-band 
absolute magnitude, $M_r$, and $M_{HI}$ against distance. A weight (or volume correction) 
cut of 60 applied on $S_{opt}$ results in $M_r$ brighter than $\sim -16$, 
whereas applied on $S_{HI}$ results in 
$M_{HI} \gtrsim 10^{8.2}~{\rm M_\odot}$. 
Though the two samples probe similar $M_r$ ranges (panel e), the SDSS is deeper than ALFALFA, 
as evident in the distance distribution (panels d). 
No bias against edge-on galaxies in the $S_{HI}$ population is seen in panel (f). 

(A color version of this figure is available in the online journal.) 
}
\label{fig:basic}
\end{figure*}

Figure \ref{fig:basic} illustrates the comparison of quantities relevant to sample selection 
between the $S_{HI}$ and the $S_{opt}$ populations. In panels (a-c), red points denote galaxies in $S_{opt}$, 
whereas blue points denote galaxies in $S_{HI}$. In the histograms, red lines illustrate the distribution of 
$S_{opt}$ and blue lines trace $S_{HI}$; above each histogram, separate panels show the fraction of $S_{opt}$ galaxies 
that are cross-matched to $\alpha.40$ in each bin. The numbers in each subset are indicated in panels (a) and (b). 

The top row contains two Spaenhauer diagrams showing, respectively, the $r$-band absolute magnitude (panel a) and HI mass (panel b) versus distance. 
The solid vertical line represents the $cz$ cutoff, 15000~km~s$^{-1}$. 
We use the SDSS redshifts to derive distances for the $S_{opt}$ and ALFALFA HI velocity for the $S_{HI}$. As discussed also by \citet{Martin2010}, a survey must sample sufficient volume 
to detect very massive galaxies in either stellar (panel a) or HI (panel b) mass. 
ALFALFA for the first time provides a full census of HI-bearing 
objects over a cosmologically significant volume of the local universe. 

As evident in  Figure \ref{fig:basic}(a), SDSS is volume limited to $M_r \sim -19$ mag. 
The sharp lower edge of the $S_{opt}$ distribution above $D \sim 60$ Mpc
results from the magnitude limit of the SDSS main galaxy redshift sample; as noted
before, the adopted weight cutoff
corresponds to $M_r \sim -16$ mag (horizontal dashed line). 
Since no limit on any optical quantity is applied to the $S_{HI}$ subset, 
many blue points from $\alpha.40$ show up faintwards of the lower edge of $S_{opt}$, 
as faint as $M_r\sim-14$. 
We note that the blue points lying faintwards of the lower edge of $S_{opt}$ and above 
$D \sim 60$~Mpc are still detected by the SDSS, but most often only 
as photometric objects; hence their optical redshifts are generally unknown. 
To enable SED-fitting however, all the $S_{HI}$ galaxies are required to be detected 
in the SDSS; the very rare ``dark'' HI clouds without identified OCs included in $\alpha.40$ 
are outside the scope of this work and are not included in this discussion. 

Figure \ref{fig:basic}(b) shows how $M_{HI}$ increases with distance. 
The HI measures of $S_{opt}$ all come from the $\alpha.40$ catalog 
(5653 out of 16817 $S_{opt}$ galaxies, i.e., a cross-match rate of 34\%). 
The weight cutoff applied to the $S_{HI}$ sample results in an approximate 
limit of
$M_{HI} \simeq 10^{8.2}~{\rm M_\odot}$ (horizontal dashed line). 
The red points below $10^{8.2}~{\rm M_\odot}$ are still detected by $\alpha.40$, 
but are excluded from the $S_{HI}$ simply because of their high weights. 
Because of the way distances are derived \citep{Haynes2011} from the observed redshifts, 
an insignificant difference (0.1 Mpc) of the median distance arises for the same galaxies in the 
$S_{opt}$--$S_{HI}$ overlap sample. This difference is mainly due to the fact that the
$S_{HI}$ galaxies may be assigned membership in group whereas such information has not been
applied to the $S_{opt}$ sample. 
A few red outliers below the main distribution indicate objects without robust SDSS redshifts. 

Panel (c) shows the sky distribution of the two samples. 
Besides the large scale structure, the required availability of GALEX 
photometry also contributes to the pattern. For example, the Virgo region is densely 
covered by at least MIS level visits, whereas the patches of sky blank in either 
sample and with regular edges arise from the lack of GALEX coverage in FUV and/or NUV. 
The distribution of distance is shown in panel (d). The two samples roughly coincide within 
$\sim$100 Mpc. Above the histogram, the fraction of the $S_{opt}$ galaxies that 
are cross-matched to $\alpha.40$ is shown. 
The first peak in number density coincides with the Virgo cluster
($\sim$16.5 Mpc), where $S_{opt}$ clearly out numbers $S_{HI}$. Since the two distributions
agree with each other again at larger distances $\sim$50 Mpc, 
the disagreement at the Virgo distance indicates a real underdensity 
of gas-rich detections in $S_{HI}$, reflecting 
the well-known HI deficiency \citep{Davies1973, Giovanelli1985, Solanes2002}. 
Beyond $\sim$100 Mpc, 
$S_{opt}$ significantly overtakes $S_{HI}$, though the shapes of peaks or gaps 
still agree. This suggests that ALFALFA is capable of detecting HI massive objects 
at large distances, although the survey is not as deep as SDSS. Although in a given 
distance bin, the least massive objects contained in $S_{HI}$ are even fainter 
than those in $S_{opt}$ (i.e., the blue points below the lower edge of 
red distribution in panel a), ALFALFA is not as complete as SDSS at large distances. 

The distributions of $M_r$ for both samples are shown in panel (e). The vertical 
solid line denotes the equivalent weight cut applied to $S_{opt}$. 
While $S_{opt}$ peaks at a slightly fainter 
$M_r$ than $S_{HI}$, 
the two samples probe a similar $M_r$ range so that their comparison is valid.

The distributions of axial ratio, given by the SDSS pipeline measures of the 
exponential fit $a/b$ in $r$-band, are shown in panel (f). Because the ALFALFA sensitivity depends on the HI line profile width (see \S\ref{sample}), 
$S_{HI}$ might be expected to be biased against edge-on galaxies with high $a/b$ values. 
For example, \citet{West2010} demonstrated that their Parkes Equatorial Survey 
\citep[ES, a search through HIPASS cubes;][]{Garcia-Appadoo2009} -- SDSS common 
sample is slightly biased towards face-on galaxies, relative to an SDSS DR4 sample, 
with the mean $\log a/b$ equal to 0.17 and 0.21 for their ES--SDSS and DR4 samples respectively. 
However, panel (f) shows no such obvious bias. 
Both $S_{HI}$ and $S_{opt}$ have the same $\langle \log a/b\rangle =0.28$. 
Furthermore, the cross-match rate 
even slightly rises for high $a/b$ galaxies, with only a mild drop in the very last bin. 
Visual inspection shows that shredding can cause large errors in the $a/b$ measures 
by the SDSS pipeline. 
The $S_{opt}$ sample contains more galaxies with bulges making their $a/b$ 
values appear to be smaller; in contrast, $S_{HI}$ is biased against such galaxies. 

\subsubsection{Internal extinction in HI-selected galaxies}
\label{ext}

\begin{figure*}
\center{
\includegraphics[scale=0.9]{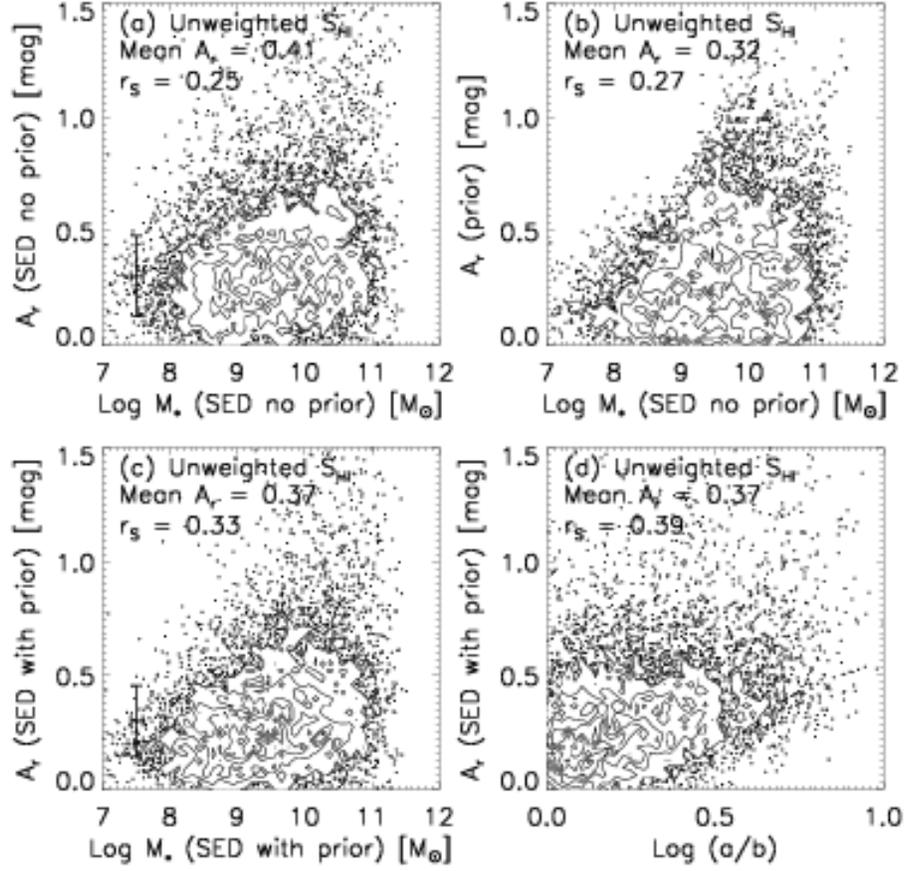}
}
\caption[]{
{\it Panels (a-c) -} The $r$-band internal extinction versus stellar mass. 
	The values in panel (a) are derived by SED-fitting without the application of a prior 
	$\tau_V$ distribution. 
	The mean of the prior distribution is plotted in panel (b), 
	based on equation (12) in \citet{Giovanelli1997}. 
	The SED-fitting has been improved by adopting
     the prior $\tau_V$ distribution as evident in panel (c). 
	Internal extinction is a weakly increasing function of $M_*$ with the greatest correlation,
      with a coefficient = 0.33, shown in panel (c). 
{\it Panel (d) -} Internal extinction is systematically higher in more inclined galaxies, as expected. 
}
\label{fig:Oext}
\end{figure*}

\begin{figure*}
\center{
\includegraphics[scale=0.9]{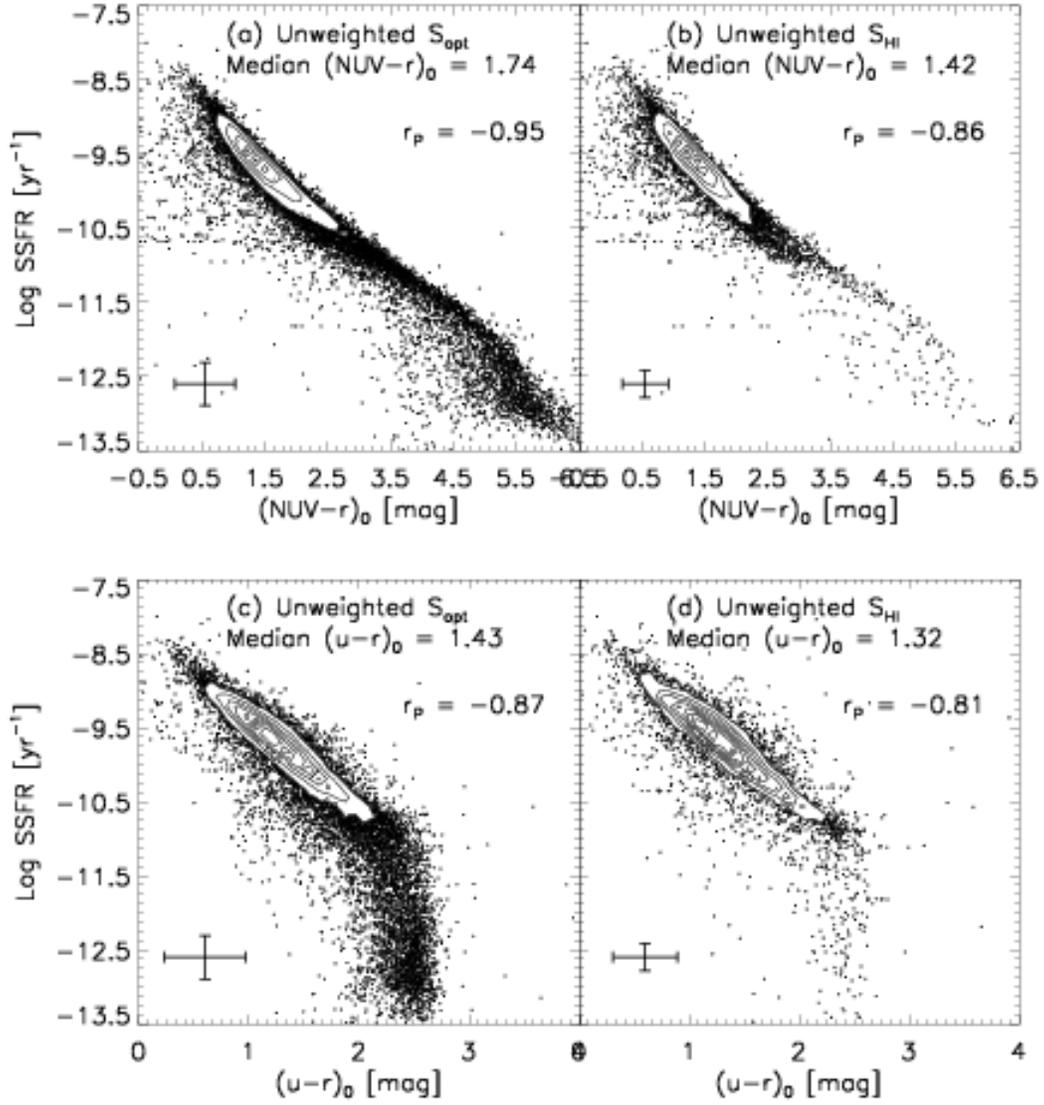}
}
\caption[]{$SSFR$ versus intrinsic colors, denoted by the subscript zero, 
after correction for internal extinction. 
The left column shows results for $S_{opt}$ while the right shows $S_{HI}$. 
Pearson correlation coefficients and typical error bars are also shown and the 
median intrinsic colors are indicated. 
In addition to the blueward shift of the distribution in panel (b) 
the scatter here is also greatly reduced compared to the similar plot 
shown in Figure \ref{fig:SF}(f) where no internal extinction correction has been applied. 
The ($NUV-r$) color breaks down the degeneracy of ($u-r$) in the red range when inferring the 
SFH. The red and low SSFR tail in $S_{opt}$ disappears in $S_{HI}$. 
$S_{HI}$ is on average intrinsically bluer than $S_{opt}$ in both colors.  
}
\label{fig:SFS6b}
\end{figure*}

\begin{figure*}
\center{
\includegraphics[scale=1]{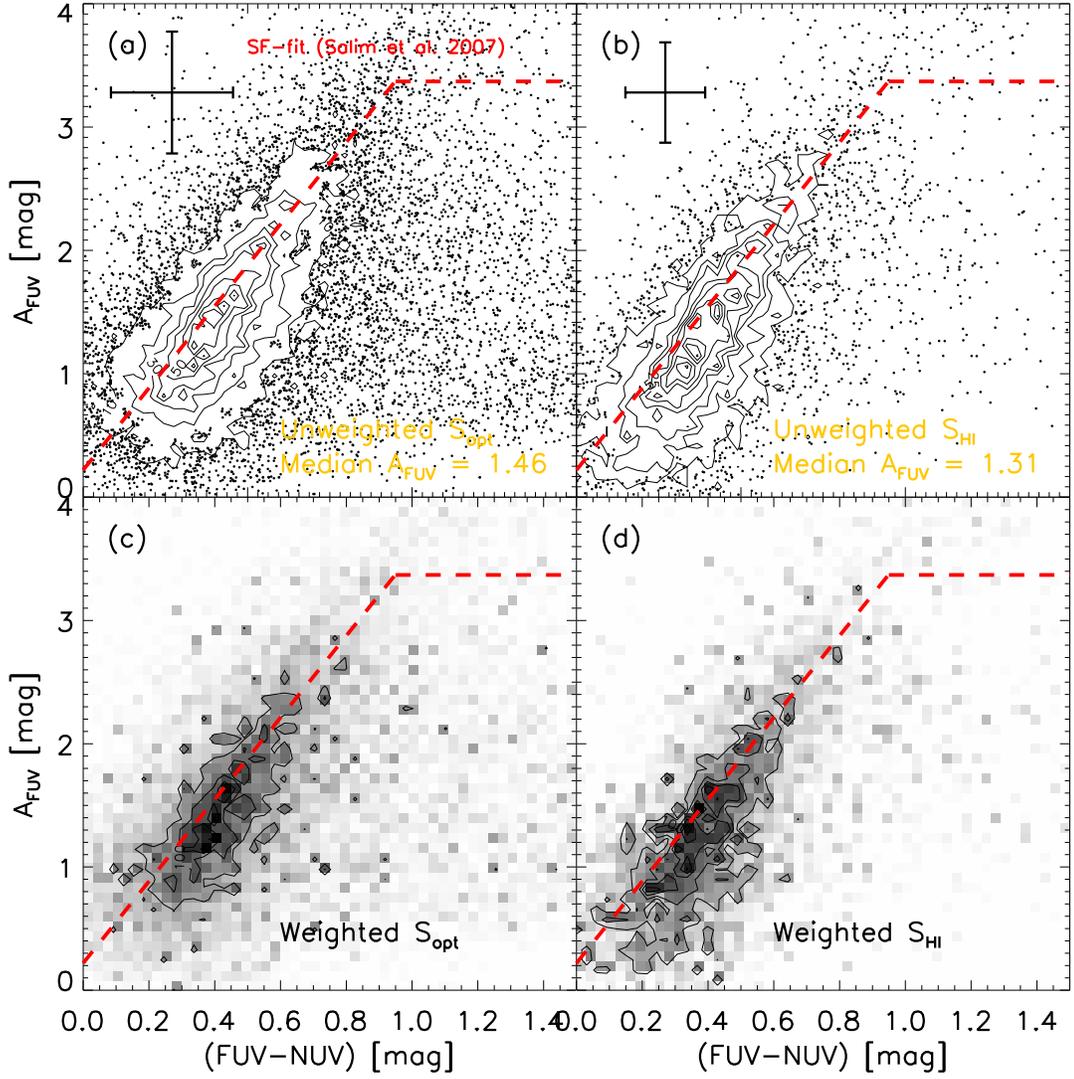}
}
\caption[]{
Computed $A_{FUV}$ -- UV color diagrams for 
$S_{opt}$ in the left column and $S_{HI}$ in the right column. 
The upper row shows the results for the galaxies themselves, without weighting 
while the bottom row shows results after applying the $V_{sur}/V_{max}$ 
weighting scheme. 
Typical error bars are shown in the corners of panels (a) and (b), 
together with the median $A_{FUV}$ values listed for both the samples. 
The red dashed line corresponds 
to the fit to the star-forming galaxies derived by \citet{Salim2007}, based on a 
typical local SDSS-GALEX cross-matched catalog. It is in agreement with 
$S_{opt}$. In contrast, the $S_{HI}$ distribution is offset from the fit, with lower $A_{FUV}$ at a fixed $(FUV-NUV)$ 
color, which may due to lower metallicity, different SFH, 
and/or dust geometry of the HI-selected galaxies. 
}
\label{fig:UVext}
\end{figure*}

Previously, and in many analyses of SDSS derived samples, internal extinction is ignored. 
However, while the outer parts of galaxy disks are transparent, it is well established 
that the inner regions are optically thick at short wavelengths. Therefore, the neglect of
internal extinction in disk-dominated galaxies is likely to introduce systematic
inclination-dependent effects. In this section, we discuss (a) how internal extinction 
varies with stellar mass, 
(b) how internal extinction may introduce scatter into relationships involving colors and 
(c) how the extinction characteristics of the $S_{HI}$ HI-selected galaxies compare to those
derived from an optically selected sample $S_{opt}$. In the three figures associated with
this section, Figure \ref{fig:Oext} - \ref{fig:UVext}, typical error bars on individual 
points are plotted in selected panels in 
as well as the Spearman's rank correlation coefficients. 

Estimates of internal extinction are derived from UV/optical SED-fitting as before.
The two-component dust model \citep{Charlot2000} is incorporated into the 
construction of the library of model SEDs \citep{Gallazzi2005}; 
the process accounts for both the diffuse interstellar medium (ISM) and short-lived 
(10~Myr) giant molecular clouds. 
Such estimates have overall larger uncertainties among the red-sequence galaxies 
relative to the blue cloud ones \citep[e.g.][]{Saintonge2011b}, 
and contribute to the SFR uncertainties. 
Furthermore, \citet{Salim2007} identified differences between the effective 
optical depth in $V$ band, $\tau_V$, derived from emission-line fitting and that derived from SED-fitting, 
as a function of stellar mass. 
Specifically, at lower masses, the SED-fitting-derived value is systematically higher than the 
line-fitting-derived one, but the situation is reversed at the high mass end. 
Therefore, we applied a Gaussian prior distribution of $\tau_V$ for each model, given the 
absolute magnitude and axial ratio of the individual galaxies (see \S\ref{SED}). 
The mean of the prior distribution is given by equation (12) in \citet{Giovanelli1997}, 
which depends on the axial ratio and absolute magnitude, 
i.e. more luminous edge-on galaxies have larger extinctions. 

Figures \ref{fig:Oext} (a-c) show plots of $r$-band internal extinction versus stellar mass for the $S_{HI}$ galaxies. 
Despite the large uncertainty, internal extinction is a weakly
increasing function of $M_*$ in all these panels. 
Panel (a) shows SED-fitting-derived values before the prior distribution applied. 
A population of low mass red galaxies have unrealistically high $A_r$ (SED no prior), 
because of the age-extinction degeneracy. 
The mean of the prior distribution of internal extinction, $A_r$ (prior), 
is in panel (b). Although the $A_r$ (prior) values of low mass galaxies are confined to 
low values, a population of massive galaxies have unrealistically low $A_r$ (prior) likely
due to the underestimate of $a/b$. Visual inspection shows that 
shredding tends to describe sources as rounder in edge-on galaxies; dominant bulges,
dust lanes and seeing effects will also lead to systematic underestimates of the axial ratio.
Panel (c) plots the SED-fitting-derived values after the prior distribution is applied, $A_r$ (SED with prior). 
Both a lack of high mass galaxies with low $A_r$ as well as of low mass ones with high $A_r$ 
are evident in panel (c). Combining the distributions in panel (a) and (b), the 
$A_r$ (SED with prior)--$M_*$ correlation is the tightest of the three, with a correlation coefficient $r_S=0.33$. 
Whereas applying the prior reduces the systematic offsets of the $A_r$ estimates 
by SED-fitting, as well as the $SFR$ values, it has little effect on the stellar mass values \citep{Huang2012}. Figure \ref{fig:Oext} (d) demonstrates that, as expected, the
derived values of $A_r$ (SED with prior) are higher in more edge-on galaxies. 
Neglect of corrections for internal extinction will lead to the systematic 
underestimate of luminosity, so that hereafter we apply the SED-fitting with prior
corrections. 

For the subset of the $S_{HI}$ galaxies (6164/7157) which are included in the MPA-JHU DR7 release of SDSS 
spectral measurements \citep[http://www.mpa-garching.mpg.de/SDSS/DR7/, ][]{Brinchmann2004}, 
we have verified that at $M_* \lesssim 10^{10}M_\odot$ the $A_r$ inferred from the Balmer decrement 
and from the SED fitting using an $A_r$ prior are in good agreement and the offset observed in \citet{Salim2007} is 
reduced. 
Above this mass the Balmer decrement leads to larger $A_r$ values, 
but this is not unexpected as the SDSS spectra observe only a small region, typically towards the center of the 
galaxy where metallicities and dust attenuations are higher.

Another way to explore the importance of extinction correction in a population 
involves examining the scatter in the $\log SSFR$ -- color diagram, as shown in 
Figure \ref{fig:SFS6b}. Results for individual points (unweighted) for the 
$S_{opt}$ sample are shown in the 
left panels and for the $S_{HI}$ galaxies on the right, using the colors
$(NUV-r)_0$ in the top row and $(u-r)_0$ in the bottom, respectively.
The subscript `0' in the labels indicates that the colors are 
corrected for internal extinction.
As demonstrated in \S\ref{SFS}, $SSFR$ is an intrinsically strong function of ($NUV-r$), 
because NUV traces the SFR and the $r$-band luminosity is related to the stellar mass. 
For the HI-selected population, comparison can be made directly of 
the volume-limited $S_{HI}$ sample shown in Figure \ref{fig:SFS6b}(b) with the
corresponding result, uncorrected for internal extinction, shown in Figure \ref{fig:SF}(f)
for the full $\alpha$.40-SDSS-GALEX. As evident by inspection, in addition to a 
shift bluewards of the points in Figure \ref{fig:SFS6b}(b), 
the dispersion about the mean relation is greatly reduced when the extinction correction
is applied, and the Pearson correlation coefficient likewise improves from $r_P=-0.78$ to $r_P=-0.86$. 
This analysis indicates that the scatter in Figure \ref{fig:SF}(f) is substantially
amplified by the lack of a correction for dust extinction. Other factors causing scatter
include different SFHs, metallicities, 
as well as different population synthesis models and even IMFs \citep{Gunawardhana2011}. 
However, dust extinction dominates among these factors. 
Furthermore, the star-forming $S_{HI}$ galaxies have on average 
bluer $(NUV-r)_0$ colors than the $S_{opt}$ ones (median 1.42 versus 1.74 mag). 
Meanwhile, the tail of red and low SSFR galaxies in panel (a) disappears in panel (b), 
again reflecting the bias present in the HI-selected population. An additional
result of the HI-selection is that the typical error bar in panel (b) is slightly smaller
than in (a). 

The lower panels in Figure \ref{fig:SFS6b} examine the optical SDSS colors $(u-r)_0$. 
As discussed previously in \S\ref{xmatch}, on average $\delta(u-r)/\delta(NUV-r) \sim 0.6$.
Figure \ref{fig:SFS6b}(c) shows that galaxies with $SSFR \lesssim 10^{-11}~{\rm yr^{-1}}$ 
have similar ($u-r$) colors, 
forming a vertical tail in these plots; this red tail is much less pronounced in panel (d). 
The adoption of ($NUV-r$) as the color index breaks down the degeneracy of ($u-r$) 
in the red range ($u-r \gtrsim 2.3$) when inferences on the SFH are inferred.
Furthermore, the SSFR correlates more linearly with $(NUV-r)_0$ than with 
$(u-r)_0$, e.g. $r_P=-0.95$ versus $-0.87$ in $S_{opt}$. 

Given a simple assumption of the dust and stellar geometry, a well calibrated $IRX-\beta$ 
relation ($IRX$, infrared-excess defined as $L_{TIR}/L_{FUV}$; $\beta$, the UV spectral slope) 
is sufficient to predict the $A_{FUV}$ (tightly correlated with $IRX$) from the UV color 
(characterizing the UV spectral slope, $\beta$), in starburst galaxies \citep{Calzetti1994}. 
However, the $IRX-\beta$ relation in normal star-forming galaxies has a shallower slope with larger 
scatter \citep{Salim2007}. This result may be due to differences in the SFH \citep{Kong2004}, 
or dust geometry \citep{Cortese2006}. 
Therefore, the loci of galaxies on an $A_{FUV}$ versus $(FUV-NUV)$ plot give an indication of 
the dust extinction behavior. 

Following this approach, the two samples $S_{HI}$ and $S_{opt}$ are compared in Figure \ref{fig:UVext}, with results for
$S_{opt}$ on the left and $S_{HI}$ on the right, both
before and after applying the weight correction (see \S\ref{sample}) in the upper and bottom rows respectively. 
The red dashed line corresponds to the fit to star-forming galaxies derived in \citet{Salim2007}, 
based on a typical local SDSS-GALEX cross-matched catalog. 
Despite the large scatter, it is on average in close agreement with the distribution of the $S_{opt}$ galaxies. 
Galaxies closer to the fit have overall higher weights so that the correlation appears to be slightly 
tighter in panel (c) than in (a). However, the distribution of the $S_{HI}$ galaxies is offset from the fit 
in panels (b) and (d), toward the lower $A_{FUV}$ side, i.e. for a given $(FUV-NUV)$ color, the HI-selected galaxies on average have lower extinctions. This result is also confirmed by the generally lower 
$A_{FUV}$ of the $S_{HI}$ sample overall, with a median value of  
1.31 mag, relative to that of $S_{opt}$, 1.46 mag.
Such a deviation of the main trend from the SF-fit of \citet{Salim2007} 
suggests that the HI-selected galaxies have different SFHs or dust geometries.

Unfortunately, the metallicity of the stellar population is poorly constrained by the SED-fitting. 
Nevertheless, given the correlation between dust and metallicity \citep[e.g.][]{Draine2007}, 
the lower extinction infers that the $S_{HI}$ galaxies have lower metallicity. 
Besides the well-known mass-metallicity (gas-phase) relation \citep[e.g.][]{Tremonti2004}, 
\citet{Mannucci2010} demonstrate that at lower $M_*$($<10^{10.9} \rm M_\odot$), metallicity 
decreases sharply with increasing SFR, while at high stellar mass, 
metallicity does not depend on the SFR. Given the bias towards high SFR of the $S_{HI}$ sample 
at a given $M_*$ (see \S\ref{SF}), the bias towards low metallicity and low extinction is expected. 
The gas-phase metallicity measures (oxygen abundance) from the MPA-JHU DR7 release 
\citep{Tremonti2004} are available for 4211/7157 of the $S_{HI}$ galaxies and 10311/16817 
of the $S_{opt}$ galaxies. 
In addition to the caveat of the small SDSS fiber aperture, the requirement of being an SDSS spectroscopic target 
may reduce the difference between the two subsets. 
The mean $12+\log(O/H)$ is only slightly higher among the $S_{opt}$ galaxies (8.74) than in the $S_{HI}$ 
galaxies (8.71). 
The overall lower extinction among the $S_{HI}$ galaxies is consistent with the idea that 
HI-selected galaxies are relatively gas rich and less evolved, with more likely gas 
infall, lower SFE and metallicity, less dust and thus lower extinction. 

\subsubsection{Distribution in the intrinsic CMD}
\label{CMD}

The bimodal distribution in the optical CMD \citep{Baldry2004} has been interpreted 
as an evolutionary sequence, with the blue-cloud galaxies growing through mergers and 
the consumption of gas and later migrating to the red sequence. This evolutionary scenario 
also predicts that galaxies evolve from a state of low stellar mass, high SSFR
and high HI fraction to the opposite. 
As discussed in \S\ref{MHIMs}, the HI fraction is lower in redder galaxies
leading to the result that an HI-selected sample like $S_{HI}$ is biased against red galaxies
whereas those are commonly included in, and may even dominate, an optically-selected one like $S_{opt}$. 
Here, we use the CMD to quantify this bias. 

\begin{figure*}
\center{
\includegraphics[scale=0.9]{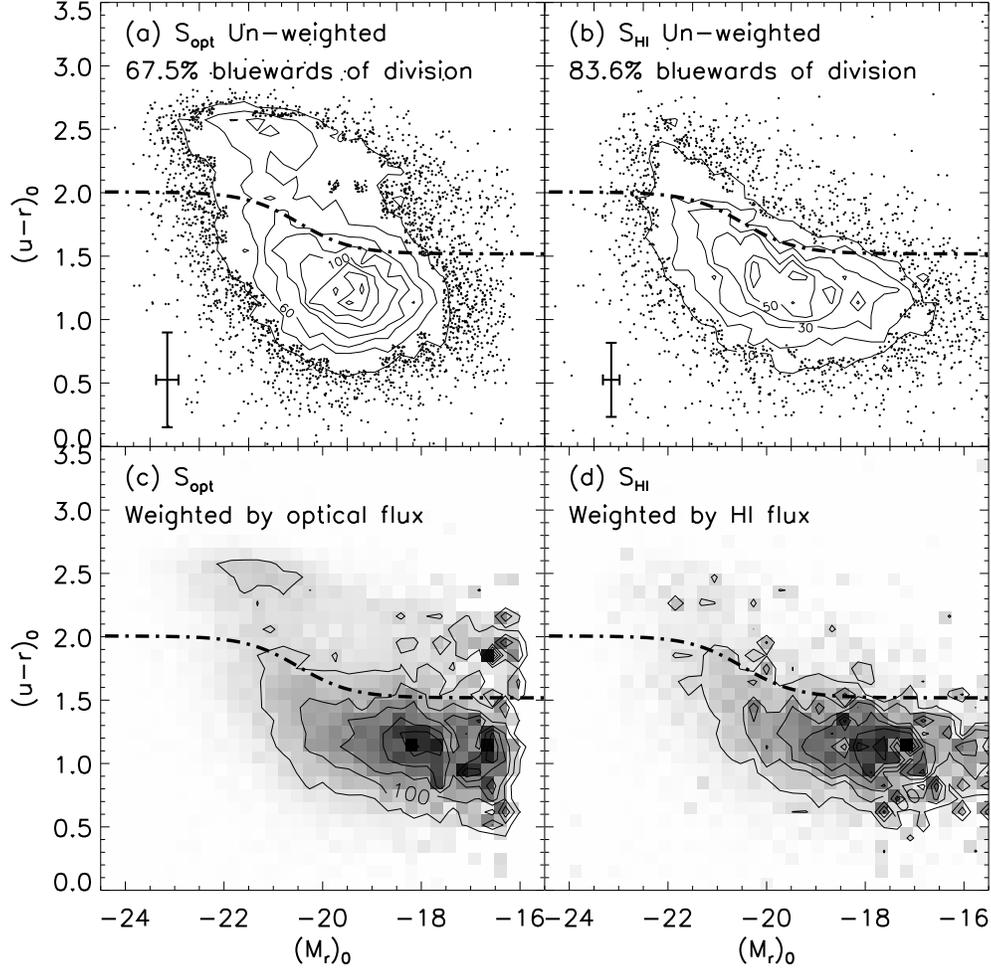}
}
\caption[]{Intrinsic optical color-magnitude diagram of $S_{opt}$ (left panels) and $S_{HI}$ (right panels); 
before (upper) and after (lower) applying the $V_{sur}/V_{max}$ weight correction. 
The dash-dotted curve is derived in \citet{Baldry2004} as the best fit to the division of red 
sequence and blue cloud, shifted for the extinction corrections. 
Typical error bars including the uncertainties in the extinction corrections are plotted in the 
lower left corner of panels (a) and (b). 
The bimodel distribution is more evident in the $S_{opt}$ representations; colors become generally bluer 
in fainter galaxies for both populations. 
The weighted panels better represent the luminosity function, which predicts more 
faint galaxies relative to the bright ones in the blue cloud. However, a second 
peak of number density at the faint end is seen on the red sequence, which suggests 
a diversity of SFH within the dwarf galaxies. 
A higher percentage of galaxies lie bluewards of the division in $S_{HI}$ 
(84\%) relative to $S_{opt}$ 
(67.5\%). HI surveys such as ALFALFA are highly biased against red sequence galaxies. 
}
\label{fig:Baldry}
\end{figure*}

Figure \ref{fig:Baldry} shows the CMDs of the $S_{opt}$ (left panels) and the $S_{HI}$ 
(right panels); before and after applying weight corrections in the upper and bottom rows respectively. 
In these plots, absolute magnitudes $(M_r)_0$ and $(u-r)_0$ colors are corrected for internal extinction 
as given by the SED-fitting with the prior applied. 
A similar CMD for the whole $\alpha.40$-SDSS population but without corrections can be found in 
Figure 7 of \citet{Haynes2011}. 
The dash-dotted curve is based on the best fit to the division of the red and blue populations 
derived by \citet{Baldry2004} but with shifts toward bluer colors and brighter $M_r$ applied in the 
amount of the mean extinction of the $S_{opt}$ sample. 

As is obvious in panels (a) and (b), the 
red sequence is far more pronounced in $S_{opt}$ than in $S_{HI}$. 
Among the $S_{opt}$ galaxies, only
68\% lie on the blue side of the division, 
whereas the percentage of blue galaxies in the HI-selected $S_{HI}$ population is as high as 
84\%. The small population of `red' galaxies with HI represents
candidate objects which (i) have recently migrated onto the 
red sequence retaining some residual gas, or (ii) will transit back to the blue cloud 
via late gas re-accretion \citep{Cortese2009}. Note that most of the red HI-bearing galaxies 
optically luminous. The HI mass of the faint red galaxies is usually so low that 
their HI line flux densities are below the sensitivity limit of ALFALFA. 

In the $V_{sur}/V_{max}$ weight corrected diagrams (panels c and d), the peaks of number density shift 
to fainter $(M_r)_0$ in both samples. Galaxies densely occupy the faintest $(M_r)_0$ bins,
especially in the $S_{opt}$ sample. 
This result is consistent with the optical luminosity function of the blue population \citep{Baldry2004}. 
A peak of the number density on the red sequence around $(M_r)_0 \sim -21$ also 
coincides with the maxima of the luminosity function derived by \citet{Baldry2004}. 
However, a second peak appears on the red sequence at the faint end. 
Note that the data points in \citet{Baldry2004} also suggest a rise in number 
density in the faintest bins of the red sequence luminosity function (see their Figure 7), 
though those authors ignore these objects when they fit the Schechter function. 
In agreement with this, \citet{Hogg2003} reported a non-monotonic density trend 
along the red sequence, seen as a dip in the typical density for intermediate-mass 
red-sequence galaxies despite higher densities at higher and lower masses. 
Similar to the finding of a sudden broadening in the SSFR distribution faintwards of 
$M_B \sim -15$ \citep{Lee2007}, a sudden broadening in the color distribution 
faintwards of $(M_r)_0 \sim -16$ is seen in the $S_{opt}$ sample. 
This increased spread suggests that, unlike the galaxies of intermediate mass in which the SF 
is mainly regulated by stellar mass and for which the blue population dominates in number, 
the gas in dwarfs can be easily removed so that SF suddenly quenches driving their migration
onto the red sequence. 

Figure \ref{fig:Baldry}(c) shows a CMD for an optically-selected population 
similar to Figure 2 of \citet{Baldry2004}. However, $S_{opt}$ is limited to $cz < 15000~{\rm km~s^{-1}}$ 
whereas the low-redshift sample \citet{Baldry2004} studied has $1200~{\rm km~s^{-1}}< cz < 24000~{\rm km~s^{-1}}$. 
The imposition of the volume cut here clearly decreases the presence of luminous red galaxies; 
the local volume is dominated by blue star-forming galaxies. 
This effect is also enhanced by the additional UV selection applied to the $S_{opt}$ sample. 

The near-absence of the red sequence among the $S_{HI}$ population is clear 
in panel (d); as we have noted before, 
a blind HI survey like ALFALFA is highly biased against red sequence galaxies. 
Additionally, in both populations galaxies appear to be bluer 
as their $M_r$ gets fainter, though with large scatter. Such a trend is weaker but 
still visible after internal extinction corrections are applied. 
Therefore, the slope is not only due to extinction, but is intrinsic. 

Besides comparing the CMDs of $S_{opt}$ and $S_{HI}$, we can also study the impact of HI 
selection through an examination of the fraction of $S_{opt}$ galaxies that are cross 
matched to $\alpha.40$. As was discussed in \S\ref{sample}, 34\% of the $S_{opt}$ 
galaxies are cross-matched to the $\alpha.40$ catalog. 
However, we note that such a fraction is a lower limit of the HI-detection rate of 
optical-UV selected galaxies in $\alpha.40$,  
because some $S_{opt}$ objects may be a shredded 
photometric object of a gas bearer while another piece is cross-matched to the $\alpha.40$ entry. 
The fraction of these objects should however not be large 
\citep[see][]{Haynes2011}. 

Figure \ref{fig:dtrt} explores the fraction of the $S_{opt}$ population included in the $\alpha.40$ 
within the CMD (left) and the $SSFR$--$M_*$ diagram (right). In the left panel, the
NUV-to-optical CMD with internal extinction corrections is color coded by the fraction of galaxies 
in the optically-selected $S_{opt}$ sample which are also included in the $\alpha.40$; the
contours trace the total number density of the $S_{opt}$ galaxies. Clearly, an HI survey like
ALFALFA is more efficient at detecting blue galaxies than red ones. 
Especially at very bright end of the blue cloud (lower left corner), 
the detection rate is close to 100\%. 
Furthermore, the ALFALFA population also detects the majority of optically
faint blue galaxies with the detection rate again approaching
$\sim$100\% (lower right corner). 
The latter have the highest HI fractions (see also Figure \ref{fig:ccfgas}). 
HI survey is most efficient in detecting galaxies with large diameter and high $M_{HI}/L$ values 
\citep{Garcia-Appadoo2009}. 
The general shape of the color scale variation above a cross-match rate of $\sim$40\% 
resembles that of the number density variation on the bluer side of the blue cloud. 
In contrast, starting from the red edge of the blue cloud and throughout the whole 
red sequence region, the rate of detection by ALFALFA drops below 10\% and even to $\sim$0\%. 
These results again demonstrate that HI-selection is highly biased against 
galaxies on the red sequence and furthermore, 
that such bias begins to become pronounced on the redder portions of what still
is considered the blue cloud. The precise location of this limit to some extent
reflects the `fast' nature of
ALFALFA (only 40 seconds integration per beam); a deeper survey of comparable solid
angle would probe to lower gas fractions and hence deliver a higher cross-match rate.

\subsubsection{Star-formation sequence}
\label{SF}

The scaling relation that the SFR increases with $M_*$ among the $\alpha.40$ galaxies is demonstrated in 
\S\ref{SFS} (Figure \ref{fig:SF}a). Figure \ref{fig:SF}(d) similarly shows the well confined 
star forming sequence in the $SSFR$--$M_*$ diagram. 
Based on the ES--SDSS sample, \citet{West2009} suggested that the significant offset towards bluer colors, 
induced by HI-selection can be explained by enhanced recent bursts of star formation. 
Furthermore, rising SFHs, i.e. high $b-$parameter, are required to explain the colors of HI-selected galaxies 
bluer than $g-r<0.3$, which may result from gas infall and enhanced SF subsequently. 
Here, we compare the distributions of the $S_{HI}$ and $S_{opt}$ galaxies, 
on and off the star-forming sequences to explore further the impact of HI-selection. 

Figure \ref{fig:SFS1} examines the $SFR$ versus $M_*$ scaling relation for the optical (left) 
and HI (right) selected samples. 
The upper row shows the individual galaxies while the lower traces the $V_{sur}/V_{max}$ 
weighted distributions. 
The cyan dashed line at $M_* \geq 10^{10}~{\rm M_\odot}$ shows 
the fit to this relation obtained by \citet{Schiminovich2010}, based on the high $M_*$ GASS sample in all four panels: 
\begin{eqnarray*}
	\log \langle SFR(M_*) \rangle = 0.15 \log M_* - 1.5,~\log M_* > 10.0.  
\end{eqnarray*} 
Like $S_{opt}$, GASS is essentially an SDSS selected sample but only includes the high stellar mass end. 
The blue diamonds and lines represent $\log \langle SFR \rangle$ derived for $S_{opt}$ or $S_{HI}$ in each stellar 
mass bin. 
Note these are different from the average trends in Figure \ref{fig:gas} or Figure \ref{fig:SF}. 
The values are weighted averages in the weighted panels. 
The numbers of galaxies in each stellar mass bins are listed at the top of panels (a) and (b). 
Typical error bars are shown in panels (a, b). 

\begin{figure*}
\center{
\includegraphics[scale=0.8]{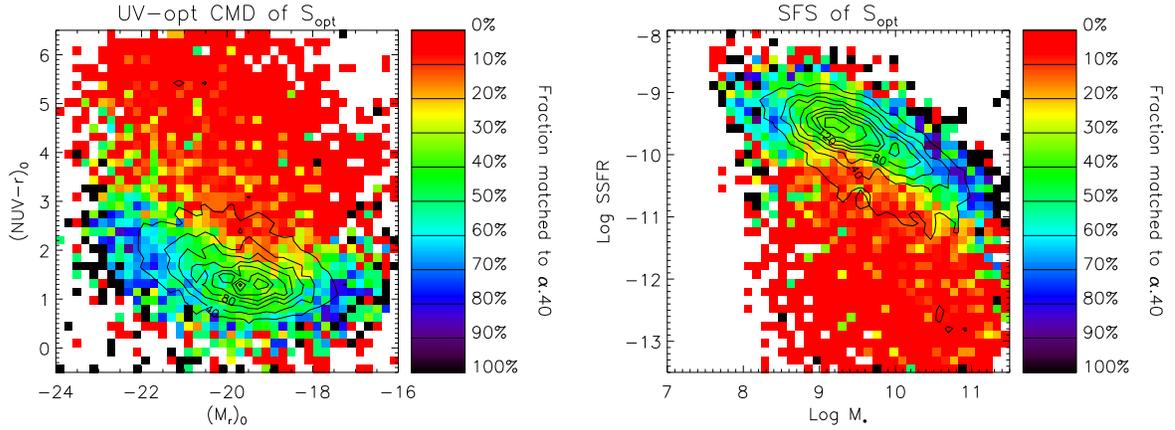}
}
\caption[]{Diagrams with shade scale showing the fraction of galaxies 
in $S_{opt}$ that are cross-matched to the $\alpha.40$ in each grid, 
which is close to the HI detection rate of the $S_{opt}$ by $\alpha.40$. 
The $S_{opt}$ number density is indicated by the contours. 
The cross-match fraction is the lower limit of the detection rate of $S_{opt}$ galaxies by $\alpha.40$; 
its overall average is 34\%.  
{\it Left panel -} Intrinsic UV-to-optical CMD of $S_{opt}$. ALFALFA is more efficient in detecting 
blue galaxies, especially (i) the very bright and blue galaxies with huge gas reservoirs and (ii) the 
galaxies with the highest HI fraction lie on the faint end of the blue cloud. 
However, starting from the redder edge of the blue cloud, the cross-match rate drops below 10\% and 
to even $\sim$0\% throughout most of the red sequence. 
{\it Right panel -} SSFR versus stellar mass for the $S_{opt}$ galaxies. 
The cross-match rate is the highest among the galaxies with high SSFRs at 
both high and low stellar mass ends and is close to the overall average throughout the 
high number density region along the star forming sequence. 
However, it drops to below $\sim$20\% from the lower edge of the sequence to 
even $\sim$0\% in the low SSFR regime. 

(A color version of this figure is available in the online journal.) 
}
\label{fig:dtrt}
\end{figure*}

\begin{figure*}
\center{
\includegraphics[scale=0.9]{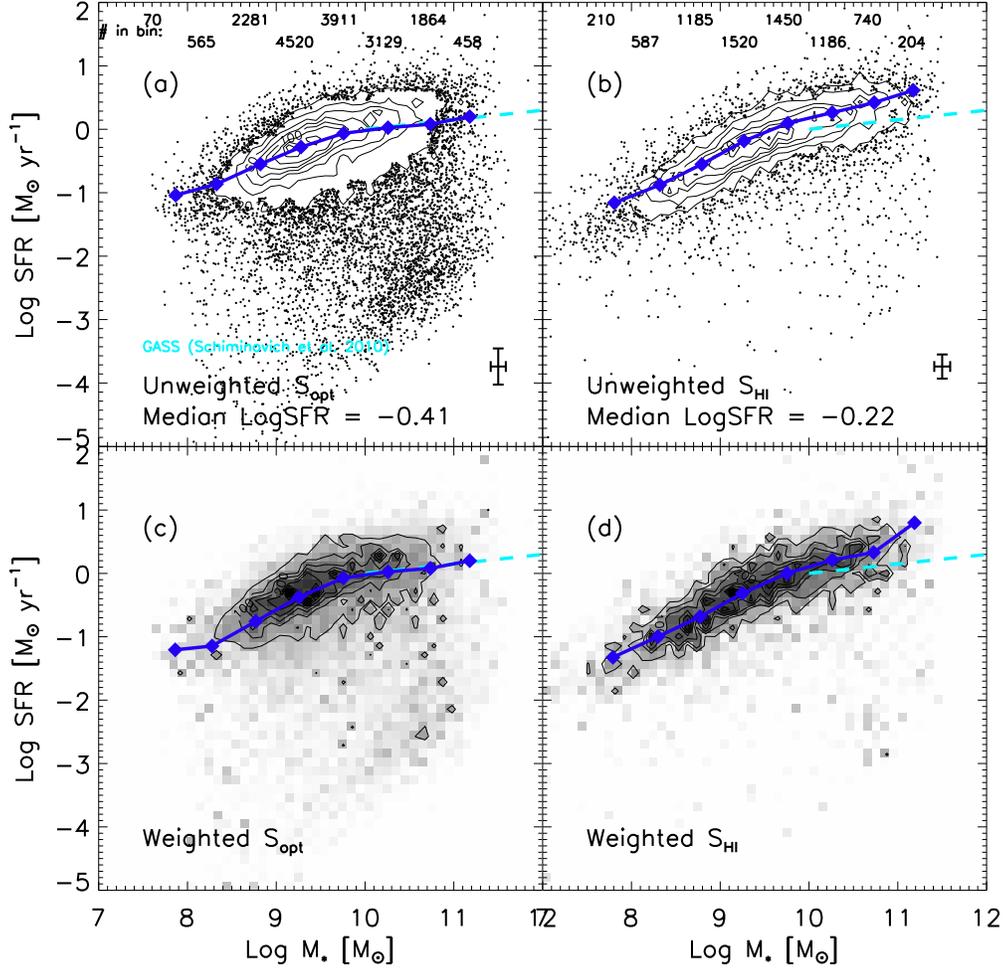}
}
\caption[]{$SFR$ versus $M_*$ for the $S_{opt}$ (left) and the $S_{HI}$ (right) samples; 
upper panels show the individual galaxies while the low ones show the results of applying the 
$V_{sur}/V_{max}$ weighting. 
The cyan dashed lines above $M_* = 10^{10} M_\odot$ show the 
fit to this relation obtained by \citet{Schiminovich2010}, based on the GASS sample. 
The blue diamonds and lines represent the corresponding $\log \langle SFR \rangle$ of our datasets. 
The number of galaxies in each stellar mass bin is listed at the top in panels (a) and (b), 
together with the typical error bars plotted in the lower right corners. 
The median SFRs for both the samples are indicated. 
The GASS fit is consistent with the distribution seen for the $S_{opt}$ sample 
but is systematically below the average of the HI-selected galaxies. 
The slope of the relation appears to steepen below $M_* \sim 10^{9.5} M_\odot$. 
The $S_{HI}$ sample has a higher overall $SFR$ value than the $S_{opt}$ galaxies, and probes to slightly 
lower $M_*$ ranges with generally lower SFRs. 
The data points reveal a concentration of massive low SFR galaxies below the main trend
in $S_{opt}$, which are largely absent from the $S_{HI}$ population. 
An HI survey samples the star-forming population. 
The flattening of the trend in the lowest mass bin in the $S_{opt}$ plots is artificial, 
due to the weight cut applied. 
}
\label{fig:SFS1}
\end{figure*}

The main distribution, consisting of the star forming galaxies, 
is associated with the parameter space inside the contours.  
In the un-weighted diagrams, the GASS fit is consistent with $S_{opt}$ in the high 
mass range (panel a), but falls below the average of $S_{HI}$ (panel b). 
This offset indicates that even when only the star forming sequence is considered, 
an HI-selected population has a higher SFR overall, because $S_{HI}$ is not complete even within the 
blue population (see also Figure \ref{fig:dtrt}). 
Similarly, the blue diamonds in the weighted $S_{opt}$ plot roughly coincide with the GASS fit (panel c). 
At the same time, the $\log \langle SFR \rangle$ of $S_{HI}$ galaxies becomes systematically larger than the GASS fit 
with increasing $M_*$ (panel d). 

Below the locus of star forming galaxies, the $S_{opt}$ sample contains a large population of 
passive galaxies visible as the cloud of points extending to very low SFRs (panel a).  
However, this population is barely seen in the same diagram for the $S_{HI}$ galaxies (panel b), 
i.e. an HI-selected sample is highly biased against non-star-forming galaxies. 

Because $S_{opt}$ is truncated at the low mass end 
due to the applied $V_{sur}/V_{max}$ weight limit (see \S\ref{sample}), 
$S_{HI}$ probes to lower $M_*$ and thus lower SFRs on average. 
Despite this dwarf tail, the average SFR is still slightly higher in $S_{HI}$. 
The median SFR is 
$0.4~{\rm M_\odot~yr^{-1}}$ in $S_{opt}$ versus 
$0.6~{\rm M_\odot~yr^{-1}}$ in $S_{HI}$. 
In the weighted plots, the correlation extending to the low $M_*$ range is better illustrated (panel d). 
There is a hint of a change of slope in the $SFR$--$M_*$ scaling relation below the SF
transition mass 
$M_* \simeq 10^{9.5} M_\odot$ noted earlier (\S\ref{SFS} and equation \ref{eqa:SFS}), 
with the slope at high $M_*$ being shallower, especially evident in the $S_{opt}$ plots. 
We note that the trend appears to flatten again in the lowest $M_*$ bin in panel (c) 
because of the weight cut. 
To be more specific, at a given $M_r$, the bluer galaxies with higher SSFRs have on average 
lower stellar masses, according to the mass-to-light ratio versus color relation. 
Therefore, the uniform $M_r$ cutoff applied on the $S_{opt}$ sample makes the distribution of $M_*$ extend to 
lower limits in higher SSFR bins; this effect is visible in the $S_{opt}$ diagrams 
(Figure \ref{fig:SFS1}a, c; see also the right panel of Figure \ref{fig:dtrt}). 
This effect artificially results in the higher $\log \langle SFR \rangle$ value in the lowest stellar mass bin. 
Furthermore, the lowest $M_*$ bin only contains 
70 galaxies, so that the average is even less reliable 
for statistical reasons. 
Such a sudden flattening is not seen within the $S_{HI}$ population (panel d), 
which probes to lower $M_*$ and with better sampling. 

\begin{figure*}
\center{
\includegraphics[scale=0.9]{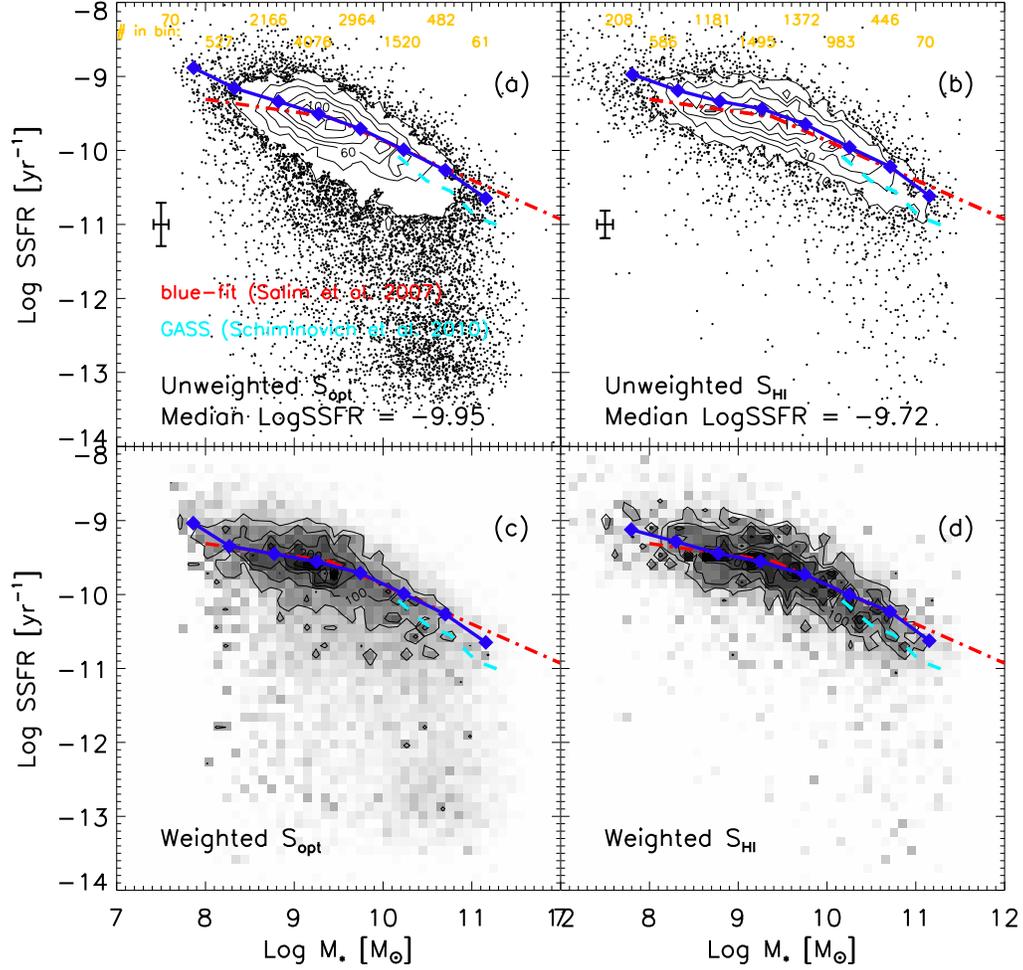}
}
\caption[]{Similar diagrams of $SSFR$ versus $M_*$. 
The numbers of galaxies in each stellar mass bin are listed, as well as the median SSFRs for both samples.  
The contours outlying the high number density region roughly trace the star forming sequence. 
The red dashed line shows the fit to such a sequence of the blue galaxies 
with $(NUV-r) < 4$ \citep{Salim2007}, derived from a typical local 
SDSS-GALEX cross-matched catalog, 
with the majority confined to the stellar mass range $10^8-10^{10}~{\rm M_\odot}$. 
The blue diamonds and lines are obtained by applying the same color criteria to the $S_{opt}$ or 
$S_{HI}$ galaxies.
The cyan dashed line comes from \citet{Schiminovich2010} based on the high $M_*$ GASS sample; 
it well represents the contours of $S_{opt}$, but lies systematically below the average of $S_{HI}$. 
When only blue galaxies are considered in both samples, the discrepancy between 
the main trend of $S_{opt}$ and $S_{HI}$ is small. 
Furthermore, both agree well with the red dashed line. 
Galaxies selected by HI criteria have on average higher SSFRs than optically-selected ones. 
The breakdown of the star forming sequence 
above $M_* \sim 2 \times 10^{10} M_\odot$ is only evident among the $S_{opt}$ sample, whereas 
the $S_{HI}$ galaxies are strongly biased against massive and low SSFR galaxies. 
}
\label{fig:SFS2}
\end{figure*}

Figure \ref{fig:SFS2} shows similar plots of the $SSFR$ - $M_*$ correlation. 
The cyan dashed line again denotes the $\log \langle SSFR \rangle$ values tabulated in \citet{Schiminovich2010}, 
derived from the high $M_*$ GASS sample.  
The contoured region roughly represents the star forming sequence defined in \citet{Salim2007}. 
The red dashed line is the fit of the sequence to the blue galaxies ($NUV-r < 4$) only as derived by 
those authors for a typical local SDSS-GALEX cross-matched catalog, 
with the majority falling in the stellar mass range of $10^8 - 10^{10}~{\rm M_\odot}$: 
\begin{eqnarray*}
	\log SSFR = 
	\left\{
	\begin{array}{lr}
	 	-0.17 (\log M_*-10) - 9.65,~\log M_* \leq 9.4 \\
	 	-0.53 (\log M_*-10) - 9.87,~\log M_* > 9.4. 
	\end{array}
	\right.
\end{eqnarray*}
For comparison, the blue diamonds in Figure \ref{fig:SFS2} are derived for our datasets but 
also considering only the blue galaxies ($NUV-r < 4$). 
Note these fits are different from the one given as equation \ref{eqa:SFS}, shown in Figure \ref{fig:SF}(d). 

Similar to what was seen in Figure \ref{fig:SFS1}, the $S_{HI}$ population has on average higher 
SSFRs than the $S_{opt}$ galaxies. The median SSFR is 
$10^{-9.95}~{\rm yr^{-1}}$ for $S_{opt}$ versus 
$10^{-9.72}~{\rm yr^{-1}}$ for $S_{HI}$. The breakdown of the star forming sequence 
above stellar mass $\sim 2\times 10^{10}~{\rm M_\odot}$ is only evident 
among the $S_{opt}$ population, whereas comparably low SSFR galaxies are almost entirely absent from 
the $S_{HI}$ sample. 
The GASS result agrees with the contours of $S_{opt}$ above $10^{10}~{\rm M_\odot}$ (panel a and c),  
but lies below the average of $S_{HI}$ (panel b and d). 

In panel (c), the contours, associated with the high number density regions, that fall far below the main 
star forming sequence ($\log SSFR < -11$) mostly show up only at the highest and lowest $M_*$ ends, 
suggesting that the scatter in the SSFR distribution 
is larger at both $M_*$ ends than in intermediate mass range. 
However, the effective $M_r$ cutoff applied to $S_{opt}$ inhibits us from 
a more convincing conclusion on the possible breakdown of the star forming sequence 
at the low mass end. 

At the high $M_*$ end, the averaged SSFRs are systematically higher in the $S_{HI}$ population than in the $S_{opt}$ one. 
However, when only blue galaxies are considered in either sample, the discrepancy between 
the two is reduced so that the trends indicated by the blue diamonds are 
comparable in both. 
The blue diamonds are in good agreement with the definition of the star forming sequence in 
\citet{Salim2007}, particularly for the high mass galaxies ($M_* \gtrsim 10^{9.4}~{\rm M_\odot}$) in the weighted plots, 
whereas the $S_{HI}$ distribution is slightly offset to higher SSFRs in an un-weighted plot (panel b). 
In addition, note that we obtain a steeper slope than the blue fit in \citet{Salim2007} 
for low mass galaxies, even in the $S_{opt}$ sample. 
Again because of the cutoff applied to $S_{opt}$, the averaged SSFR 
value in the lowest mass bin is artificially high. 

Finally, we inspect the cross-match rate of $S_{opt}$ to $\alpha.40$ over the  $SSFR$ - $M_*$ diagram 
in the right panel of Figure \ref{fig:dtrt}. The detection rate is the highest among galaxies with high SSFRs 
at both high and low $M_*$ ends (above $\sim 60\%$) and is close to the overall average ($\sim 40\%$) 
throughout the high number density region along the star forming sequence. 
However, it drops to below $\sim$20\% from the lower edge of the star-forming sequence, 
and the slope of the low match rate division, which corresponds to a cross-match rate of $\sim 25\%$ 
(yellow band in this plot), coincides well with that of the star-forming sequence. 
The SF properties of a galaxy have a significant impact on its likelihood of detection by an HI survey. 

We again note that the blue diamonds in Figure \ref{fig:SFS2}(b) are slightly above the red
dot-dashed line suggesting that not all star-forming galaxies are detected by ALFALFA.
From the fact that the HI detection rate of the galaxies in the SF sequence is 
far from 100\% (see right panel of Figure \ref{fig:dtrt}), one may naively conclude 
that not all star-forming galaxies contain HI. However, this conclusion is not necessarily correct.
For example, given the beam size of the ALFA frontend ($\sim$3.5\arcmin), confusion can contribute to a 
lower effective detection rate, e.g. when a close pair of star-forming HI bearing galaxies is 
unresolved. In fact, the red dot-dashed line is in excellent agreement with the blue diamonds in 
Figure \ref{fig:SFS2}(d), indicating
that the mean trend of the star forming sequence can be well reproduced by the
$\alpha.40$ galaxies {\it after} volume correction. An
HI-selected sample can effectively trace the average of the star forming sequence despite
the non-detection of some star-forming galaxies by ALFALFA (which is by design a short-integration
time survey).
In order to predict the SSFR of an HI-selected sample, one should still use equation 
(\ref{eqa:SFS}) rather than the average scaling relations derived in this section. The 
latter relations, as well as the star forming sequence
derived in \citet{Salim2007} rely on additional information, e.g., UV magnitudes or emission lines,
to define the star-forming galaxies. However, the $(NUV-r)<4$ (or similarly $\log SSFR < -11$~yr$^{-1}$) is
crude, since the SSFR generally decreases with increasing $M_*$. Furthermore, there are actually 
some very low SSFR
galaxies detected in HI \citep[e.g.][]{Hallenbeck2012}. Direct adoption of the SF sequence defined in
\citet{Salim2007} will overpredict the SSFR of an HI-selected sample.

\section{The host halos of HI-selected galaxies}
\label{lambda}

The results of the previous sections suggest that HI blind surveys detect in abundance the
star-forming population but are highly biased against red sequence galaxies. Conversely, 
nearly all star-forming galaxies contain HI. Because ALFALFA is both wide area and sensitive,
it samples the HI mass function over 4 orders of magnitude $ 7 < \log M_{HI}  < 11$ with
a mean of 9.56 dex 
\citep{Martin2010} and, while not as deep as the SDSS, both surveys probe the same range
of $M_r$. The brightest and reddest galaxies are missing from $\alpha.40$, but an HI-selected
sample provides an important perspective on the star forming sequence. While galaxies in
rich clusters are well known to be HI deficient because of the secular evolutionary
processes (ram pressure stripping, thermal viscosity, harassment), HI-bearing galaxies
dominate the extragalactic population in the low density field. As discussed in \S\ref{SFS},
the ALFALFA galaxies have, on average, lower SFEs, and equivalently therefore, longer
$t_R$ than the optically selected population. Both this work and the study of the lowest
HI mass dwarf population by \citep{Huang2012} suggest that overall, the HI-selected population
is less evolved and has higher SFR and SSFR but lower SFE and extinction than one selected
by optical brightness or stellar mass. 

Perhaps the most surprising result of the
ALFALFA survey has been the detection of a large number of galaxies with very high 
HI masses, log $M_{HI} > 10.3$, including a significant number with abnormally high HI 
gas fractions (for their stellar mass), dubbed the HIgh HI Mass (HIghMass) galaxies. 
The existence of this population, albeit rare, begs the question: how can such massive disks 
retain their huge gas reservoirs {\it without} forming stars? One explanation is that, while
recent inflow of HI gas has taken place, SF in that gas has not yet been triggered. 
Alternatively, the overall low SFEs characteristic of the ALFALFA population
may be explained by the semi-analytic models of galaxy formation in \citet{Boissier2000}, 
which predict that more extended disks with larger scale length and lower stellar surface density
are associated with dark matter halos of high spin parameter $\lambda$. 
Selected examples of very massive but gas-rich galaxies have been studied in recent years 
\citep[e.g.,][]{Portas2010}, with most belonging to the extreme category of low surface brightness 
(LSB) galaxies known as the ``crouching giants" \citep{Disney1987} or ``Malin 1 cousins" 
\citep{Bothun1987, Impey1997}. Their characteristic huge size, low optical surface brightness, 
star formation rate and low metal abundances may be explained by the 
large angular momentum and thus the low surface density of their ISM; 
star formation in such disks is suppressed according to canonical star formation laws. 
However, we note that many of the HIghMass galaxies are not crouching giants. 
Instead of being quiescent objects like Malin 1, most of the HIghMass galaxies have blue disks 
and exhibit healthy on-going SF."

The discussion throughout this paper is based on the scenario of hierarchical galaxy 
formation through mergers. Alternatively, cold mode accretion \citep{Keres2005} 
can build up gaseous galaxies rapidly at high redshift ($z>1$). In
the local universe volume probed by ALFALFA, it is likely that a large fraction of the gas accretion 
happens in a slower ``hidden'' mode, e.g., from the ionized hot corona or driven by 
the galactic-fountain process \citep{Marinacci2010}. 
In a future work, we will investigate the angular momentum distribution in the HIghMass 
galaxies to investigate whether the majority of the HI in 
local gas-rich galaxies reaches the assembling halo through filaments of cold flow at 
high redshift or results from gradual cooling out of the hot corona.

According to the \citet{Boissier2000} model, 
for a given halo mass, galaxies whose halos are characterized by
different values of $\lambda$ have similar stellar masses and current SFRs. 
However, galaxies in high $\lambda$ halos exhibit higher gas fractions, lower metallicities 
and bluer colors than those in low $\lambda$ halos.
In this picture, the halo mass controls current absolute values while the spin parameter 
determines mainly the shape of the SFH. 
Blue colors indicate high SSFRs, i.e., suppressed SF in the past relative to the current SFR. 
Massive compact disks have the shortest time scales of gas infall, 
rapid early SF, and thus are dominated by old stars today.
In contrast, the low mass galaxies reside in high $\lambda$ halos can even have a 
rising SFH instead of an exponentially decaying one; 
their timescales of gas infall, as well as SF, are long. 
This picture agrees with the downsizing scenario of galaxy formation (\S\ref{SFS}). 

Meanwhile, the low stellar surface density found in gas-rich systems
results in a weaker gravitational field, and, by extension,
a lower mid plane gas pressure and a higher fraction of 
diffuse HI gas \citep{Ostriker2010}. 
The model constructed by \citet{Fu2010}, tracking both the atomic and
molecular gas in disk galaxies, predicts a low HI to H$_2$ conversion efficiency in the 
high $\lambda$ galaxies. 
The HI, stellar and SF properties 
of the $\alpha.40$-SDSS-GALEX galaxies, relative to the other local samples, therefore
suggest that the HI-selected galaxies are biased towards ones in high $\lambda$ halos, 
and thus are blue, inefficient in SF and less evolved. Nevertheless, they have comparable 
or even slightly higher current SFRs relative to the ones in the low $\lambda$ halos with similar 
halo mass (see \S\ref{SF}). In this section, we test this hypothesis. 

\subsection{Spin parameters derived from the Tully-Fisher relation}
\label{lambda1}

In their analysis,
\citet{Boissier2000} adopted the scaling properties derived by \citet{Mo1998}, in the framework of 
the CDM scenario for galaxy formation. In this scenario, primordial density fluctuations give rise to 
dark matter halos of maximum rotational velocity $V_{halo}$, with the mass of the halo $M_{halo}\propto V_{halo}^3$. 
Inspired by theoretical studies to break the degeneracy, a second parameter, the spin parameter $\lambda$ , 
is introduced to describe the halo \citep[e.g.][]{Mo1998}, defined as $\lambda = J|E|^{1/2}G^{-1}M_{halo}^{-5/2}$, 
where $J$ is the angular momentum and $E$ is the total energy of the halo. 
Observationally, $\lambda$ is seen to determine properties such as color, disk thickness and 
bulge-to-disk ratio \citep{Hernandez2006}. 
Within the halo, baryonic gas condenses later and forms the stellar disk with mass $M_*$ and characterized by 
an exponential surface density profile with disk scale length $R_d$. 
Those authors also assumed that the density profile of the dark matter halo is isothermal and
responsible for establishing the flat disk rotation curve $V_{rot}$. 
Under the further assumptions that the potential energy of the galaxy is dominated by that of the halo 
and that it is a virialized gravitational structure, etc., \citet{Boissier2000} related the quantities 
describing the halo to those describing the disk, and express the spin parameter as 
\begin{equation}
\label{eqa:lambda1}
	\lambda = \frac{\sqrt{2}V_{rot}^2 R_d}{GM_{halo}}. 
\end{equation}
See also \citet{Hernandez2006} for similar derivation of equation (\ref{eqa:lambda1}). 
The only unobservable quantity $M_{halo}$ can be replaced by $M_{baryon}/F$ where $F$ 
is the baryonic fraction, or, alternatively, $M_*/F_d$ where $F_d$ is the stellar disk mass fraction. 

Based on this framework 
\citet{Hernandez2007} analyzed the empirical $\lambda$ distribution of samples taken 
from the SDSS. Because their sample lacked direct measures of $V_{rot}$, 
those authors invoked the Tully-Fisher relation to infer $V_{rot}$ from the disk luminosity. 
Furthermore, a disk mass Tully-Fisher relation and a constant disk mass fraction $F_d = 0.04$ 
are assumed to eliminate the $M_{halo}$ term. 
As a first approach for 
direct comparison with \citet{Hernandez2007}, we repeat the method used in that work to 
derive the distributions, separately, of the 
$\lambda$ spin parameter for the 
the HI-selected $S_{HI}$ and the optically-selected $S_{opt}$ galaxies. We adopt the
$\lambda$ estimator proposed by \citet{Hernandez2007}, 
\begin{equation}
\label{eqa:lambda2}
	\lambda = 21.8 \frac{R_d[{\rm kpc}]}{(V_{rot}[{\rm km~s^{-1}}])^{3/2}}. 
\end{equation}
\citet{Hernandez2007} adopted the $R$-band Tully-Fisher relation
derived by \citet{Barton2001} to assign $V_{rot}$ to the observed galaxies; since this relation
is valid only over the absolute magnitude range $-20 > M_R > -22.5$, they restrict their
analysis to that range. To minimize errors due to internal extinction, they trim the 
sample to leave only spiral galaxies \citep{Park2005} having $b/a>0.6$.

The $R$-band luminosity is inferred from the SDSS bands based on Lupton (2005), 
\begin{eqnarray*}
	(M_R)_0 = r_0 - 0.2936\times(r - i)_0 - 0.1439. 
\end{eqnarray*}
We applied the same absolute magnitude cut
$-20 > (M_R)_0 > -22.5$ to both the $S_{HI}$ and the $S_{opt}$ samples. 
Within our distance range, the SDSS is volume limited to $M_r \sim -19$ mag 
(Figure \ref{fig:basic}a). We also require that the $r$-band 
light profile should be better fitted by an exponential profile than a 
deVaucouleurs profile to be sure we were including mainly disk galaxies. 
Since \citet{Hernandez2007} ignored internal extinction but we have corrected for it
using the SED fitting estimate of $A_R$, the $b/a$ requirement in our case 
is less strict, $b/a>0.35$. 
Such a cut is adopted so that the $b/a$ distribution is close to being flat above 0.35. 
Like \citet{Hernandez2007}, we convert the SDSS
$r$-band axial ratio to inclination $i$,  
\begin{eqnarray*}
	\cos^2 i = \frac{(b/a)^2-q_0^2}{1-q_0^2}, 
\end{eqnarray*}
adopting an intrinsic axial ratio of the 
disk of $q_0=0.18$ as proposed by \citet{Courteau1997}. We note, however, that since
that study was restricted to a relatively small number of Sb-Sc galaxies, a more
conventional value of $q_0=0.2$ may be more appropriate (see \S\ref{lambda2}).

First, we confirm that the TF relation \citet{Hernandez2007} used is systematically 
consistent with the $S_{HI}$ galaxies, albeit with significant scatter. The
HI line width of the $S_{HI}$ galaxies, $W_{50}$, is converted to $V_{rot}$ through 
$V_{rot}=(W_{50}/2/\sin i)$, where the small broadening effect 
($\sim 5~{\rm km~s^{-1}}$) of turbulence and non-circular motions on the
observed HI linewidths is ignored. 
The average trend evident between $(M_R)_0$ and $V_{rot}$ agrees with the relation, 
i.e., applying such a TF relation should have little effect on the mean value 
of the $\lambda$ distribution. 
The difference of the mean $\lambda$ value derived here as compared with that derived 
next in \S\ref{lambda2} will be mainly due to the different assumptions of the
baryon fraction $F$. 

Next, we obtain the $\lambda$ distributions of the $S_{HI}$ and the $S_{opt}$ samples, 
according to equation (\ref{eqa:lambda2}), and following \citet{Hernandez2007}, we confirm
that they both are well fit by a lognormal function:
\begin{eqnarray*}
	P(\lambda_0, \sigma_{\lambda};\lambda) {\rm d}\lambda = 
	\frac{1}{\sigma_{\lambda}\sqrt{2\pi}}
	\exp\left[-\frac{\ln^2(\lambda/\lambda_0)}{2\sigma_{\lambda}^2}\right]
	\frac{{\rm d}\lambda}{\lambda}. 
\end{eqnarray*}
The best fit parameters are $\lambda_0 = 0.0525$, 
$\sigma_\lambda = 0.422$ for the $S_{HI}$ distribution of the HI selected galaxies $S_{HI}$ 
and $\lambda_0 = 0.0489$, $\sigma_\lambda = 0.446$ for the SDSS-selected 
$S_{opt}$ distribution. Because we correct $M_R$ for extinction, 
our $\lambda_0$ value is slightly below the value found by \citet{Hernandez2007},
$\lambda_0 = 0.0585$. While suggestive, this difference of $\lambda_0$ between 
the two samples $S_{HI}$ and $S_{opt}$, {\it under the assumption of a constant disk-to-total 
mass fraction} $F_d$, is small. The various cuts and selection effects applied to both 
samples, e.g. detected in UV, better fit by an exponential profile, small dynamic
range of $M_r$, may have already minimized the possible difference in $\lambda_0$
within the full sample. Given that 
the remaining galaxies in the $S_{HI}$ sample are brighter on average than those 
in the $S_{opt}$ sample, i.e., $V_{rot}$ is higher on average, the slightly higher $\lambda_0$ 
found by this analysis is due to the fact that $S_{HI}$ galaxies have on average larger $R_d$ values 
(see equation \ref{eqa:lambda2}). 
HI selection instills a bias towards more extended galaxies.

\subsection{$\lambda$ distribution of the parent population}
\label{lambda2}

The SDSS galaxies analyzed by \citet{Hernandez2007} lack direct measurements of $V_{rot}$, 
whereas the $\alpha.40$--SDSS--GALEX sample has homogeneous $W_{50}$ measurements for all its
galaxies. We hence improve the derivation of the $\lambda$ distribution for all the 
$\alpha.40$--SDSS--GALEX galaxies 
in this section. Assuming that an HI-selected sample has little contamination from galaxies which
are not disk dominated, we drop the requirement of a higher probability of an exponential fit and
adopt an intrinsic $q_0=0.2$. Because HI line widths are available,
we no longer rely on the Tully-Fisher relation 
to infer $V_{rot}$, and for that reason, the $(M_r)_0$ limit is also dropped. 
Most importantly, we estimate $\lambda$ directly from equation (\ref{eqa:lambda1}) 
assuming a non-constant $F$ to derive $M_{halo}$. 

\begin{figure*}
\center{
\includegraphics[scale=0.8]{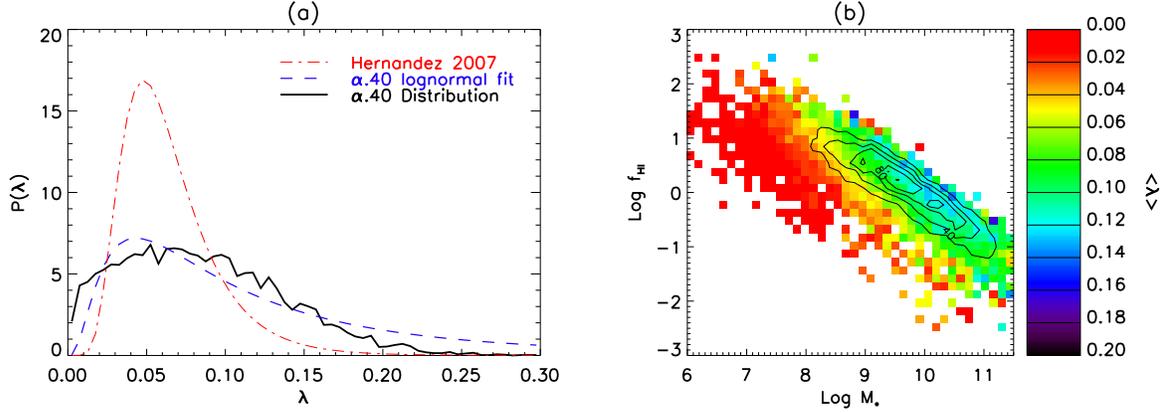}
}
\caption[]{Spin parameter $\lambda$ distribution obtained by assigning the $M_{halo}$ through a 
$V_{rot}$--$V_{halo}$ relationship. 
{\it Panel (a)} - The black solid line is the normalized PDF of the $\lambda$ distribution of 7459 galaxies from the 
	$\alpha.40$--SDSS-GALEX sample. The blue dashed line is the best lognormal fit to the distribution. 
	The red dash-dotted line is the lognormal fit to a sample of spiral galaxies from the SDSS, assuming 
	constant baryonic mass fraction $F=0.04$ \citep[$\lambda_0 = 0.0585$][]{Hernandez2007}. 
	The $\lambda$ distribution of the $\alpha.40$ galaxies has a higher mean $\lambda$ (0.0852) 
	and a wider dispersion, mainly arising from the adopted distribution of baryonic fraction. 
{\it Panel (b)} - $f_{HI}$ vs. $M_*$ diagram color coded by the mean $\lambda$ values of the galaxies 
	in each grid. Below $M_* \sim 10^{10.5}~{\rm M_\odot}$, $\langle \lambda \rangle$ 
	increases with increasing $f_{HI}$ at a given $M_*$, but remains almost constant along 
	constant $M_{HI}$ lines. Such a trend reflects the adopted variation of $F$: 
	HI-selected gas-rich galaxies favor high $\lambda$ halos. 

(A color version of this figure is available in the online journal.)
}
\label{fig:fitdis}
\end{figure*}

Following \citet{Hernandez2007}, the preceding assumes that all halos are associated with the
same disk mass fraction $F_d = 0.04$. However, based on abundance matching between the 
observed stellar mass function and dark matter halo mass function derived from CDM simulations, 
it is quite well established that $F_d$ is not a constant 
\citep[][and references therein]{Behroozi2010}. Instead, the distribution of
$F_d$ as a function of $M_{halo}$ 
peaks around $M_{halo} \sim 10^{12}~{\rm M_\odot}$ \citep[e.g.][]{Guo2010}, i.e., 
low mass galaxies, corresponding to $M_* \lesssim 10^{10.5}~{\rm M_\odot}$, 
retain fewer baryons as their halo mass declines.
The trend turns over above $M_* \sim 10^{10.5}~{\rm M_\odot}$. 
Based on the $\alpha.40$ catalog, \citet{Papastergis2011} have studied how galaxies 
with different $V_{rot}$ populate dark matter halos, by applying similar abundance 
matching to the velocity width function, under the assumption that rotational
velocity well traces the dark matter mass. We adopt the $V_{halo}$--$V_{rot}$ relation from Table 1 of 
\citet{Papastergis2011}, assuming $V_{halo,max} = 360~{\rm km~s^{-1}}$. 
The resulting $V_{halo}$ is converted into the virial mass of the halo following the simulation 
result of \citet{Klypin2011}: $V_{halo} = 2.8\times 10^{-2} (M_{halo}h)^{0.316}$. 
For the disk scale length $R_d$, we use that measured by the SDSS pipeline in the $r$-band, 

Among the 9417 $\alpha.40$--SDSS-GALEX galaxies, 1829 are dropped because of the
adopted axial ratio cut ($b/a>0.35$ as in \S\ref{lambda1}), and a further 
130 are dropped because they lie outside of the valid range of the $V_{rot}$--$V_{halo}$ 
relation ($16~{\rm km~s^{-1}}<V_{rot}<431~{\rm km~s^{-1}}$). In the end, 
$\lambda$ values are estimated for 7458 $\alpha.40$ galaxies; their distribution 
is shown in the Figure \ref{fig:fitdis}(a). 
The black solid line traces the normalized PDF for the $\alpha.40$ galaxies.
The red dash-dotted line in Figure \ref{fig:fitdis}(a) shows the best log-normal
fit with 
$\lambda_0 = 0.0585$ and $ \sigma_{\lambda} = 0.446$ from \citet{Hernandez2007}
and the blue dashed line is the best fit to the $\alpha.40$ galaxies with 
$\lambda_0 = 0.0929$ and $ \sigma_{\lambda} = 0.875$. 
\citet{Hernandez2007} noted that values they found for both
$\lambda_0$ and $\sigma_{\lambda}$ 
are in good agreement with results from cosmological simulations. 
However, the $\lambda$ distribution of the $\alpha.40$-SDSS-GALEX galaxies has a mean 
well above this value, $\lambda = 0.0852$. 
Furthermore, the distribution for the $\alpha.40$--SDSS--GALEX sample
has a much wider dispersion than the previous result, arising mainly from 
the fact that we adopt different $F$ values according to the $V_{rot}$,  
rather than treating it as a constant. 
As a result, the distribution of spin parameters 
of the HI-selected population is no longer well fit by 
a lognormal function. 
The KS statistic implies that the probability that the 
$\lambda$ distribution for the HI-selected population is
drawn from the same underlying distribution with a PDF of the best-fit lognormal 
function with $\lambda_0 \sim 0.0929$ is negligible. 

The technique of assigning $M_{halo}$ is valid only in a statistical sense, 
rather than as a concrete measure for individual galaxies. For this reason, we 
look only for a general trend in the dependence of the luminous component on 
the $\lambda$ of the halos. 
Figure \ref{fig:fitdis}(b) shows a $f_{HI}$--$M_*$ diagram 
similar to Figure \ref{fig:gas}(c), but color coded by the mean $\lambda$ value of galaxies 
in each grid. 
The black contours outline the distribution of the 
7458 $\alpha.40$ galaxies in the 
high number density region. A trend is evident that, in a given $M_*$ bin, the 
gas rich galaxies with higher $f_{HI}$ on average reside in halos with higher $\lambda$. 
Meanwhile, along lines of constant $M_{HI}$ (straight lines with slope 
of `$-1$' in the $f_{HI}$--$M_*$ diagram), $\langle \lambda \rangle$ barely changes with increasing $M_*$. 
Taking into account the mean scaling relation defined by the $\alpha.40$ galaxies 
(blue diamonds in Figure \ref{fig:gas}c) relative to the similar relations obtained for other 
samples, this result reinforces the hypothesis that the HI-selected galaxies favor 
high $\lambda$ halos. 
The $f_{HI}$--$M_*$ relations confined by the `HI-normal' or the `outside Virgo' galaxies in 
Figure \ref{fig:gas}(c) are close to the well accepted $\lambda \sim 0.05$ region, 
the yellow band in Figure \ref{fig:fitdis}(b). 

We note that such a clear trend is only weakly due to the overall larger $R_d$, i.e., 
lower surface brightness, which the more gas-rich galaxies have in a given $M_*$ bin, 
but is instead largely due to the rapidly increasing 
$F = (M_{HI}+M_*)/M_{halo}$ with $M_{HI}$ in a given $M_*$ bin.
However, it is known that the increase in $F$ with $M_{halo}$ is reversed near 
$M_* \sim 10^{10.5}~{\rm M_\odot}$. In fact, the variation of $\langle \lambda \rangle$ along the 
$f_{HI}$ axis disappears around $M_* \sim 10^{10.5}~{\rm M_\odot}$ in 
Figure \ref{fig:fitdis}(b). 
In addition, we note that within the luminosity range of $-20 > (M_R)_0 > -22.5$, 
the stellar disk fraction, $F_d$, inferred by this method spreads over a range with 
a median of 0.04; the $M_{halo}$--$F$ relation crosses the turnover around 
$(M_R)_0 \sim -21.5$. Despite the systematic variation of $F_d$ we adopt, 
the median is consistent with the assumption followed in \S\ref{lambda1}

Because the $V_{rot}$--$V_{halo}$ 
relation we have used is less constrained at the highest masses, it barely reproduces the 
trend of decreasing $F$ with $M_{halo}$ above $M_* \sim 10^{10.5}~{\rm M_\odot}$. 
Alternatively, we can estimate the $M_{halo}$ following the $M_*$--$M_{halo}$ relation 
\citep{Guo2010}, 
{\small
\begin{eqnarray*}
	\frac{M_*}{M_{halo}} = 0.129 
	\left[ \left( \frac{M_{halo}}{10^{11.4}~{\rm M_\odot}}\right)^{-0.926}
	+\left(\frac{M_{halo}}{10^{11.4}~{\rm M_\odot}}\right)^{0.261}
	\right] ^{-2.440}.
\end{eqnarray*}
}
The fit is valid in the $M_{halo}$ range of $10^{10.8}$ to $10^{14.9}~{\rm M_\odot}$, 
and thus describes better the most massive $\alpha.40$ galaxies. However, it
cannot constrain the low mass HI-bearing halos. As expected, 
such a $M_*$--$M_{halo}$ relation, with a
turnover $M_{halo} \sim 10^{12}~{\rm M_\odot}$, produces a similar trend of the 
$\langle \lambda \rangle$ variation as seen in Figure \ref{fig:fitdis}(b) for galaxies with 
$M_* \lesssim 10^{10.5}~{\rm M_\odot}$. 
However, for the most massive galaxies with $M_* \gtrsim 10^{10.5}~{\rm M_\odot}$, 
such a trend disappears, and galaxies with similar $M_*$ have similar $\lambda$, 
regardless of their $f_{HI}$. 
Meanwhile, $\langle \lambda \rangle$ turns over to decrease with increasing $M_*$ with approximately
constant $M_{HI}$.  

As a result, the pattern of colors in Figure \ref{fig:fitdis}(b) predicts that 
the most gas rich ones with $M_*$ near $10^{10}~{\rm M_\odot}$ should
be associated with halos with the highest $\lambda$ parameters. We are testing this
prediction by studying the stellar, gas and dark matter components of 
a sample of very high HI mass ($M_{HI} > 10^{10}~{\rm M_\odot}$), high gas fraction 
(for their stellar mass)
galaxies, the HIghMass sample. Visual inspection of the HIghMass galaxies shows that 
the ones with intermediate $M_*$ shows the strongest color gradient (bluer in outer regions), 
as predicted for high $\lambda$ galaxies \citep{Hernandez2006}.
Future work will constrain the spin parameters using velocity fields derived from 
HI synthesis maps for 20 of the HIghMass galaxies. Furthermore, the HI distributions and
sites of star formation will be examined to yield more detail on the SFL and possible
mechanisms which inhibit the formation of stars in these massive HI disks.

The HI, stellar and SF properties of the ALFALFA population as exemplified
by the $\alpha.40$-SDSS-GALEX galaxies suggest that 
the HI-selected population is biased towards extended disks which
reside in high $\lambda$ dark matter halos. As such, their disks
are currently forming stars but in an inefficient manner. Despite the low SFE, 
such galaxies can have 
comparable or even slightly higher current SFRs relative to the ones in the 
low $\lambda$ halos of similar halo mass (see \S\ref{SF}). This combination of
on-going star formation and inefficiency in the conversion of gas
into stars causes their disks to be extended, blue and of low metallicity,
suggestive of their being less evolved. Remembering too that
the $\alpha.40$ galaxies are among the least clustered population \citep{Martin2012}, 
we are reminded that these systems likely follow a quite different, quiescent
evolutionary history relative to galaxies residing in higher density volumes. 
Although the details of the involved processes
still elude us, the study of the ALFALFA population strongly suggests that
where a galaxy is born and resides determines
to a large extent how and when it converts its gas into stars.

\section{Conclusion}
\label{con}

Given the importance of proposed future
extragalactic HI surveys at high redshift by the SKA, it is critical to develop
a full understanding of the characteristics of gas rich galaxies at the present epoch. 
ALFALFA is an on-going blind HI survey, for the first time providing a full census of HI-bearing objects 
over a cosmologically significant volume of the local universe. Building on the $\alpha.40$ catalog 
\citep{Haynes2011}, we use the corresponding photometry available from the SDSS DR7 and 
GALEX GR6 catalogs to explore the population of galaxies detected by ALFALFA. 
The combined $\alpha.40$-SDSS-GALEX sample includes 9417 galaxies. 
SED-fitting to the seven UV and optical bands yields the stellar properties
of the HI-selected galaxies, including their stellar masses, SFRs and internal extinctions. 
The lack of a correction for internal extinction can lead to systematic underestimates of the
optical luminosities and SFRs. In order to reduce the overestimate of internal extinction 
and thus the SFR with decreasing $M_*$
while still accounting for the effect of dust in these mostly disk-dominated systems, we apply
a prior distribution of $\tau_V$ in the fitting process. 
Although extinction is even more of an issue at the short
wavelengths, the addition of photometry in the UV bands is critical to the diagnostic power 
of the SFH because the UV measures break the degeneracy evident in optical-only colors, 
particularly among the red sequence galaxies. 

ALFALFA offers a complete statistical sampling of the full range of $M_{HI}$ and $M_*$, 
from the most massive giant spirals with $M_{HI} > 10^{10}~{\rm M_\odot}$ 
to the lowest HI mass dwarfs with log $M_{HI} < 10^{7.5}~{\rm M_\odot}$,
thereby providing a firmer basis for the derivation of
the fundamental scaling relations linking the global properties. 
First, we confirm the existence of the relations which have been
found in typical local optical-UV catalogs \citep[e.g.][]{Salim2007}. 
For example, (1) SFRs increase but SSFRs decrease with increasing $M_*$, 
with different slopes in the high and low $M_*$ ranges, with the transition
occurring at $M_* \sim 10^{9.5}~{\rm M_\odot}$; 
(2) the SSFR is tightly correlated with the ($NUV-r$) color, especially after the
latter is corrected for internal extinction. 
Second, we investigate similar relations involving the HI mass. For example, 
(3) SFRs also increase but SSFRs decrease with increasing $M_{HI}$, though with larger scatter. 
The HI gas contributes a significant fraction of the baryons in HI-selected galaxies, and their SFRs show 
a good correlation with $M_{HI}$, suggesting that a global, atomic, volumetric SFL applies 
in HI-selected systems.
(4) The HI fraction, $f_{HI} \equiv M_{HI}/M_*$, nicely correlates with 
$M_*$, but a change in the slopes of the relation is evident near $M_* \sim 10^{9}~{\rm M_\odot}$. 
(5) Galaxies with bluer colors in general have higher $f_{HI}$. 
(6) The star formation efficiency, $SFE \equiv SFR/M_{HI}$, mildly increases with stellar mass
with a slightly steeper relation
for $M_*\lesssim10^{9.5}~{\rm M_\odot}$. 
In \S\ref{MHIMs}, we give the best linear fits to the principal 
scaling relations among the ALFALFA population, 
including the $M_{HI}$--$M_*$ and $SSFR$--$M_*$ correlations, as well as the fundamental planes 
of $f_{HI}$--$(NUV-r)$--$\mu_*$, etc. In particular, we argue that 
Equation (\ref{eqa:fgas1}-\ref{eqa:fgas6}) provides the most robust predictor based on 
optical properties of the detection rate by future HI line surveys with the SKA and its pathfinders. 

Besides the scaling relations themselves, the combined $\alpha.40$-SDSS-GALEX dataset, as a large 
and homogeneous sample with HI measures, enables the quantitative appraisal of the scatter in relations and 
a deeper understanding of the role HI plays in the galaxy evolution. 
In particular, the decreasing $f_{HI}$ with increasing $M_*$ is related to the star-forming sequence 
identified in the $SSFR$--$M_*$ diagram, 
or the evolutionary tracks of galaxies on the CMD, i.e. as galaxies assemble their stellar masses, 
they evolve gradually to relatively red and gas poor regimes, and also show lower SSFRs. 
Furthermore, {\it only} evident among the low mass galaxies ($M_*\lesssim 10^{9.5}~{\rm M_\odot}$), 
the galaxies with higher $f_{HI}$ on average also have higher $SSFR$ {\it in a given $M_*$ bin}.
Similarly, within a given $M_r$ bin, higher $f_{HI}$ on average indicates bluer ($NUV-r$) color.
However, the corresponding trend that the HI-rich galaxies are more likely to be blue starburst 
galaxies with high SSFRs is weak among the high mass galaxies, i.e., the regulation of SF by HI 
is weaker in more massive systems. 
Similarly, the dispersion of the color distribution in a given $M_r$ bin and the dispersion of the 
$SSFR$ distribution in a given $M_*$ bin are both at a minimum near $M_* \sim 10^{9.5}~{\rm M_\odot}$, and
both increase at masses lower than that. It appears that SF is no longer regulated principally
by the stellar mass in low mass systems. In their shallow potential wells,
gas can be easily removed so that the SF quenches causing the galaxy to migrate onto the red sequence.

We also focus on the nature of the population detected by the ALFALFA survey, 
in the context of populations better understood through observations at 
other wavelengths, e.g. optical or UV. 
The $\alpha.40$ galaxies on average have higher HI fractions in a given stellar mass bin, 
compared to the optically-selected samples, 
with an overall average of 
$f_{HI}\sim 1.5$. 
95.6\% of the $\alpha.40$-SDSS-GALEX galaxies have $(NUV-r)<4$ and belong to the blue cloud 
on a UV-to-optical CMD. 
The red ALFALFA detections include early type galaxies with quenched SF but unusually high HI masses, 
suggesting a recent acquisition of HI. 
The very blue HI-rich galaxies may be attributed to a SFH that steadily rises to the present day 
with little integrated past SF. 
The SFEs of the HI-selected galaxies are lower on average, or equivalently, 
their gas depletion timescales are longer (average $t_R=8.9$~Gyr), 
compared to the high stellar mass galaxies included in the GASS survey \citep{Schiminovich2010}.  
Given the fact that the overall SFRs of $\alpha.40$ galaxies are even higher than those 
in GASS at a given $M_*$, the low SFEs found for HI-selected galaxies are caused by their high HI masses rather than 
by lower SFRs. 
This result is consistent with the idea that HI-selected sample is biased towards the most
gas-rich galaxies and that the SFE is low in HI-dominated systems. 
A bottleneck may exist in the HI to H$_2$ conversion, or the process of SF from H$_2$ may obey an unusual SFL 
with low efficiency in the very HI-dominated galaxies. 
For a given ($FUV-NUV$) color, HI-selected galaxies have on average lower extinctions, 
suggesting that they have different SFHs or dust geometries. 

To quantify better the bias of the HI-selected population relative to the optically-selected galaxies, 
we constructed two volume-corrected control samples, 
starting from the $\alpha.40$ and the SDSS DR7 catalogs, which we designate $S_{HI}$ and 
$S_{opt}$ respectively. 
The HI-selected $S_{HI}$ sample is found to be biased against red-sequence galaxies
as well as massive but low SFR, low SSFR galaxies. 
However, if only the blue cloud galaxies with $(NUV-r) < 4$ are considered, both samples 
define similar SF sequences, i.e., an HI survey well samples the star forming population. 
For the SDSS-selected volume-limited $S_{opt}$ sample, the rate of detection by ALFALFA
decreases towards redder colors. Virtually all very blue $S_{opt}$ galaxies at both the
high and low stellar mass ends are detected by ALFALFA; however, among the blue population,
the HI detection rate drops to $\sim40\%$ throughout the high number density region along 
the SF sequence. At the same time, only a very few of the galaxies which lie below 
the SF sequence in an $SSFR$ vs. $M_*$ diagram are detected by ALFALFA. Furthermore,
ALFALFA misses a substantial fraction of the optical galaxies lying on the redder 
side of the blue cloud. The volume-corrected optically-selected $S_{opt}$ sample 
well reproduces various scaling relations derived from the high stellar mass GASS sample
\citep{Catinella2010,Schiminovich2010}. However, at the highest stellar masses,
the HI-selected $S_{HI}$ galaxies show systematically larger discrepancies in their SF 
properties from the GASS results as the fraction of the total population which is detected 
by ALFALFA in a given stellar mass bin declines. In comparison with optically-selected samples,
HI-selected galaxies that have high gas fractions are relatively less evolved and 
have, on average, bluer colors, higher SFRs and SSFRs, but lower SFEs, extinctions and 
metallicities.

Previous authors, notably \citet{Boissier2000}, have proposed that the overall low SFEs found
in gas-rich systems may be explained if, their disks are characterized by large disk scale lengths
and lower stellar surface densities because their host dark matter haloes have high
spin parameters $\lambda$. We explore this hypothesis comparing the spin parameter distributions
of the volume-limited $S_{HI}$ and $S_{opt}$ samples following the approach outlined in 
\citet{Hernandez2007} which estimates the $V_{rot}$ using the Tully-Fisher relation. As
presented by \citet{Hernandez2007}, this estimate of $\lambda$ assumes that all galaxies are
characterized by the same disk mass fraction $F_d = 0.04$. Under that (unlikely) assumption,
we find a spin parameter distribution close to that found by \citet{Hernandez2007} for their
SDSS disk subsample and well-fit by a log normal distribution in agreement with numerical 
simulations. There is a slight hint that the HI-selected population $S_{HI}$ has
a slightly higher $\lambda_0$ than the 
$S_{opt}$ sample which could reflect the bias that the HI-selected sample
is characterized by more extended disks. 
However, abundance matching between the observed stellar mass functions and the 
CDM halo mass functions derived from simulations strongly suggests that that $F$ is not a constant
but rather peaks around a halo mass of
$M_{halo} \sim 10^{12}~{\rm M_\odot}$ \citep[e.g.][]{Guo2010}. 
At the low mass end, baryon depletion again
grows in the shallow potential wells \citep{Hoeft2006}.
Because the $\alpha.40$ catalog contains HI line widths, we calculate spin parameters
from them adopting the $V_{halo}$--$V_{rot}$ relation from Table 1 of 
\citet{Papastergis2011}.  
While this method of assigning $M_{halo}$ is only valid in a statistical sense, 
the result is clear: the distribution of $\lambda$ is no longer log normal and has a 
mean value well in excess of the expectation, strongly reinforcing the hypothesis that 
the ALFALFA population favors high $\lambda$ dark matter hosts.

In the future, our multiwavelength program
to study the HIghMass sample of high gas fraction, high HI mass galaxies will explore the
star formation process and the interrelationships of the stellar, gas, dust and dark matter
components within this set of exceptionally massive HI disks. The Survey of HI in Extremely
Low mass Dwarfs (SHIELD) is exploring similar relationships among the lowest HI mass
galaxies detected by ALFALFA \citep{Cannon2011} and has already uncovered evidence
that the SFL in those galaxies is unusual.

\acknowledgements
The authors acknowledge the work of the entire ALFALFA collaboration team 
in observing, flagging, and extracting the catalog of galaxies used in this work. 
The ALFALFA team at Cornell is supported by NSF grants AST-0607007 and AST-1107390 to RG and MPH and 
by grants from the Brinson Foundation. We thank Manolis Papastergis for useful discussions.

The Arecibo Observatory is operated by SRI International under a cooperative agreement with the 
National Science Foundation (AST-1100968), and in alliance with Ana G. M\'endez-Universidad 
Metropolitana, and the Universities Space Research Association.

GALEX is a NASA Small Explorer, launched in 2003 
April. We gratefully acknowledge NASA's support for construction, operation and science 
analysis for the GALEX mission, developed in cooperation with the Centre National 
d'Etudes Spatiales of France and the Korean Ministry of Science and Technology. 
SH, RG and MPH acknowledge support for this work 
from the GALEX Guest Investigator program under NASA grants NNX07AJ12G, NNX07AJ41G,
NNX08AL67G, NNX09AF79G and NNX10AI03G.

Funding for the SDSS and SDSS-II has been provided by the Alfred P. Sloan Foundation, 
the participating institutions, the National Science Foundation, the US Department 
of Energy, the NASA, the Japanese Monbukagakusho, the Max Planck Society and 
the Higher Education Funding Council for England. The SDSS Web Site is 
http://www.sdss.org/.
The SDSS is managed by the Astrophysical Research Consortium for the 
participating institutions. The participating institutions are the American 
Museum of Natural History, Astrophysical Institute Potsdam, University of Basel, 
University of Cambridge, Case Western Reserve University, University of Chicago, 
Drexel University, Fermilab, the Institute for Advanced Study, the Japan 
Participation Group, Johns Hopkins University, the Joint Institute for Nuclear 
Astrophysics, the Kavli Institute for Particle Astrophysics and Cosmology, the
 Korean Scientist Group, the Chinese Academy of Sciences (LAMOST), Los Alamos 
National Laboratory, the Max Planck Institute for Astronomy, the MPA, New Mexico 
State University, Ohio State University, University of Pittsburgh, University 
of Portsmouth, Princeton University, the United States Naval Observatory and 
the University of Washington.

\newpage
\begin{thebibliography}{}
\bibitem[Abazajian et al.(2009)]{Abazajian2009} Abazajian, K.~N., 
	Adelman-McCarthy, J.~K., Ag{\"u}eros, M.~A., et al.\ 2009, \apjs, 182, 543 
\bibitem[Abdalla \& Rawlings(2005)]{Abdalla2005} Abdalla, F.~B., \& Rawlings, S.\ 2005, \mnras, 360, 27
\bibitem[Baldry et al.(2004)]{Baldry2004} Baldry, I.~K., Glazebrook, K., Brinkmann, J., et al.\ 2004, \apj, 600, 681
\bibitem[Barnes et al.(2001)]{Barnes2001} Barnes, D.~G., 
	Staveley-Smith, L., de Blok, W.~J.~G., et al.\ 2001, \mnras, 322, 486 
\bibitem[Barton et al.(2001)]{Barton2001} Barton, E.~J., Geller, 
	M.~J., Bromley, B.~C., van Zee, L., \& Kenyon, S.~J.\ 2001, \aj, 121, 625
\bibitem[Behroozi et al.(2010)]{Behroozi2010} Behroozi, P.~S., 
	Conroy, C., \& Wechsler, R.~H.\ 2010, \apj, 717, 379
\bibitem[Bell et al.(2003)]{Bell2003} Bell, E.~F., McIntosh, D.~H., Katz, N., \& Weinberg, M.~D.\ 2003, \apjs, 149, 289
\bibitem[Bigiel et al.(2008)]{Bigiel2008} Bigiel, F., Leroy, A., Walter, F., et al.\ 2008, \aj, 136, 2846
\bibitem[Bigiel et al.(2010a)]{Bigiel2010} Bigiel, F., Leroy, A., Walter, F., et al.\ 2010, \aj, 140, 1194
\bibitem[Bigiel et al.(2010b)]{Bigiel2010b} Bigiel, F., Leroy, A., Seibert, M., et al.\ 2010, \apjl, 720, L31
\bibitem[Bigiel et al.(2011)]{Bigiel2011} Bigiel, F., Leroy, A.~K., Walter, F., et al.\ 2011, \apjl, 730, L13
\bibitem[Blanton \& Roweis(2007)]{Blanton2007} Blanton, M.~R., \& Roweis, S.\ 2007, \aj, 133, 734
\bibitem[Blanton \& Moustakas(2009)]{Blanton2009} Blanton, M.~R., \& Moustakas, J.\ 2009, \araa, 47, 159
\bibitem[Blanton et al.(2011)]{Blanton2011} Blanton, M.R., Kazin, E., Muna, D., Weaver, B.A., \& Price-Whelan, A.
         2011, \aj, 142, 31
\bibitem[Boissier \& Prantzos(2000)]{Boissier2000} Boissier, S., \& Prantzos, N.\ 2000, \mnras, 312, 398
\bibitem[Boselli et al.(2001)]{Boselli2001} Boselli, A., Gavazzi, G., Donas, J., \& Scodeggio, M.\ 2001, \aj, 121, 753
\bibitem[Bothun et al.(1987)]{Bothun1987} Bothun, G.~D., Impey, C.~D., Malin, D.~F., \& Mould, J.~R.\ 1987, \aj, 94, 23
\bibitem[Bothwell et al.(2009)]{Bothwell2009} Bothwell, M.~S., 
	Kennicutt, R.~C., \& Lee, J.~C.\ 2009, \mnras, 400, 154 
\bibitem[Brinchmann et al.(2004)]{Brinchmann2004} Brinchmann, J., Charlot, S., White, S.~D.~M., et al.\ 2004, \mnras, 351, 1151
\bibitem[Calzetti et al.(1994)]{Calzetti1994} Calzetti, D., Kinney, 
	A.~L., \& Storchi-Bergmann, T.\ 1994, \apj, 429, 582
\bibitem[Cannon et al.(2011)]{Cannon2011} Cannon, J.~M., 
	Giovanelli, R., Haynes, M.~P., et al.\ 2011, \apjl, 739, L22
\bibitem[Cardelli et al.(1989)]{Cardelli1989} Cardelli, J.~A., 
	Clayton, G.~C., \& Mathis, J.~S.\ 1989, \apj, 345, 245 
\bibitem[Catinella et al.(2008)]{Catinella2008} Catinella, B., Haynes, M.~P., Giovanelli, R., Gardner, J.~P., 
	\& Connolly, A.~J.\ 2008, \apjl, 685, L13
\bibitem[Catinella et al.(2010)]{Catinella2010} Catinella, B., Schiminovich, D., Kauffmann, G., et al.\ 2010, \mnras, 403, 683
\bibitem[Chabrier(2003)]{Chabrier2003} Chabrier, G.\ 2003, \pasp, 115, 763
\bibitem[Cortese et al.(2006)]{Cortese2006} Cortese, L., Boselli, A., Buat, V., et al.\ 2006, \apj, 637, 242
\bibitem[Cortese \& Hughes(2009)]{Cortese2009} Cortese, L., \& Hughes, T.~M.\ 2009, \mnras, 400, 1225
\bibitem[Cortese et al.(2011)]{Cortese2011} Cortese, L., Catinella, 
	B., Boissier, S., Boselli, A., \& Heinis, S.\ 2011, arXiv:1103.5889
\bibitem[Courteau(1997)]{Courteau1997} Courteau, S.\ 1997, \aj, 114, 2402
\bibitem[Cowie et al.(1996)]{Cowie1996} Cowie, L.~L., Songaila, A., Hu, E.~M., \& Cohen, J.~G.\ 1996, \aj, 112, 839
\bibitem[Charlot \& Fall(2000)]{Charlot2000} Charlot, S., \& Fall, S.~M.\ 2000, \apj, 539, 718
\bibitem[Davies \& Lewis(1973)]{Davies1973} Davies, R.~D., \& Lewis, B.~M.\ 1973, \mnras, 165, 231
\bibitem[Disney \& Phillipps(1987)]{Disney1987} Disney, M., \& Phillipps, S.\ 1987, \nat, 329, 203
\bibitem[Disney et al.(2008)]{Disney2008} Disney, M.~J., Romano, 
	J.~D., Garcia-Appadoo, D.~A., et al.\ 2008, \nat, 455, 1082
\bibitem[Donovan et al.(2009)]{Donovan2009} Donovan, J.L., Serra, P., van Gorkom, J.H., et al.\ 2009, \apj, 137, 5037
\bibitem[Draine et al.(2007)]{Draine2007} Draine, B.~T., Dale, D.~A., Bendo, G., et al.\ 2007, \apj, 663, 866
\bibitem[Fabello et al.(2011)]{Fabello2011} Fabello, S., Catinella, B., 
	Giovanelli, R., et al.\ 2011, \mnras, 411, 993
\bibitem[Fu et al.(2010)]{Fu2010} Fu, J., Guo, Q., Kauffmann, G., \& Krumholz, M.~R.\ 2010, \mnras, 409, 515
\bibitem[Gallazzi et al.(2005)]{Gallazzi2005} Gallazzi, A., Charlot, 
	S., Brinchmann, J., White, S.~D.~M., \& Tremonti, C.~A.\ 2005, \mnras, 362, 41
\bibitem[Garcia-Appadoo et al.(2009)]{Garcia-Appadoo2009} Garcia-Appadoo, 
	D.~A., West, A.~A., Dalcanton, J.~J., Cortese, L., \& Disney, M.~J.\ 2009, \mnras, 394, 340
\bibitem[Gavazzi et al.(1996)]{Gavazzi1996} Gavazzi, G., Pierini, D., \& Boselli, A.\ 1996, \aap, 312, 397
\bibitem[Gavazzi et al.(2012a)]{Gavazzi2012a} Gavazzi1, G., Fumagalli, M., Galardo, V., et al.\ 2012, in preparation
\bibitem[Gavazzi et al.(2012b)]{Gavazzi2012b} Gavazzi1, G., Fumagalli, M., Galardo, V., et al.\ 2012, in preparation
\bibitem[Giovanelli \& Haynes(1985)]{Giovanelli1985} Giovanelli, R., \& Haynes, M.~P.\ 1985, \apj, 292, 404
\bibitem[Giovanelli et al.(1995)]{Giovanelli1995} Giovanelli, R., Haynes, M.~P., Salzer, J.~J., et al.\ 1995, \aj, 110, 1059
\bibitem[Giovanelli et al.(1997)]{Giovanelli1997} Giovanelli, R., Haynes, M.~P., Herter, T., et al.\ 1997, \aj, 113, 22
\bibitem[Giovanelli et al.(2005a)]{Giovanelli2005a} Giovanelli, R., 
	Haynes, M.~P., Kent, B.~R., et al.\ 2005a, \aj, 130, 2598
\bibitem[Giovanelli et al.(2005b)]{Giovanelli2005b} Giovanelli, R., 
	Haynes, M.~P., Kent, B.~R., et al.\ 2005b, \aj, 130, 2613
 \bibitem[Giovanelli et al.(2007)]{Giovanelli2007} Giovanelli, R., 
	Haynes, M.~P., Kent, B.~R., et al.\ 2007, \aj, 133, 2569
\bibitem[Giovanelli et al.(2010)]{Giovanelli2010} Giovanelli, R., 
	Haynes, M.~P., Kent, B.~R., \& Adams, E.~A.~K.\ 2010, \apjl, 708, L22
\bibitem[Gunawardhana et al.(2011)]{Gunawardhana2011} Gunawardhana, 
	M.~L.~P., Hopkins, A.~M., Sharp, R.~G., et al.\ 2011, \mnras, 415, 1647
\bibitem[Guo et al.(2010)]{Guo2010} Guo, Q., White, S., Li, C., \& Boylan-Kolchin, M.\ 2010, \mnras, 404, 1111 
\bibitem[Hallenbeck et al.(2012)]{Hallenbeck2012} Hallenbeck, G., Papastergis, E., Huang, S., et al.\ 2012, submitted
\bibitem[Haynes et al.(2011)]{Haynes2011} Haynes, M.~P., Giovanelli, R., Martin, A.~M., et al.\ 2011, \aj, 142, 170
\bibitem[Hernandez \& Cervantes-Sodi(2006)]{Hernandez2006} Hernandez, X., \& Cervantes-Sodi, B.\ 2006, \mnras, 368, 351
\bibitem[Hernandez et al.(2007)]{Hernandez2007} Hernandez, X., Park, C., Cervantes-Sodi, B., \& Choi, Y.-Y.\ 2007, \mnras, 375, 163
\bibitem[Hoeft et al.(2006)]{Hoeft2006} Hoeft, M., Yepes, G., Gottl{\"o}ber, S., \& Springel, V.\ 2006, \mnras, 371, 401
\bibitem[Hogg et al.(2003)]{Hogg2003} Hogg, D.~W., Blanton, 
	M.~R., Eisenstein, D.~J., et al.\ 2003, \apjl, 585, L5
\bibitem[Huang et al.(2012)]{Huang2012} Huang, S., Haynes, M.~P., Giovanelli, R., et al.\ 2012, \aj, 143, 133
\bibitem[Impey \& Bothun(1997)]{Impey1997} Impey, C., \& Bothun, G.\ 1997, \araa, 35, 267
\bibitem[Kannappan(2004)]{Kannappan2004} Kannappan, S.~J.\ 2004, \apjl, 611, L89
\bibitem[Kannappan et al.(2009)]{Kannappan2009} Kannappan, S.~J., 
	Guie, J.~M., \& Baker, A.~J.\ 2009, \aj, 138, 579
\bibitem[Kauffmann et al.(2003)]{Kauffmann2003} Kauffmann, G., Heckman, T.~M., White, S.~D.~M., et al.\ 2003, \mnras, 341, 33
\bibitem[Kennicutt(1998)]{Kennicutt1998} Kennicutt, R.~C., Jr.\ 1998, \apj, 498, 541
\bibitem[Kere{\v s} et al.(2005)]{Keres2005} Kere{\v s}, D., 
	Katz, N., Weinberg, D.~H., \& Dav{\'e}, R.\ 2005, \mnras, 363, 2 
\bibitem[Klypin et al.(2011)]{Klypin2011} Klypin, A.~A., Trujillo-Gomez, S., \& Primack, J.\ 2011, \apj, 740, 102
\bibitem[Kong et al.(2004)]{Kong2004} Kong, X., Charlot, S., 
	Brinchmann, J., \& Fall, S.~M.\ 2004, \mnras, 349, 769
\bibitem[Krumholz et al.(2008)]{Krumholz2008} Krumholz, M.~R., McKee, C.~F., \& Tumlinson, J.\ 2008, \apj, 689, 865
\bibitem[Krumholz et al.(2009a)]{Krumholz2009} Krumholz, M.~R., McKee, C.~F., \& Tumlinson, J.\ 2009, \apj, 693, 216
\bibitem[Krumholz et al.(2009b)]{Krumholz2009b} Krumholz, M.~R., McKee, C.~F., \& Tumlinson, J.\ 2009, \apj, 699, 850
\bibitem[Krumholz et al.(2011a)]{Krumholz2011} Krumholz, M.~R., Dekel, A., \& McKee, C.~F.\ 2011, arXiv:1109.4150
\bibitem[Krumholz et al.(2011b)]{Krumholz2011b} Krumholz, M.~R., Leroy, A.~K., \& McKee, C.~F.\ 2011, \apj, 731, 25
\bibitem[Lee et al.(2007)]{Lee2007} Lee, J.~C., Kennicutt, 
	R.~C., Funes, S.~J., Jos{\'e} G., Sakai, S., \& Akiyama, S.\ 2007, \apjl, 671, L113 
\bibitem[Leroy et al.(2009)]{Leroy2009} Leroy, A.~K., Walter, F., 
	Bigiel, F., et al.\ 2009, \aj, 137, 4670
\bibitem[Mannucci et al.(2010)]{Mannucci2010} Mannucci, F., Cresci, 
	G., Maiolino, R., Marconi, A., \& Gnerucci, A.\ 2010, \mnras, 408, 2115
\bibitem[Marinacci et al.(2010)]{Marinacci2010} Marinacci, F., Binney, J., Fraternali, F., et al.\ 2010, \mnras, 404, 1464
\bibitem[Martin et al.(2010)]{Martin2010} Martin, A.~M., Papastergis, E., Giovanelli, R., et al.\ 2010, \apj, 723, 1359
\bibitem[Martin et al.(2012)]{Martin2012} Martin, A.~M., Giovanelli, R., Haynes, M.~P., \& Guzzo, L.\ 2012, \apj, 750, 38
\bibitem[Masters et al.(2010)]{Masters2010} Masters, K.~L., Mosleh, M., 
	Romer, A.~K., et al.\ 2010, \mnras, 405, 783
\bibitem[Meyer et al.(2004)]{Meyer2004} Meyer, M.~J., Zwaan, M.~A., 
	Webster, R.~L., et al.\ 2004, \mnras, 350, 1195 
\bibitem[Mo et al.(1998)]{Mo1998} Mo, H.~J., Mao, S., \& White, S.~D.~M.\ 1998, \mnras, 295, 319
\bibitem[Moran et al.(2010)]{Moran2010} Moran, S.~M., Kauffmann, G., Heckman, T.~M., et al.\ 2010, \apj, 720, 1126
\bibitem[Morganti et al.(2006)]{Morganti2006} Morganti, R., de Zeeuw, P.~T., Oosterloo, T.~A., et al.\ 2006, \mnras, 371, 157
\bibitem[Morrissey et al.(2007)]{Morrissey2007} Morrissey, P., Conrow, T., Barlow, T.~A., et al.\ 2007, \apjs, 173, 682
\bibitem[Obreschkow et al.(2009)]{Obreschkow2009} Obreschkow, D., Kl{\"o}ckner, H.-R., Heywood, I., Levrier, F., 
	\& Rawlings, S.\ 2009, \apj, 703, 1890
\bibitem[Oosterloo et al.(2010)]{Oosterloo2010} Oosterloo, T.~A., Morganti, R., Crocker, A. et al.\ 2010, \mnras, 409, 500
\bibitem[Ostriker et al.(2010)]{Ostriker2010} Ostriker, E.~C., McKee, C.~F., \& Leroy, A.~K.\ 2010, \apj, 721, 975
\bibitem[Papastergis et al.(2011)]{Papastergis2011} Papastergis, E., Martin, A.~M., 
	Giovanelli, R., \& Haynes, M.~P.\ 2011, \apj, 739, 38
\bibitem[Park \& Choi(2005)]{Park2005} Park, C., \& Choi, Y.-Y.\ 2005, \apjl, 635, L29
\bibitem[Roberts(1963)]{Roberts1963} Roberts, M.~S.\ 1963, \araa, 1, 149
\bibitem[Portas et al.(2010)]{Portas2010} Portas, A.~M., Brinks, E., Filho, M.~E., et al.\ 2010, \mnras, 407, 1674
\bibitem[Saintonge et al.(2011a)]{Saintonge2011} Saintonge, A., Kauffmann, G., Kramer, C., et al.\ 2011, \mnras, 415, 32
\bibitem[Saintonge et al.(2011b)]{Saintonge2011b} Saintonge, A., Kauffmann, G., Wang, J., et al.\ 2011, \mnras, 415, 61
\bibitem[Salim et al.(2005)]{Salim2005} Salim, S., Charlot, S., Rich, R.~M., et al.\ 2005, \apjl, 619, L39
\bibitem[Salim et al.(2007)]{Salim2007} Salim, S., Rich, R.~M., Charlot, S., et al.\ 2007, \apjs, 173, 267
\bibitem[Sandage(1986)]{Sandage1986} Sandage, A.\ 1986, \aap, 161, 89
\bibitem[Schiminovich et al.(2007)]{Schiminovich2007} Schiminovich, D., Wyder, T.~K., Martin, D.~C., et al.\ 2007, \apjs, 173, 315
\bibitem[Schiminovich et al.(2010)]{Schiminovich2010} Schiminovich, D., Catinella, B., Kauffmann, G., et al.\ 2010, \mnras, 408, 919
\bibitem[Schlegel et al.(1998)]{Schlegel1998} Schlegel, D.~J., 
	Finkbeiner, D.~P., \& Davis, M.\ 1998, \apj, 500, 525
\bibitem[Schruba et al.(2011)]{Schruba2011} Schruba, A., Leroy, A.~K., Walter, F., et al.\ 2011, \aj, 142, 37
\bibitem[Serra et al.(2012)]{Serra2012} Serra, P., Oosterloo, T., Morganti, R., et al.\ 2012, \mnras, 422, 1835
\bibitem[Solanes et al.(2002)]{Solanes2002} Solanes, J.~M., 
	Sanchis, T., Salvador-Sol{\'e}, E., Giovanelli, R., \& Haynes, M.~P.\ 2002, \aj, 124, 2440
\bibitem[Springob et al.(2005)]{Springob2005} Springob, C.~M., Haynes, M.~P., \& Giovanelli, R.\ 2005, \apj, 621, 215
\bibitem[Strauss et al.(2002)]{Strauss2002} Strauss, M.~A., 
	Weinberg, D.~H., Lupton, R.~H., et al.\ 2002, \aj, 124, 1810
\bibitem[Toribio et al.(2011)]{Toribio2011} Toribio, M.~C., 
	Solanes, J.~M., Giovanelli, R., Haynes, M.~P., \& Martin, A.~M.\ 2011, \apj, 732, 93
\bibitem[Tremonti et al.(2004)]{Tremonti2004} Tremonti, C.~A., Heckman, T.~M., Kauffmann, G., et al.\ 2004, \apj, 613, 898
\bibitem[van Driel et al.(1988)]{vanDriel1988} van Driel, W., van Woerden, H., Schwarz, U.~J., \& Gallagher, J.~S., III 1988, \aap, 191, 201
\bibitem[Walter et al.(2008)]{Walter2008} Walter, F., Brinks, E., de Blok, W.~J.~G., et al.\ 2008, \aj, 136, 2563 
\bibitem[Wang et al.(2011)]{Wang2011} Wang, J., Kauffmann, G., Overzier, R., et al.\ 2011, \mnras, 412, 1081
\bibitem[Wardle \& Knapp(1986)]{Wardle1986} Wardle, M., \& Knapp, G.~R.\ 1986, \aj, 91, 23
\bibitem[West et al.(2009)]{West2009} West, A.~A., Garcia-Appadoo, D.~A., Dalcanton, J.~J., et al.\ 2009, \aj, 138, 796
\bibitem[West et al.(2010)]{West2010} West, A.~A., Garcia-Appadoo, D.~A., Dalcanton, J.~J., et al.\ 2010, \aj, 139, 315
\bibitem[Wolfire et al.(2010)]{Wolfire2010} Wolfire, M.~G., Hollenbach, D., \& McKee, C.~F.\ 2010, \apj, 716, 1191
\bibitem[Wong et al.(2006)]{Wong2006} Wong, O.~I., Ryan-Weber, 
	E.~V., Garcia-Appadoo, D.~A., et al.\ 2006, \mnras, 371, 1855 
\bibitem[Wyder et al.(2007)]{Wyder2007} Wyder, T.~K., Martin, D.~C., Schiminovich, D., et al.\ 2007, \apjs, 173, 293
\bibitem[Zhang et al.(2009)]{Zhang2009} Zhang, W., Li, C., Kauffmann, G., et al.\ 2009, \mnras, 397, 1243 
\end{thebibliography}
\end{document}